\documentclass[12pt]{report}
\usepackage[utf8]{inputenc}
\usepackage[margin=2.5cm]{geometry}
\usepackage{setspace}
\usepackage{graphicx}
\usepackage[toc,page]{appendix}
\graphicspath{ {Images/} }
\usepackage[colorlinks=true, linkcolor  = blue,urlcolor=blue,citecolor=violet]{hyperref}
\usepackage{xcolor}
\usepackage{amsmath, amssymb}
\usepackage{float}
\usepackage{physics}
\usepackage{subfigure}
\usepackage{cite}
\usepackage[font=small,labelfont=bf]{caption}
\usepackage[bottom]{footmisc}

\author{Eleanor Harris}
\date{June 2023}

\begin{document}

\begin{titlepage}
    \begin{center}
        \vspace*{1cm}
            
        \Huge
        \textbf{Quantum Features of the Cosmological Horizon}
            
        \vspace{1.5cm}
        \LARGE
        Eleanor Harris
            
        \vfill

        \large
        A thesis presented for the degree of\\
        Doctor of Philosophy
            
        \vspace{0.8cm}
            
        \includegraphics[width=0.3\textwidth]{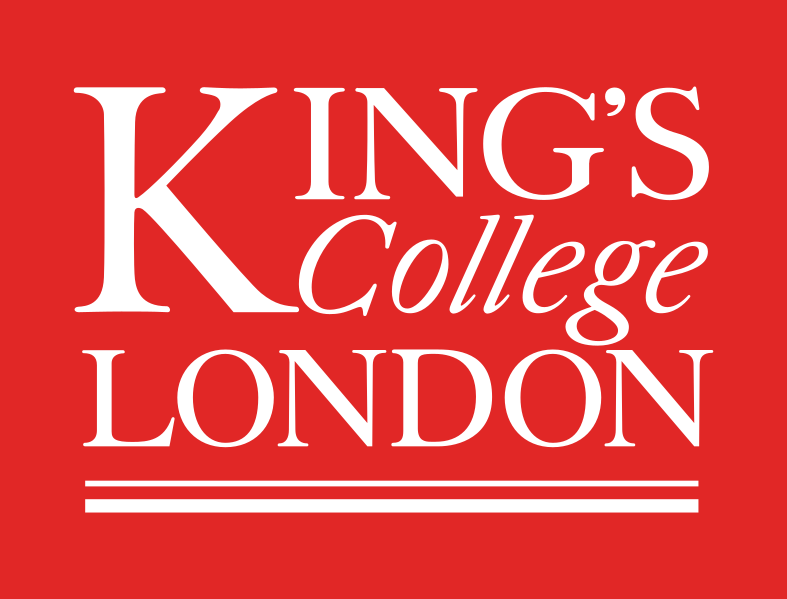}\\
        \vspace{0.5cm}
        \large
        Department of Mathematics\\
        King's College London\\
        United Kingdom\\
        June 2023
            
    \end{center}
\end{titlepage}

\chapter*{Abstract}
This thesis explores the thermodynamics of the cosmological horizon, aiming to make progress towards a better understanding of the microscopic nature of its entropy. We utilise the constrained nature of low-dimensional theories of gravity to do so and investigate timelike boundaries in these theories, with an emphasis on the stretched horizon holographic picture. Throughout, we make use of the Euclidean continuation of de Sitter to the sphere. \\ 
\\
Given the success of the AdS/CFT correspondence, one might attempt to embed a piece of de Sitter inside an anti-de Sitter geometry, and then describe the expanding region from the AdS boundary. In two dimensions, there exist dilaton potentials that give rise to such geometries and, by studying this problem in the presence of a timelike boundary, we demonstrate how to stabilise such solutions against thermal fluctuations. We then propose a dual matrix model living at the AdS boundary that should describe the interior spacetime, including the expanding region. Having proposed such a duality, a key method for testing it would be to compare correlation functions in the bulk and boundary. However, for de Sitter this poses a puzzle: in the saddle point approximation two-point functions can be calculated as a weighted sum over geodesic lengths. Such correlation functions are known to exist for any two points in de Sitter, but geodesics do not exist between arbitrary points. The puzzle is resolved by including complex length saddles that appear upon analytic continuation from the sphere. \\
\\
Additionally, one can use the semi-classical relationship between three-dimensional gravity and Chern-Simons theory to explore thermodynamic contributions to the de Sitter horizon. The Euclidean gravitational path integral provides the exact, all-loop quantum corrected de Sitter entropy. To better understand the microscopic origin of this entropy, one hopes to find an analogous Lorentzian calculation that produces this result. This takes the form of an edge-mode theory living close to the cosmological horizon with a complexified gauge group, leading to an unbounded spectrum. To make progress, we study an analogous Abelian theory and see that the there is an entanglement character to the entropy of the horizon. Finally, we summarise some new results on edge-mode theories arising from general boundary conditions.

\chapter*{Acknowledgements}
I would like to begin by expressing my immeasurable gratitude to my supervisor, Dionysios Anninos. Your seemingly infinite passion for physics has truly inspired me throughout the last four years, and whenever I’m in any doubt, a conversation with you always reminds me to just focus on the physics and remember that it’s incredible that we get paid to play around with maths all day. I would like to thank you for your kind and patient approach to mentorship; for always making yourself available for silly questions, and for telling me never to apologise for asking them. In a poorly-constructed de Sitter pun, there seems to be no boundary to the amount of support and encouragement you give your students. I am furthermore very appreciative for the fun and interesting projects that you have proposed over the course of this doctorate, which as well as being engaging have certainly expanded my horizons into new areas of physics. I could absolutely not have done this without your supervision. \\
\\
Secondly, I would like to thank my amazing collaborators who worked on the various projects that contributed to this thesis; Shira Chapman, Kévin Nguyen, Sameer Sheorey, and David Vegh. You were all wonderful and very fun to work with. I would like to express particular gratitude to Damián Galante for always making himself available to answer questions and give advice. More broadly, I would also like to express my appreciation for the theory group at King’s College London; it has been a pleasure to conduct research in such a friendly and open environment. I would further like to thank STFC for financially supporting this doctorate. \\
\\
Of course, it would be remiss of me not to acknowledge my fellow PhD students at King’s, whose unfaltering willingness to go to the pub at the end of a long day was vital to the completion of this PhD. The group is too big to list all your names individually, but you all know who you are. A special shout-out to Ben Pethybridge though for offering to proofread parts of this thesis for me. \\
\\
I would also like to thank my family: to my mum for endlessly supporting me and always knowing the right words to say at the end of a tough week; to my dad for believing in me and keeping me inspired with a constant stream of wonderful pop-science articles; and to my aunt Rosie, whose encouragement and kindness was integral to getting me to the end of this PhD. \\
\\
Moreover, I also would like to express my gratitude to my partner Mike, for doing a very good impression of someone who is interested in hearing about niche topics in theoretical physics for four years straight. In all seriousness, your calm and collected attitude to every hurdle and your untiring encouragement and belief in me has been invaluable in helping me complete this PhD. To say I could not have done this without you is a huge understatement. \\
\\
And last, but definitely not least, I would like to thank Abby. You may not have known much about the expansion of the universe, but my world is certainly much bigger for having had you in it. 

\tableofcontents

\chapter*{Introduction}
\addcontentsline{toc}{chapter}{\Large Introduction}  
For more than twenty years we have had observational evidence that the expansion of the universe is accelerating, driven by a mysterious source which we refer to as dark energy \cite{SupernovaSearchTeam:1998fmf,SupernovaCosmologyProject:1998vns,Astier:2012ba}. This accelerated expansion causes galaxies to recede away from us in all directions, with the furthest away galaxies retreating the fastest. If a galaxy is far enough away such that the light it emits cannot overcome the expansion in order to reach us, it will be hidden behind our cosmological horizon – the ultimate barrier to our observations of the cosmos. Eventually, the expansion will push more and more galaxies beyond our cosmological horizon, and future civilisations may believe that the entire universe consists of just the stars of the Milky Way and Andromeda, since these will be all that they see when they peer up at the night sky. Such a future civilisation may not even be able to detect the presence of dark energy at all since they will have no distant supernovae to use to detect it, as we do today. On long enough timescales even the radiation of the cosmic microwave background - the imprint of light from the very early universe - will dwindle behind the cosmological horizon, cutting the inhabitants of Milkdromeda off from the history of the universe. Fortunately, we happen to live in an era where we can observe this accelerated expansion, and we are therefore conscious of the fact that a complete understanding of the universe must incorporate this phenomenon.\\
\\
In fact, it seems that this is not the only occasion in the history of the universe that it has exhibited accelerated expansion. The first was the epoch directly after the Big Bang, where the universe rapidly increased in size to $10^{78}$ times its original volume in only around $10^{-33}$ seconds \cite{Dodelson:2003vq,Liddle:2003as}. We call this era inflation \cite{Guth:1980zm,Linde:1981mu,Albrecht:1982wi,Guth:2007ng}, and it is clearly a much more rapid flavour of expansion to that which we experience today, yet we are able to model both periods using the same mathematical description. A universe exhibiting accelerated expansion is described by solving Einstein’s equations with a positive cosmological constant, denoted by $\Lambda>0$, the result of which is the de Sitter solution \cite{1917KNAB...19.1217D}. After inflation ended, the expansion of the universe slowed until dark energy started to drive the uptick that we see today. Experiments indicate that the observed value of the cosmological constant in our universe is $\Lambda_{\text{Obs}} \sim 10^{-52} m^{-2}$; an incredibly small, but notably positive quantity. \\
\\
The cosmological horizon is a fascinating feature that begs to be better understood. The more familiar situation in which horizons occur in physics is in the study of black holes, which are comparatively understood in greater detail. There are certainly some key differences between the black hole and cosmological horizons, perhaps amongst the more glaring of which is their behaviour with respect to an observer. The position of the black hole horizon is independent of where we are when we measure it, while if one tries to move towards their cosmological horizon, it will always move away in such a way as to keep them in the centre. However, there are also some key similarities between the two. Bekenstein and Hawking proposed that we should associate an entropy to a black hole that is proportional to the surface area of its horizon \cite{Bekenstein:1972tm,Hawking:1975vcx}. This entropy can be thought of as quantifying the lack of information that we have about the system; it is counting the possible configurations of all the particles that are inaccessible to us behind the black hole horizon. It is therefore somewhat surprising that this entropy is proportional to the area of the black hole horizon, rather than its volume, given one would expect that all the possible configurations of matter inside the black hole would fill its entire interior. In a complete theory of quantum gravity, we should be able to perform this counting of configurations (known as microstates) precisely, and we even know how to do so in some highly symmetric setups. Gibbons and Hawking proposed that, similarly to the black hole, we should also associate an entropy to the cosmological horizon, and that this entropy should also follow an area law \cite{Gibbons:1976ue,Gibbons:1977mu}. However, it is fair to say that the question of whether a horizon should have an associated entropy is much less well understood for the cosmological case. It is therefore important to search for a theory of quantum gravity that describes expanding spacetimes, since if we could achieve this then we would be able to count the microstates of the cosmological horizon precisely and thus verify the proposal. In this thesis, we will be examining the hypothesis that such a theory exists.\\
\\
As already mentioned, accelerated expansion can be modelled by the de Sitter solution, which has $\Lambda >0$, and it is therefore quantum features of this spacetime with which this body of work is concerned. We already have a good understanding of how quantum gravity works in a very closely related spacetime, which is appropriately known as anti-de Sitter since it is the solution to Einstein’s equations with a negative cosmological constant, $\Lambda <0$. The key to understanding quantum features of anti-de Sitter space lay in uncovering the correspondence between anti-de Sitter space and conformal field theory, which is abbreviated to AdS/CFT \cite{Maldacena:1997re}. This correspondence states that a theory of quantum gravity in anti-de Sitter space can be completely described by a non-gravitating theory that resides at the boundary of AdS. This reduction in dimension of spacetime is known as holography, reflecting the fact that the volume’s worth of information in the interior is expressed as a hologram on the surface area at the boundary. This harks back to our description of the black hole as having an entropy proportional to its surface area, rather than the volume of its interior. Indeed, holography has been very successfully applied to the case of black holes by, for example, studying black holes in anti-de Sitter spacetime \cite{Banados:1992wn,Benini:2015eyy,Cabo-Bizet:2018ehj}. We are very fortunate to be able to describe quantum gravity in any theory of gravity, but somewhat unfortunate in the sense that where we have been able to do so, it is not for the spacetime that describes the universe that we inhabit. It is still very much an open question whether it is possible to describe de Sitter holographically in a similar way to that which has been achieved for anti-de Sitter. In this thesis, our aim is to take some steps towards understanding quantum features of expanding spacetimes, and hence build towards a better understanding of the microscopic origin of the de Sitter horizon entropy.\\
\\
One reason for the success of anti-de Sitter is that it possesses the virtue of a boundary at spatial infinity where gravity effectively ‘switches off’, allowing us to identify the holographic dual theory more easily. We call such a boundary timelike since it is extended in time. In de Sitter space, by contrast, the boundaries are in the far future and far past. This originally led to the dS/CFT proposal \cite{Witten:2001kn,Strominger:2001pn,Anninos:2011ui}, which states that a conformal field theory living in the far future of de Sitter should describe the physics of the interior. This was certainly a useful step forward and may have applications for describing early universe inflation, since for that de Sitter era the accelerated expansion came to an end. We therefore presently exist as ‘metaobservers’ of inflation since the far future of the early universe exists in our past today. However, this setup is not as useful for describing the period of dark energy driven expansion that we experience today since we are confined to our current cosmological horizon, and it is impossible to make measurements in the infinite future. Therefore, it is widely thought that we should search for a quantum theory of de Sitter that is intrinsic to the piece of the universe that we have access to inside our de Sitter horizon. There are two main proposals for this, which are known as worldline holography \cite{Anninos:2011af,Leuven:2018ejp} and stretched horizon holography \cite{Susskind:1993if,Bousso:1999cb,Banks:2006rx,Dong:2018cuv}. Both of these involve putting an artificial timelike boundary between us and our horizon. Worldline holography attempts to establish a holographic dual in the vicinity of an observer, who could be a person with a telescope on Earth, say. Stretched horizon holography, on the other hand, suggests that the natural placement of the holographic dual is very close to the cosmological horizon. Here, we will be investigating several versions of the latter setup. This picture has gained renewed interest in recent years \cite{Susskind:2021esx, Shaghoulian:2021cef, Shaghoulian:2022fop}, but it is fair to say that a deep understanding of holography in de Sitter is still lacking. \\
\\
When we encounter a difficult problem, our job as physicists is to simplify it as much as possible, and then once the simpler system is understood, we can attempt to gradually add details back in. When we calculate the orbits of planets, for example, we do not attempt to keep track of every atom in the solar system. The method of simplification that we will employ in this body of work is to remove some spacetime dimensions from the problem, and work in theories of low-dimensional gravity. These are theories consisting of two or even just one spatial dimension, along with one timelike dimension. This strategy has a long and fruitful history in many areas of physics \cite{Onsager:1943jn,Coleman:1973ci,tHooft:1974pnl, Witten:1988hc,Achucarro:1987vz,Belavin:1984vu,Banados:1992wn,Saad:2019lba}. The hope is that this lower dimensional system is constrained enough that we can perform experiments more concretely, and yet that we still retain some of the features that we are interested in studying in the higher dimensional theories. Indeed, this methodology has been particularly worthwhile in the analogous setup for black holes since theories of low-dimensional gravity often still exhibit horizons \cite{Saad:2019lba}. As we will see, the cosmological horizon is also retained in theories of lower-dimensional gravity, allowing us to more easily probe features that we are interested in understanding, such as the thermodynamic behaviour. Thus, eventually we may be able to uplift our findings to the real world of three space and one timelike direction. Occasionally, we will be able to extend our results to higher dimensions explicitly, but in other situations it may not yet be straightforward to do so. In such cases, it is not unreasonable to expect that the results we obtain will still be able to guide our understanding in four dimensions.\\
\\
The results of this thesis are split into two main parts. In Part \ref{Part1}, we discuss features of de Sitter in two dimensions. Recently, there has been an increased interest in applying the standard tools of AdS/CFT to probe the static patch of de Sitter. A non-comprehensive list includes \cite{Banks:2003cg, Banks:2004eb, Banks:2006rx, Parikh:2004wh, Banks:2018ypk, Geng:2019ruz,  Aalsma:2020aib, Geng:2020kxh, Aalsma:2021bit, Shyam:2021ciy, Coleman:2021nor,  Svesko:2022txo, Banihashemi:2022jys, Silverstein:2022dfj, Nomura:2017fyh, Nomura:2019qps, Murdia:2022giv}. One approach is to study geometries that flow between an anti-de Sitter boundary and a de Sitter interior. One could then attempt to probe the expanding region from the boundary. In dimensions $d>2$, solutions with dS$_d$ embedded within AdS$_d$ either defy the null energy condition, or contain black hole horizons which cloak the de Sitter piece of the geometry from being examined using the boundary theory \cite{Farhi:1989yr, Freivogel:2005qh}. However, in two dimensions there are no such impediments, and these solutions have been dubbed interpolating geometries \cite{Anninos:2017hhn, Anninos:2018svg}. The thermodynamics of such theories in the presence of a timelike boundary are examined in Chapter \ref{Interpolating}, where we demonstrate how to stabilise the solution under thermodynamic fluctuations. This progress in establishing a thermodynamically stable realisation of dS$_2$ inside AdS$_2$ enables us to propose concrete microscopic models that could live at the AdS$_2$ boundary. These are matrix models whose degrees of freedom we expect will capture some of the physics of the expanding piece of the bulk geometry. In Chapter \ref{ComplexGeodesics}, we look at quantum fields in the presence of a fixed de Sitter background. It is interesting to study whether scalar fields behave differently in the vicinity of a cosmological horizon compared to black holes. We utilise the worldline formalism and show that in order to correctly reproduce the large mass expansion of the correlator, it is necessary to include saddles of complex length in the geodesic approximation. These complex geodesics result in oscillatory behaviour in the two-point function, a behaviour which is not seen in the example of the AdS black hole in two dimensions. \\
\\
In Part \ref{Part2}, we study the de Sitter horizon in three dimensions, making use of the semi-classical relationship between Chern-Simons theory and three-dimensional gravity to explore thermodynamic contributions to the de Sitter horizon very precisely. In this setup, the Euclidean Chern-Simons path integral evaluated on the round three-sphere saddle gives the exact all-loop quantum corrected entropy of the de Sitter horizon. In Chapter \ref{SoftDeSitter}, the goal is to investigate a corresponding Lorentzian computation for this entropy. To make progress, we can begin with a simpler Abelian model that has a complexified gauge group structure similar to that required for three-dimensional gravity with a positive cosmological constant. It follows that there exists an edge-mode theory living on a timelike boundary close to the de Sitter horizon. The previously discovered topological entanglement entropy \cite{Kitaev:2005dm,Levin:2006zz} in Chern-Simons theory then has a natural interpretation as a contribution to the entropy of the de Sitter horizon. We subsequently consider more general boundary conditions for the same setup in Chapter \ref{GeneralBCs} and show that in fact there is no choice one can make to tame the unboundedness of the spectrum.\\
\\
The remainder of this introduction is dedicated to reviewing some general features of de Sitter that will be useful for the rest of the thesis. The results of Part \ref{Part1} first appeared in \cite{Anninos:2022hqo} and \cite{Chapman:2022mqd}. The content of Part \ref{Part2} first appeared in \cite{Anninos:2021ihe} and Chapter \ref{GeneralBCs} contains unpublished work with D. Anninos and K. Nguyen. At the beginning of each part, we will outline some more specific and technical details of gravity in two or three dimensions. The appendices contain further details of some of the calculations in the main text.
\newpage
\section*{Geometry of de Sitter}
\begin{figure}[H]
	\centering
	\includegraphics[width=6cm]{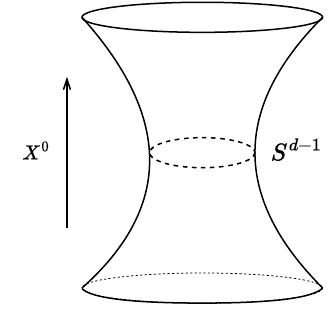}
	\caption{The $d$-dimensional hypersurface embedded in $\mathbb{R}^{1,d}$ that describes dS$_d$.}
	\label{fig:hyperbaloid}
\end{figure}

\begin{figure}[H]
	\centering
	\includegraphics[width=9.5cm]{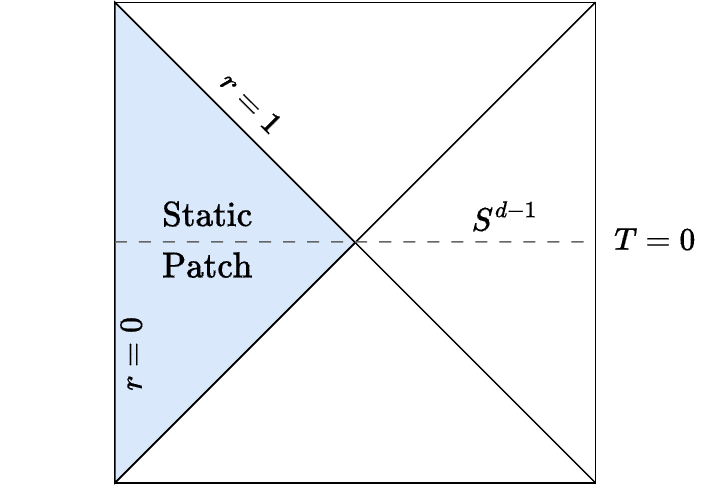}
	\caption{The Penrose diagram for de Sitter. Highlighted in blue is the static patch region which is casually accessible to an inertial observer at $r=0$. This observer experiences a cosmological horizon at $r=1$. Global time slices are $(d-1)$-spheres of radius $\ell \cosh T$, and the $T=0$ time slice has been indicated explicitly.}
	\label{fig:DdimPenrose}
\end{figure}

It will be useful to review the geometry of de Sitter in general dimensions before specialising to two and three spacetime dimensions, as required later on. The de Sitter geometry is the maximally symmetric solution to Einstein's equations with positive cosmological constant, the value of which depends on the spacetime dimension $d$ as
\begin{equation}
\Lambda = \frac{(d-2)(d-1)}{2\ell^2} \, .
\end{equation} 
Here $\ell$ characterises the curvature scale and is referred to interchangeably as the de Sitter length or the de Sitter radius. When $d=2$ we set $\Lambda = 1/\ell^2$. This solution has constant positive curvature given by
\begin{equation} \label{ricci}
    R = \frac{d(d-1)}{\ell^2} \, . 
\end{equation}
One can view $d$-dimensional de Sitter space dS$_d$ as the hypersurface embedded in $\mathbb{R}^{1,d}$ that is given by the equation 
\begin{equation} \label{ds embedding}
-X_0^2 + \sum_{i=1}^{d} X_i^2 = \ell^2 \, ,
\end{equation}
This hyperboloid is shown in Figure \ref{fig:hyperbaloid}. The hypersurface preserves an $SO(d,1)$ subgroup of the Poincar{\'e} symmetries of $\mathbb{R}^{1,d}$, and this subgroup constitutes the isometry group of dS$_d$. The de Sitter metric induced on this hypersurface is given by specific parameterisations of the $X_{\mu}$ coordinates in the flat Minkowski metric,
\begin{equation} \label{flatmetric R31}
ds^2  = -dX_0^2 + \sum_{i=1}^d \, dX_i^2 \, . 
\end{equation}
For example, the global coordinate patch is given by setting
\begin{equation} \label{globalembedding}
    X_0 = \ell \,\sinh T \, , \qquad X_i = \ell \, \omega_i \cosh T \, ,
\end{equation}
where $T \in \mathbb{R}$ is the global time coordinate and $\omega_i$ parametrises $S^{d-1}$. Plugging these coordinates into the $(d+1)$-dimensional Minkowski metric \eqref{flatmetric R31} then leads to the global metric of $d$-dimensional de Sitter space,
\begin{equation} \label{ddimGlobal}
\frac{ds^2}{\ell^2} = -dT^2 + \cosh^2 T \, d\Omega_{d-1}^2 \,,
\end{equation}
where $d\Omega_{d-1}^2$ is the round metric on unit $(d-1)$-sphere. One can see that constant $T$ slices are spheres that grow in size in the infinite past and future. 
\noindent An inertial observer in this spacetime does not have access to the full geometry and is instead confined inside a cosmological event horizon. The causal region that such an observer has access to is described by the metric,
\begin{equation} 
\label{ddimStatic}
\frac{ds^2}{\ell^2} = - (1-r^2) dt^2 + \frac{dr^2}{1-r^2} + r^2 d\Omega^2_{d-2} \,,
\end{equation}
where $t \in \mathbb{R}$ and $0 \leq r \leq 1$ for $d>2$. However, when $d=2$, the range of the radial coordinate is instead $-1 \leq r \leq 1$. The inertial observer sits at $r=0$ in this spacetime and experiences a cosmological horizon at $r = 1$, and it is this horizon which we would like to characterise the features of. Note that this metric does not explicitly depend on the coordinate $t$, and hence \eqref{ddimStatic} is referred to as the static patch metric. It is given by the choosing the embedding space coordinates in \eqref{flatmetric R31} to be
\begin{equation} \label{SPembedding}
    X_0 = \ell\sqrt{1-r^2} \sinh{t} \, , \qquad X_a = \ell r \omega_a \, , \;  a = 1 , \dots, d-1 \, , \qquad X_d = \ell \sqrt{1-r^2} \cosh{t} \, .
\end{equation}
The static patch region seen by an observer at $r=0$ is indicated by the shaded region on the Penrose diagram in Figure \ref{fig:DdimPenrose}. As already mentioned, we will be interested in timelike boundaries in the static patch, which correspond to lines of constant $r$ in \eqref{ddimStatic}. In particular, the stretched horizon will be a surface of constant $r$ that lies close to $r=1$.\\
\\
Of key importance to this body of work is the analytic continuation of de Sitter to Euclidean signature. Wick rotating $X_0 \rightarrow -i X_0$ in the embedding space description of de Sitter, equation \eqref{ds embedding} becomes the equation of an $S^d$ embedded in $\mathbb{R}^{d+1}$, and the isometries become $SO(d+1)$. Similarly, analytically continuing $T \to - i \theta$ in \eqref{ddimGlobal} gives 
\begin{equation} \label{d-dimsphere}
\frac{ds^2}{\ell^2} = d\theta^2 + \cos^2 \theta \, d\Omega_{d-1}^2 \,,
\end{equation}
which is the round metric on $S^d$. Surprisingly, analytically continuing equation \eqref{ddimStatic} by taking $t \rightarrow - i \tau$ also leads to the round metric on $S^d$ in a different coordinate system,
\begin{equation} \label{euclsp}
\frac{ds^2}{\ell^2} =  (1-r^2) d\tau^2 + \frac{dr^2}{1-r^2} + r^2 d\Omega^2_{d-2} \,,
\end{equation}
where in order to avoid a conical singularity, we must also periodically identify $\tau \sim \tau + 2 \pi$. The fact that this is still the metric on $S^d$ can be seen most clearly by performing this Wick rotation (along with $X_0 \rightarrow - i X_0$) in the choice of parameterisation in \eqref{SPembedding}, then verifying that these obey the equation of the $d$-sphere. Therefore, despite describing seemingly very different regions of the Penrose diagram in Figure \ref{fig:DdimPenrose}, the static patch and global metrics analytically continue to the same compact object in Euclidean signature, and upon evaluating the volume element, one can verify that they also have the same volume. The ‘observer dependence’ of the de Sitter horizon manifests itself in Euclidean signature by the fact that any $S^{d-1}$ path on the $S^d$ can be viewed as the Euclidean de Sitter horizon. \\
\\
In analogy with the fact that the Euclidean continuation of a black hole spacetime leads to a thermal description with characteristic temperature given by the periodicity of the time coordinate, here we can read off the temperature of de Sitter to be $T_{\text{dS}} = 1/2\pi$. This is the temperature as measured by an inertial observer at $r = 0$.  

\section*{Thermodynamics of de Sitter}
Given that de Sitter seems to have an associated temperature $T_{\text{dS}}$, we can ask about the thermal properties of this spacetime. This works in close analogy to black hole thermodynamics. Consider the usual form of the thermal partition function in the canonical ensemble, 
\begin{equation}
    \log \mathcal{Z} = S - \frac{E}{T}\, ,
\end{equation}
where $S$ is the entropy, $E$ is the total energy and $T$ is the temperature of the system which is kept fixed. Energy in general relativity is defined using the ADM formalism, which states that energy is given in terms of the asymptotic form of the metric. For the case of de Sitter, spatial slices are compact and therefore the ADM energy vanishes. This led to Gibbons and Hawking's proposal that the entropy of the de Sitter horizon should be given by the logarithm of the Euclidean gravitational path integral \cite{Gibbons:1976ue,Gibbons:1977mu}, 
\begin{equation} \label{GibHawk}
    \mathcal{Z}_{\text{grav}} =  \sum_{\mathcal{M}} \, \int \mathcal{D}g \, e^{-S_E[g_{\mu \nu}]} \, ,
\end{equation}
where $\mathcal{M}$ are compact manifolds and $S_E$ is the Euclidean Einstein-Hilbert action,
\begin{equation}
\label{E-Haction}
    S_E[g_{\mu \nu}] = - \frac{1}{16\pi G} \int d^d x \sqrt{g} \left(R - 2 \Lambda \right) \, . 
\end{equation}
Here $R$ is the Ricci scalar \eqref{ricci}, the metric $g$ is that of Euclidean de Sitter (the round $d$-sphere), and $G$ is Newton's constant. One can also include matter fields $\phi$ in the path integral, so that the above becomes 
\begin{equation} \label{GibHawk+matter}
    \mathcal{Z}_{\text{grav + matter}} =  \sum_{\mathcal{M}} \, \int \mathcal{D}g \, \mathcal{D} \phi \, e^{-S_E[g_{\mu \nu}, \phi ]} \, .
\end{equation}
This proposal has also been more recently explored in \cite{Anninos:2020hfj,Law:2020cpj,David:2021wrw}. We do not know how to compute this gravitational path integral exactly. However, equation \eqref{GibHawk} can be expanded in a semi-classical approximation around the round sphere saddle point solution, such that
\begin{equation}
    \log \mathcal{Z}_{\text{grav}} \approx -S_E = S_{\text{dS}} \, ,
\end{equation}
where $S_{\text{dS}}$ is the tree-level contribution to the de Sitter entropy, which is parametrically large in the semiclassical limit. We can use the semi-classical expansion to perform a simple check on the Gibbons-Hawking proposal. For example, for $d=4$, the Euclidean action \eqref{E-Haction} can be evaluated on-shell, leading to 
\begin{equation}
    S_{\text{dS}} = \frac{\pi \ell^2}{G} = \frac{\text{Area}}{4 G} \, ,
\end{equation}
matching the classical result that the entropy should be proportional to the area of the horizon. It is thought that the entropy of the cosmological horizon in our universe is of order $S_{\text{dS}} \approx 10^{122}$. Therefore, given that the microscopic origin of this entropy is still an open question, we are in search of a huge number of microstates.\\
\\
As previously mentioned, in this thesis we will be interested in the thermodynamics of timelike boundaries in de Sitter. In this case, the relevant temperature will be the Tolman temperature, which in the static patch coordinates \eqref{euclsp} has inverse
\begin{equation}
    \beta_T = \beta \sqrt{1-r_b^2} \, ,
\end{equation}
where $\beta = 1/T_{\text{dS}} = 2\pi$ is the periodicity of the $\tau$ coordinate and $-1 \leq r_b \leq 1$ is the position of the timelike boundary. We will explore the thermodynamics with respect to this temperature in the remainder of this body of work.

\part{Two spacetime dimensions} 
\label{Part1}
\chapter*{Introduction}
Given the enormous success we have had in describing quantum features of anti-de Sitter spacetimes, one might ask whether it is possible to embed a piece of de Sitter within an anti-de Sitter geometry, and then use the tools of the AdS/CFT correspondence to capture the physics of this internal, expanding region. In dimensions greater than two, it has been shown that such a set-up either defies the null energy condition or that this geometry can only be realised if the de Sitter region is surrounded by a black hole horizon \cite{Farhi:1989yr,Freivogel:2005qh}, making it difficult to probe its microscopic structure. However, in two dimensions such geometries have been constructed and are known as either interpolating geometries or as Centaur geometries \cite{Anninos:2017hhn,Anninos:2018svg, Chapman:2021eyy}. These geometries obey the null energy condition when viewed as pieces of higher-dimensional solutions \cite{Anninos:2020cwo} and offer the possibility of a holographic interpretation for the static patch of de Sitter. They can arise as dilaton-gravity models for certain choices of the dilaton potential. The solutions permit an asymptotically near-AdS$_2$ boundary of the type studied in \cite{Maldacena:2016upp}. For such a boundary, when the interpolating geometry ends at a dS$_2$ `black hole' horizon one must grapple with a horizon of negative specific heat. Alternatively, one can choose a dilaton profile that decreases towards the near-AdS$_2$ boundary \cite{Anninos:2017hhn} leading to a dS$_2$ `cosmological' horizon with positive specific heat, but now at the cost of a non-standard boundary behaviour for the dilaton. Neither of these features is an insurmountable obstacle in the search of a microscopic completion, but each adds an additional layer of difficulty. We will be focusing on the case where the AdS and dS regions are of the same dimension, but it is worth noting that another possibility is to consider geometries that interpolate between AdS$_d$ and dS$_D$ with $d \neq D$. Such a set-up was shown in \cite{Anninos:2022ujl} to obey the null energy condition for the case $d=2$ and $D=4$. The idea of rearranging the microphysical Hilbert space of quantum gravity in AdS to obtain de Sitter microstates also plays an interesting role in the approach of \cite{Gorbenko:2018oov,Coleman:2021nor,Shyam:2021ciy}, and \cite{Susskind:2021esx,Ecker:2022vkr}. 
\\ \\
In Chapter \ref{Interpolating}, we explore mechanisms for stabilising a portion of the dS$_2$ static patch within a near-AdS$_2$ world with ordinary boundary behaviour for the dilaton. We study Dirichlet boundary conditions at a finite proper distance from the horizon, following the method of York \cite{York:1986it,Whiting:1988qr} applied to the de Sitter case \cite{Hayward:1990zm,Wang:2001gt,Anninos:2011zn,Banihashemi:2022jys,Banihashemi:2022htw}.\footnote{Ordinary JT gravity with Dirichlet boundary conditions was considered, for example, in \cite{Gross:2019ach,Gross:2019uxi,Iliesiu:2020zld,Stanford:2020qhm, Griguolo:2021wgy}.} The temperature is fixed to be the Tolman temperature \cite{Tolman:1930zza}, which is the proper length of the Euclidean boundary $S^1$. We begin by considering the thermodynamics of the dimensional reduction of the near-Nariai solution along the lines of \cite{Anninos:2017hhn,Anninos:2018svg,Svesko:2022txo}. Throughout, we will be in Euclidean signature, in which the near-Nariai geometry analytically continues to the round $S^2$ with a running dilaton. Because the geometry caps off in this way, it is a natural setting to ask about the effects of a Dirichlet boundary cutting the two-sphere. When we select the solution for which the dilaton has standard boundary behaviour, the piece of the geometry containing the `black hole' dS$_2$ horizon is thermodynamically stable. The piece containing the `cosmological' dS$_2$ horizon, which has a decreasing dilaton profile, is thermodynamically unstable. \\
\\
We then consider the Dirichlet problem for the interpolating geometry, with standard boundary behaviour for the dilaton. We see that although for an asymptotic boundary the specific heat is negative, as the Dirichlet boundary is brought in a transition occurs rendering the specific heat positive, akin to a Hawking-Page phase transition \cite{Hawking:1982dh}. To preserve a locally thermally stable interpolating solution that is asymptotically near-AdS$_2$, with a large portion of Euclidean de Sitter in the interior, we construct interpolating geometries which include an additional piece of the Euclidean AdS$_2$ in the deep interior. Although the solution removes both Euclidean horizons of dS$_2$, we note that we can tune the dilaton potential in such a way that the dS$_2$ portion has a region that is parametrically close to one with a cosmological horizon, evoking a stretched horizon picture. In Euclidean signature, this excision corresponds to removing a small disc surrounding the Euclidean horizon. As such, we end up with a geometry that is both thermally stable and encodes a stretched `cosmological' dS$_2$ horizon, while being asymptotically near-AdS$_2$. The stability of this geometry allows us to posit the form of a dual theory living at the asymptotically AdS$_2$ boundary by considering the interpolating geometry to be a deformation of JT gravity \cite{Witten:2020wvy, Maxfield:2020ale,Eberhardt:2023rzz}. \\ 
\\
In Chapter \ref{ComplexGeodesics}, we shift focus slightly to study scalar two-point correlation functions in a fixed de Sitter background. Such correlation functions can give us an insight into how quantum fields behave in the presence of the cosmological horizon. Furthermore, a good understanding of two-point functions in de Sitter will allow us in the future to test the dual proposal put forward at the end of Chapter \ref{Interpolating}, by computing correlators in the bulk and comparing them to correlators computed in the boundary theory \cite{Jafferis:2022wez}.\\
\\
In the worldline formalism \cite{Schubert:2001he, Bastianelli:2005rc}, one can compute the correlator of a free massive scalar field between two points as a path integral over all paths connecting the points. In the large mass limit, one can take a saddle point approximation that reduces this path integral to a sum over geodesic lengths \cite{Bekenstein:1981xe, Parker:1979mf, Parker:2009uva}. This is surprising since it is known that not all points in a Lorentzian spacetime can be connected by geodesics \cite{Hawking:1973uf, Allen:1985wd}.\footnote{By the upper semi-continuity of arc-length, spacelike geodesics in Lorentzian manifolds always have (locally) maximal lengths \cite{Wald:1984rg}. However, it is possible to find scenarios where curve lengths connecting two points are unbounded from above, and in those cases, geodesics do not exist at all.}
For instance, in de Sitter space certain spacelike separated points are not connected by a geodesic. However, the two-point correlator for a free massive scalar field in the de Sitter invariant, Bunch-Davies state is known analytically for all points and all spacetime dimensions \cite{Spradlin:2001pw, Anninos:2012qw}. We solve this apparent tension by considering the analytic continuation of de Sitter to the sphere where there always exist two geodesics that form a great circle between any two points. We compute the lengths of these geodesics as well as the one-loop correction to the Euclidean correlator coming from quadratic fluctuations of the geodesic length. We show that for certain points, it is necessary to keep the contributions from both Euclidean geodesics, even though on the sphere only one has the shortest length. Upon analytically continuing back to Lorentzian signature, we reproduce the precise form of the large mass two-point function, up to an overall coefficient. The geodesics that previously seemed to be missing between certain spacelike separated points turn out to have complex length when analytically continued to Lorentzian signature from the sphere. We see that it is crucial to include these complex saddles in the geodesic approximation in order to correctly reproduce the large-mass form of the correlator. We primarily work in two dimensions for simplicity, but ultimately extend our results to any spacetime dimension. Early work on static patch correlation functions includes \cite{Balasubramanian:2002zh, Goheer:2002vf, Anninos:2011af}. Extremal surfaces in de Sitter (including geodesics in $d=3$) have been studied in \cite{Fischetti:2014uxa}, while other extended objects such as Wilson lines have been studied in \cite{Castro:2020smu}. The geodesic approximation for late time correlators in dS$_3$ has recently been studied in \cite{Hikida:2022ltr}, and \cite{Aalsma:2022eru} presents a related discussion on the role of complex geodesics in de Sitter space from the perspective of the heat kernal formalism. The geodesic approximation has also been widely used in the context of holography, starting with the work in \cite{Balasubramanian:1999zv, Louko:2000tp}, where boundary conformal correlators of heavy operators were reproduced both at zero and finite temperature from a bulk geodesic calculation. Geodesics have also been used to explore dynamical settings in holography, such as in \cite{Balasubramanian:2011ur, Aparicio:2011zy, Liu:2013iza}, and quantum chaos \cite{Shenker:2013pqa, Shenker:2013yza}.\\
\\
Chapter \ref{Interpolating} is organised as follows. In Section \ref{dilgravsec}, we introduce a general class of Euclidean dilaton-gravity theories and provide general formulae for their thermodynamic properties in the presence of both a Dirichlet boundary at finite proper distance, as well as for an asymptotically AdS$_2$ boundary.  In Section \ref{Near-Nariai geometry}, we utilise these formulae in the near-Nariai limit of the black hole in dS$_2$ which in Euclidean signature is an $S^2$. We show, as in \cite{Svesko:2022txo}, that the black hole is thermodynamically stable, while the cosmological horizon is thermodynamically unstable in this theory. In Section \ref{(A)dS$_2$ Interpolating geometries}, we study the Dirichlet problem for the interpolating geometry and show that this saddle may be both stable, and thermodynamically favoured, depending on where we place the Dirichlet wall. We also posit that a region within this geometry may be interpreted as a completion of the stretched dS$_2$ cosmological horizon to a near-AdS$_2$ boundary. In Section \ref{disec}, we introduce a dilaton-gravity theory permitting a solution which  interpolates to a dS$_2$ region while being AdS$_2$ in the deep interior as well as near the boundary. We show that this saddle has positive specific heat, while also having the virtue of being asymptotically AdS$_2$. First order phase transitions, reminiscent of the Hawking-Page transition \cite{Hawking:1982dh}, between the interpolating saddle and the AdS$_2$ black hole are discussed. Finally, in Section \ref{outlooksec}, we  discuss the possibility of realising a boundary matrix model or SYK-type model that may capture some features of the expanding piece of the geometry.  Appendices \ref{appAdS2} and \ref{Double Interpolating Geometry finite} summarise the thermodynamic properties of some further examples of dilaton potentials with finite boundaries.  \\
\\
Chapter \ref{ComplexGeodesics} is organised as follows. In Section \ref{sec_twopoint}, we review the computation of the Wightman correlator in de Sitter and compute its large mass expansion. In Section \ref{sec:geo}, we review the calculation of Lorentzian geodesics in de Sitter, recovering the result that not all spacelike separated points in de Sitter are connected by a real geodesic. In Section \ref{sec:sphere}, we move to the sphere and compute the Euclidean correlator in the geodesic approximation, including the one-loop corrections around each saddle point. We then extend our results to general dimensions. In Section \ref{lorentzian_props}, we analytically continue this result to Lorentzian signature to obtain the correct Lorentzian correlator. We apply our findings to several examples of timelike and spacelike geodesics, including those between opposite stretched horizons. In Appendix \ref{app_WKB}, we recover the asymptotic form of the two-point correlator from a WKB approximation. In Appendix \ref{path_integral_app}, we solve the quantum mechanical path integral needed to compute one-loop corrections to the Euclidean correlator. In Appendix \ref{app_ddim}, we discuss sphere geodesics in $d\geq 2$. Finally,  in Appendix \ref{app:details} we give details on the computation of correlations between stretched horizons.

\chapter{Interpolating between \texorpdfstring{dS$_2$}{dS2} and \texorpdfstring{AdS$_2$}{AdS2}}
\label{Interpolating}
In this chapter, we explore a class of dilaton-gravity models that exhibit a dS$_2$ interior region and demonstrate ways to stabilise such solutions against thermal fluctuations. In two dimensions, Newton's constant $G$ is dimensionless, and so in this chapter we choose to set $G = 1/8 \pi$ for the brevity of expressions.
\section{Two-dimensional dilaton-gravity theories}\label{dilgravsec}
In this section, we introduce the general class of models we will be interested in. These are two-dimensional dilaton-gravity models whose field content is given by a two-dimensional metric, $g_{\mu\nu}$, and the dilaton field, $\phi$. The Euclidean action governing the theory is given by
\begin{equation} \label{dilatongravity}
    S_E = S_0 -\frac{1}{2} \int_{\mathcal{M}} d^2x \sqrt{g} \left( \phi R + \frac{1}{\ell^2} V(\phi)\right) - \int_{\partial \mathcal{M}} d \tau \sqrt{h} K \phi_b\,,
\end{equation}
where the dilaton potential $V(\phi)$ will be a general function of $\phi$ unless otherwise specified. The theory is considered on a disc topology $\mathcal{M}$ with boundary $\partial \mathcal{M} = S^1$. The trace of the extrinsic curvature normal to $\partial \mathcal{M}$ is given by $K$, and $\phi_b \equiv \phi|_{\partial \mathcal{M}}$ is the boundary value of the dilaton. The first term in (\ref{dilatongravity}) is the contribution from the constant part of the dilaton
\begin{equation}
	S_0 = - \frac{\phi_0}{2} \int_{\mathcal{M}} d^2 x \sqrt{g} R - \phi_0 \int_{\partial \mathcal{M}} d\tau \sqrt{h} K  = -2\pi \phi_0 \chi\,.
\end{equation}
Due to the Gauss-Bonnet theorem, this term computes the Euler character $\chi$ of $\mathcal{M}$ and so is topological. For a disc topology one has $\chi = 1$. In what follows, will be interested in the canonical partition function for arbitrary $V(\phi)$ in the presence of a Euclidean boundary at finite proper distance from the Euclidean horizon. 

\subsection{Equations of motion and solutions}
The equations of motion stemming from (\ref{dilatongravity}) are
\begin{align}
    \nabla_\mu \nabla_\nu \phi - g_{\mu \nu} \laplacian \phi + \frac{1}{2 \ell^2} \, g_{\mu \nu} V(\phi) &= 0\,,  \label{eom1}\\
    R + \frac{1}{\ell^2} \partial_\phi V(\phi)&= 0\,. \label{Riccieom}
\end{align}
These equations can be rearranged, at least within a local neighbourhood of $\mathcal{M}$, as follows
\begin{align}
 - \laplacian \phi + \frac{1}{\ell^2} V(\phi) &= 0\,, \label{EOM1}\\
    R +\frac{1}{\ell^2} \partial_\phi V(\phi)&= 0\,, \label{EOM2} \\ 
    \nabla_\mu \xi_\nu + \nabla_\nu \xi_\mu &= 0\,, \quad \xi^\mu \equiv \varepsilon^{\mu\nu} \partial_\nu \phi\,. \label{EOM3}
\end{align}
Therefore, even though it seems that (\ref{eom1}) and (\ref{Riccieom}) are overconstrained, recasting them in the form (\ref{EOM1}), (\ref{EOM2}), and (\ref{EOM3}) shows that there are only two second order equations acting on the degrees of freedom $\phi$ and $R$. The equation (\ref{EOM3}) moreover indicates that $\xi^\mu$ is a Killing vector field of $g_{\mu\nu}$. Using the existence of $\xi^\mu$ and the two diffeomorphisms we can place a general solution of  (\ref{dilatongravity}) in the form
\begin{equation} \label{generalsolution}
   \frac{ds^2}{\ell^2} = N(r) d\tau^2 + \frac{dr^2}{N(r)}\,, \qquad \phi(r) =  r\,, \qquad \tau \sim \tau+\beta\,.
\end{equation}
Given this form of the metric, equation \eqref{EOM1} then becomes
\begin{equation}
     N'(r) = V(\phi) \,,
\end{equation}
which gives the following relation between the dilaton potential and the coefficient of the metric
\begin{equation} \label{V-Areleation}
    N(r, r_h) = \int_{r_h}^r dr' \, V(r') >  0  \,.
\end{equation}
Since we are considering Euclidean solutions, this must be positive for all $r$. Noting that $N(r_h,r_h) = 0$, the metric (\ref{generalsolution}) describes a Euclidean black hole solution with the horizon at $r_h$, and where $\phi(r_h)  \equiv \phi_h$ is the value of the dilaton at the horizon. We emphasise that $r_h$ is not an independent variable as it is fixed by the form of $N(r,r_h)$. Expanding (\ref{generalsolution}) near to the horizon $r = r_h$ and imposing periodicity of $\tau \sim \tau +\beta$ to ensure smoothness of the Euclidean solution leads to the relation
\begin{equation} \label{beta}
    \beta = \frac{4\pi}{|V(r_h)|}\,.
\end{equation}
This formula holds regardless of the asymptotic form of the potential. The usual JT gravity \cite{Jackiw:1984je,Teitelboim:1983ux} corresponds to the case where $V(\phi) = 2 \phi$. For this potential (\ref{Riccieom}) ensures that the curvature is constant and negative, giving an AdS$_2$ solution for the metric. 

\subsection{Boundary conditions} \label{bcsec}

As boundary conditions, we consider the Euclidean Dirichlet problem for which the  induced metric $h$ at $\partial\mathcal{M}$ and the corresponding proper length $\beta_T$, as well as the boundary value of the dilaton $\phi_b$ on $\partial \mathcal{M}$ are fixed. The gauge parameter $\xi_\mu$ is required to vanish at $\partial\mathcal{M}$, to ensure the boundary condition on $h$ is preserved and the location of the boundary is not disrupted (see for instance \cite{Witten:2018lgb}).
\newline\newline
The induced metric $h$ can be arranged to be constant along $\partial\mathcal{M}$ by a judicious reparameterisation of the boundary time $\tau$. In what follows, we moreover take $\phi_b$ to be constant along $\partial \mathcal{M}$. The physical motivation behind this choice of boundary condition stems from the Gibbons-Hawking prescription for Euclidean black hole thermodynamics \cite{Gibbons:1976ue,Gibbons:1977mu}, whereby one views the Euclidean solutions on the disc as contributing to the thermal partition function of the underlying theory.\footnote{Alternative boundary conditions could fix the trace of the extrinsic curvature $K$ at $\partial\mathcal{M}$ or the normal derivative of $\phi$ at $\partial\mathcal{M}$, or more general combinations thereof.} We will not necessitate that the boundary is asymptotically (near) AdS$_2$  for large part of our discussion, in line with the considerations of \cite{Svesko:2022txo}. \\
\\
Given the solution (\ref{generalsolution}) with (\ref{V-Areleation}), it is clear that a general choice of $\phi_b$ and $h$ will not permit a  real solution to our boundary value problem. For instance, let us consider the dilaton potential with $V(\phi) = -2\phi$, which yields $N(r,r_h) = r_h^2- r^2$. Smoothness of the solution at the origin of the disc moreover fixes $r_h = 2\pi/\beta$ such that the proper size of the boundary circle is $\beta_T = \sqrt{4\pi^2-\beta^2 r_b^2}$. Regardless of our choice for $\phi_b$ and assuming our solution is real, we have that $\beta_T \in (0,2\pi)$ which is only a subset of the positive half-line. More markedly, for $\phi_b=0$, only $\beta_T = 2\pi$ is permitted. In particular, going from a non-vanishing to a vanishing value of $\phi_b$ leads to a discontinuous jump in the allowed set of $\beta_T$. These restrictions on independent boundary data are a two-dimensional analogue of the obstructions arising when setting up the Dirichlet problem in four-dimensional general relativity \cite{Andrade:2015gja,An:2021fcq,Anderson:2007jpe,Witten:2018lgb} on manifolds with a boundary. In the case at hand, aside from such restrictions, the Dirichlet problem is well-posed due to the absence of locally propagating gravitational degrees of freedom. Nonetheless, we must ensure that the Dirichlet data $(\beta_T,\phi_b)$ we consider are indeed sensible.  

\subsection{Thermodynamics} \label{Thermodynamics}
We would like to study the thermodynamics of the class of theories (\ref{dilatongravity}) in the presence of  a finite boundary. One motivation for studying the finite boundary case is that certain geometries, such as the static patch of de Sitter space, do not afford an asymptotic boundary. In such circumstances, it is natural to consider a Dirichlet problem with a finite boundary. Our goal here is to present formulae for thermodynamic quantities at a finite boundary, generalising the results of \cite{Anninos:2017hhn,Grumiller:2007ju,Witten:2020ert}. \\
\\
We now evaluate the on-shell Euclidean action, which is related to the partition function by $-S_E = \log Z$. Given the Dirichlet problem under consideration, we have that $Z$ is a function of $\phi_b$ and $\beta_T$. The induced metric $h$ and $\phi_b$ at $\partial\mathcal{M}$ are given by
\begin{equation} \label{boundarymetric}
    ds^2_{\text{bdy}} = \ell^2 N(r_b ,r_h) d\tau^2 \equiv h \, d\tau^2\,, \quad\quad \beta_T = \beta \sqrt{N(r_b,r_h)}\,, \quad\quad \phi_b = r_b\,.
\end{equation}
The trace of the extrinsic curvature is  
\begin{equation}
    K = \frac{1}{h} K_{\tau \tau} = \frac{1}{2\sqrt{h}} \, \partial_{r_b} N(r_b,r_h)\,.
\end{equation}
We will further assume, for now, that the dilaton takes its minimal value at the Euclidean black hole horizon $r_h$. This is not necessarily the case; when the horizon is cosmological it will be located at the largest possible value of the dilaton, as will be discussed in the next section. The on-shell action is  
\begin{equation}
    \begin{split}
        S_E &=  \frac{\phi_0 \beta}{2} \int_{r_h}^{r_b} dr  N''(r) - \frac{\phi_0 \beta}{2} N'(r_b) - \frac{\beta}{2} \int_{r_h}^{r_b} dr \left( -  r N''(r) +   N'(r) \right) - \frac{\beta}{2}N'(r_b) \phi_b\,.
    \end{split}
\end{equation}
Integrating the first term in the bulk action by parts and using (\ref{beta}), along with the fact that for a black hole horizon $V(r_h)>0$ (otherwise (\ref{V-Areleation}) would not hold for $r$ slightly greater than $r_h$), this evaluates to 
\begin{equation} \label{Onshellresult}
  \log Z =  2 \pi\, ( \phi_0 + \phi_h ) + \beta_T  \sqrt{N(r_b,r_h)}\,.
\end{equation}
We see that the  expression (\ref{Onshellresult}) takes the suggestive form $\log Z = S- E/T$ for a canonical thermal partition function.  Indeed,   we are interested in the thermodynamics experienced by an observer at a finite cutoff $r_b$. As such, the relevant temperature is the Tolman temperature \cite{York:1986it} which is precisely $\beta_T$ in (\ref{boundarymetric}).
To evaluate thermodynamic quantities, we must vary $\log Z$ with respect  to $\beta_T$ while keeping $\phi_b$ (and hence $r_b$) fixed.
The following expressions prove to be useful
\begin{equation}
        {\partial_{\beta_T} N(r_b, r_h)} = \frac{2 V(r_h)^2 N(r_b, r_h)}{\beta_T( V(r_h)^2 + 2 V'(r_h) N(r_b , r_h) )}\,,  \quad\quad  {\partial_{r_h} N(r_b, r_h)}  =  - V(r_h) \,.
\end{equation}
The thermal energy $E$, and entropy $S$ are found to be
\begin{eqnarray}
        E &=& -  \sqrt{N(r_b,r_h)}\,, \label{energy}\\ 
         S &=& 2 \pi  ( \phi_0 + \phi_h)\,, \label{entropy}
\end{eqnarray}
while the specific heat is given by\footnote{In the case where the dilaton decreases towards the AdS$_2$ boundary, rather than growing, these formulae will be modified to have an overall minus sign.}
\begin{equation}
  C =  \frac{ \beta_T \sqrt{N(r_b,r_h)} \, V(r_h)^2}{V(r_h)^2 + 2 V'(r_h) N(r_b , r_h)}\,, \label{heatcapacity}
\end{equation}
in agreement with \cite{Grumiller:2007ju}. In the case where there are multiple possible saddles for a given $\beta_T$, the above formulae will hold at each saddle and the free energy must be computed to determine which solution is dominant. We observe that the entropy (\ref{entropy}) is a function of quantities located at the Euclidean horizon, reflecting its more universal nature. This property was already observed in the early work of York \cite{York:1986it} when placing a four-dimensional black hole in a box with Dirichlet conditions. The energy and specific heat depend more sensitively on the choice of boundary conditions.  Consequently, though the entropy is insensitive to the location of the boundary, the thermal stability of a horizon may depend on this. \\
\\
A special case of the above is when the dilaton potential at large values of $\phi$ takes the form $V(\phi) = 2 \phi +\mathcal{O}(\phi^{-\epsilon})$. Here, our geometries acquire an asymptotic near-AdS$_2$ boundary such that the metric takes the asymptotic form $N(r_b,r_h) \approx r_b^2 - b$, and it follows from (\ref{boundarymetric}) that our thermodynamic quantities are computed (after a rescaling of $r_b$) with respect to the inverse temperature $\beta = 4\pi/V(\phi_h)$. Hence, for asymptotically near-AdS$_2$ configurations, our thermodynamic quantities are more closely associated to the behaviour of the spacetime at the horizon. Specifically, taking $\phi_b \gg 1$ while keeping $\beta_T$ fixed, the quantities (\ref{energy}), (\ref{entropy}), and (\ref{heatcapacity}) become
\begin{equation} \label{asympESC}
 E_\infty =-\phi_b + \frac{b}{2 \phi_b}\,,  \quad S_\infty = 2 \pi \left(  \phi_0 +   \phi_h \right)\,, \quad C_\infty =-  \frac{2 \pi \, V(\phi_h)}{R(\phi_h)}\,,
\end{equation}
in agreement with the results of \cite{Witten:2020ert,Anninos:2017hhn,Grumiller:2007ju}. Stated otherwise, the more rigid nature of the near-AdS$_2$ boundary has allowed for the thermodynamic derivatives to occur with respect to the ordinary Gibbons-Hawking temperature $\beta$ in (\ref{beta}), which is defined as the surface gravity at the horizon with a suitably normalised timelike Killing vector $\partial_\tau$. One may also add boundary counterterms to the action by taking $K \rightarrow K - 1$ in equation \eqref{dilatongravity} which has the effect of removing the divergent $-\phi_b$ contribution from the expression for $E_\infty$.

\section{Near-Nariai thermodynamics} \label{Near-Nariai geometry}
\begin{figure}[H]
	\centering
	\includegraphics[width=10cm]{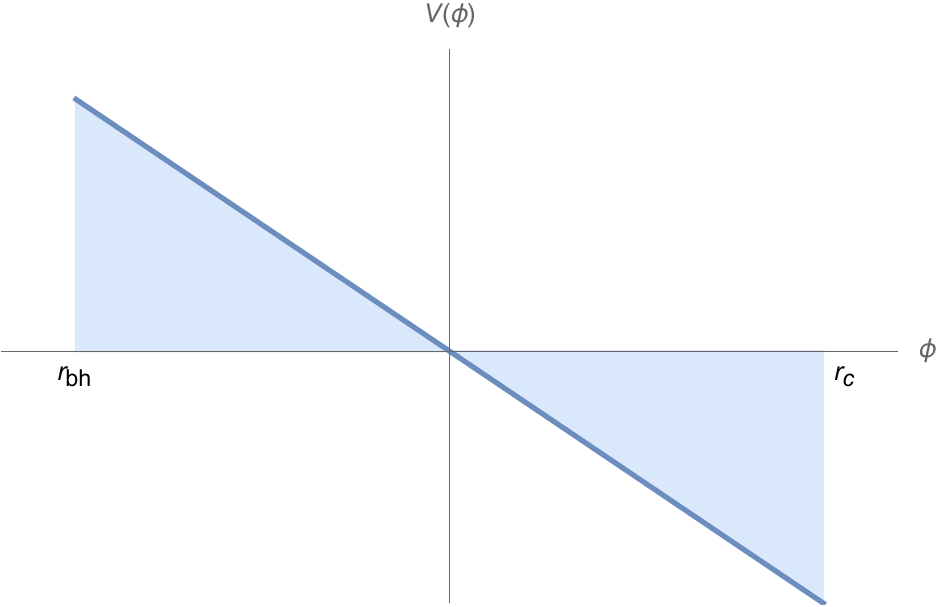}
	\caption{The dilaton potential $V(\phi) = - 2 \phi$ for the black hole in de Sitter. The solution caps off at the two horizons which have been indicated here as $r_{bh}$ and $r_c$. }
	\label{fig:dSBH-V}
\end{figure}

In this section, we compute thermodynamic properties of the dimensionally reduced near-Nariai black hole in the presence of a finite boundary. This geometry describes the near horizon limit of the Schwarzschild-de Sitter spacetime with near coincident horizons. The dilaton potential for this model is given by $V(\phi) = -2\phi$, which is plotted in Figure \ref{fig:dSBH-V}. The thermodynamics of this dilaton-gravity model were also discussed in \cite{Anninos:2017hhn,Svesko:2022txo}.

\begin{figure}[H]
	\centering
	\includegraphics[width=10cm]{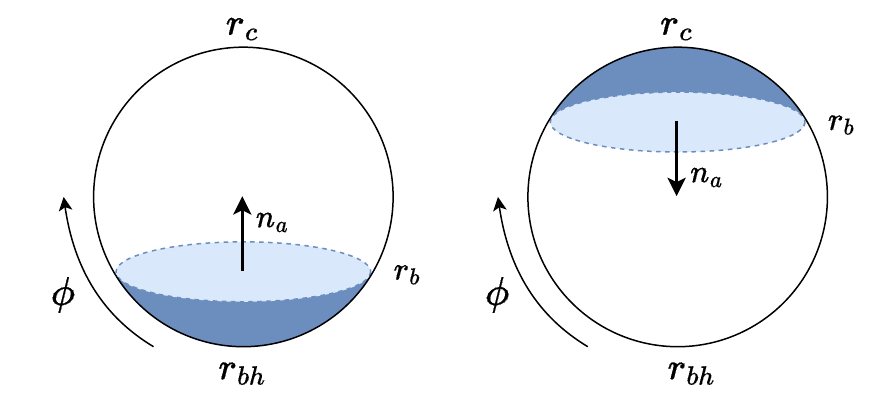}
	\caption{The Euclidean Nariai geometry is an $S^2$. The choice of sign of the outward pointing normal vector is dependent on which horizon we choose to integrate from. The direction of increasing $\phi$ is indicated, which we will always take to increase from the back hole horizon towards the cosmological horizon.}
	\label{fig:UnitNormal2}
\end{figure}

As in the previous section, we will consider the metric in Euclidean signature. Employing (\ref{V-Areleation}), we can confirm that the dilaton potential $V(\phi) = - 2 \phi$ produces the dS$_2$ geometry, whose solution reads
\begin{equation} \label{dS2metric}
    \frac{ds^2}{\ell^2} = (r_h^2 - r^2)d\tau^2 + \frac{dr^2}{(r_h^2 - r^2)}\,, \quad \tau \sim \tau + \beta\,, \quad\quad  \phi(r) =r\,,
    \end{equation}
where $r_h \leq r \leq -r_h$, with $r_h<0$ and $\tau \sim \tau + \beta$. There are two Euclidean horizons at $r = \pm r_h$ -- one is the black hole horizon which we will refer to as $r_{bh} = r_{h} < 0$, and the other is the cosmological horizon which we call $r_c = - r_{h} > 0$. In Euclidean signature the space caps off at the two horizons and so this metric coincides with the round metric on the two-sphere, as illustrated in Figure \ref{fig:UnitNormal2}. Taking $r_h = - 1$ in (\ref{dS2metric}), we retrieve equation (\ref{euclsp}) with $d=2$. Since $r_h$ is intrinsically linked to the temperature through (\ref{beta}) and we wish to consider thermal fluctuations of our system, it will be useful to keep $r_h$ explicit here. Since the dilaton is also running, it is a near-dS$_2$ static patch.
\noindent As in the previous section, $\beta$ is chosen to ensure smoothness of the Euclidean saddle. Now that there are two such potential singularities, we must have 
\begin{equation} \label{Nariaitemp}
    \beta = \frac{4\pi }{|V(r_{bh})|} = \frac{4\pi }{|V(r_{c})|} =  \frac{2\pi}{|r_h|}\,.
\end{equation}
From Figure \ref{fig:dSBH-V} we see that the condition (\ref{Nariaitemp}) is satisfied since $r_{bh} = - r_{c} = r_{h}$. We therefore have a $\mathbb{Z}_2$ symmetry $r \rightarrow -r$ in this metric which exchanges the two horizons. It is the sign of the dilaton that decides for us which is the cosmological horizon, and which is the black hole, and here we choose the dilaton to always grow away from the black hole horizon. 

\subsection{Near-Nariai with a boundary} 

In order to compute the thermal partition function, we place a boundary at some value $r=r_b \in (r_h,- r_h)$, where $\phi_b = r_b$ and $\beta_T$ are fixed. The boundary can either surround the black hole or cosmological horizon. In fact, there is no choice of saddle here. Rather, all saddles must be included provided they satisfy the same Dirichlet boundary conditions. However, of all the saddles, one will generally dominate. 
\newline\newline
Thus, we must evaluate the thermodynamic quantities for both horizons in the presence of a boundary at $r_b$. For the black hole horizon at $r=r_h$, the analysis of the previous section can be directly applied, resulting in
\begin{equation} \label{bhonshell}
    \log Z_{\text{bh}} = 2 \pi ( {\phi}_0 + r_h ) + \beta_T \sqrt{r_h^2-\phi_b^2}\,, \quad\quad r_h  = -   \frac{2\pi |\phi_b|}{\sqrt{4\pi^2-  \beta_T^2}}\,. 
\end{equation}
\newline
 An important difference for the cosmological horizon is that the outward pointing unit normal vector has the opposite sign to that of the black hole case, as shown in Figure \ref{fig:UnitNormal2}. For this reason, the sign of the boundary term must be flipped in order to have a well-posed variational problem. Furthermore, for a cosmological horizon, (\ref{V-Areleation}) becomes 
\begin{eqnarray}
	\int_{r}^{r_c} dr' \, V(r') < 0 \,. 
\end{eqnarray}
So, $V(r_c) < 0$ in order for this to hold for $r$ slightly smaller than $r_c$. 
The end result is a change in the sign of the second term in equation (\ref{Onshellresult}), such that
\begin{equation} \label{cosmoonshell}
    \log Z_{\text{cos}} = 2 \pi ( {\phi}_0 - r_h ) - \beta_T \sqrt{r_h^2-\phi_b^2}\,.
\end{equation}
We can now compute and compare various thermodynamic properties. The canonical free energies of the two saddles are
\begin{eqnarray}
F_{\text{bh}} &=&   - \frac{2 \pi  {\phi_0}}{\beta_T} +  \frac{|\phi_b|\sqrt{4 \pi ^2- \beta_T^2}}{\beta_T} \,, \\
F_{\text{cos}} &=& - \frac{2 \pi  {\phi_0}}{\beta_T} -\frac{|\phi_b|\sqrt{4 \pi ^2- \beta_T^2}}{\beta_T} \,.
\end{eqnarray} 
It follows that
\begin{equation}
\Delta F \equiv F_{\text{bh}} - F_{\text{cos}} = \frac{2  |\phi_b|}{\beta_T} \sqrt{4\pi^2 - \beta_T^2} \ge 0\,,
\end{equation}
where the inequality is saturated for $\phi_b = 0$.\\
\\
The saddle with lowest free energy is the one with the cosmological horizon. However, we must also assess whether the saddle is stable under small thermal fluctuations. We thus compute the specific heat
\begin{equation}\label{Cnariai}
C_{\text{bh}} = - C_{\text{cos}} = 
\frac{4 \pi^2 |\phi_b| \beta_T ^2}{\left(4 \pi^2-\beta_T^2\right)^{3/2}}> 0\,.
\end{equation}
We find that with Dirichlet boundary conditions, the black hole horizon of the near-Nariai geometry has positive specific heat while the cosmological horizon has negative specific heat \cite{Svesko:2022txo}.
This also resonates with the results of \cite{Draper:2022ofa,Banihashemi:2022htw}. In consequence, even though the cosmological saddle has lower free energy, it is thermally unstable in the setup under consideration. At least locally, the black hole saddle is the stable one, although it may be metastable at the non-perturbative level. \\
\\
The energy can be computed in a similar fashion via  (\ref{energy}). 
We find
\begin{equation}
    E_{\text{bh}}  = -E_{\text{cos}} = -  \frac{|\phi_b| \beta_T}{\sqrt{4\pi^2 - \beta_T^2}}\,.
\end{equation}
The energies of the black hole and cosmological horizons are equal and opposite. 
We can similarly calculate the entropy by either integrating from the black hole horizon, as in the left-hand side of Figure \ref{fig:UnitNormal2} or the cosmological one as in the right-hand side of the same figure. Using equation (\ref{entropy}), we have
\begin{eqnarray}
    S_{\text{bh}} &=& 2 \pi \left( \phi_0 +  \phi_{h} \right)\,, \\
      S_{\text{cos}} &=& 2 \pi \left( \phi_0 -  \phi_{h} \right)\,,
\end{eqnarray}
where $\phi_h  = r_h <0$ is the value of the dilaton at the horizon. \\
\\
It is interesting to note that the total energy in the near-dS$_2$ static patch, which is given by `gluing' the black hole saddle to the cosmological saddle along their common boundary yields a vanishing result
\begin{equation} \label{totalnariaienergy}
    E_{\text{tot}} \equiv E_{\text{bh}}  + E_{\text{cos}}  = 0\,. 
\end{equation}
Adding the entropies together yields the total entropy of the near-dS$_2$ static patch to be
\begin{equation} \label{totalnariaientropy}
    S_{\text{tot}} \equiv S_{\text{bh}}  + S_{\text{cos}}  = 4 \pi \phi_0\,. 
\end{equation}
We can compare (\ref{totalnariaienergy}) and (\ref{totalnariaientropy}) to the result from calculating the classical contribution to the partition function over the whole two-sphere, where now there is no longer a boundary term for the action. One finds
\begin{equation}
\begin{split}
    S_E 
    &= \frac{\phi_0 \beta}{2} \left( V(-r_h) - V(r_h) \right) - \frac{\beta \, r_h}{2} \left(V(-r_h) + V(r_h)  \right).
\end{split}
\end{equation}
Since for the Nariai geometry $V(r_h) = -2r_h$ the second term vanishes. The free energy $F$ is related to the partition function by $\log Z = - \beta_T F = S-\beta_T E$. Since there is no spatial boundary anymore, the total energy of the full sphere vanishes. We therefore have that the total entropy is
\begin{equation}
    S_{\text{full}} = - S_E = 4 \pi \phi_0\,,
\end{equation}
which exactly matches equation (\ref{totalnariaientropy}). It would be good to understand whether this agreement between the full two-sphere and the `glued' saddles persists beyond the classical level by computing quantum corrections as in \cite{Anninos:2021ene,Muhlmann:2022duj, Muhlmann:2021clm}. \\
\\
It is interesting to contrast (\ref{Cnariai}) with  the specific heat of the four-dimensional Nariai black hole, which has been reported in the literature \cite{Niemeyer:2000nq,Anninos:2010gh} to be negative leading to an evaporating black hole (like that of the Schwarzschild black hole in flat space). The four-dimensional thermodynamic quantities rely on a more local treatment of the horizon thermodynamics. The temperature $T$, for instance, is given by the surface gravity at the Killing horizon with a suitably normalised Killing vector. If one computes the area, and hence the Bekenstein-Hawking entropy, of the black hole horizon as a function of $T$, one notes that it decreases with increasing $T$ indicating a negative specific heat. For similar reasons the de Sitter horizon has been reported to have positive specific heat \cite{Anninos:2012qw}. The difference in sign with the computations above is accounted for by the different definition of temperature, which is now given by  the Tolman temperature $\beta_T$. This the near-Nariai black hole version \cite{Svesko:2022txo} of York's observation \cite{York:1986it} that when placed in a box with Dirichlet conditions, the flat space Schwarzschild black hole can have positive specific heat. 
\newline\newline
In the following section we proceed to consider a different treatment of the boundary of the near-Nariai black hole geometry and show that one can recover a negative specific heat.

\section{\texorpdfstring{(A)dS$_2$}{(A)dS2} interpolating geometries} \label{(A)dS$_2$ Interpolating geometries}
In this section, we explore an extension  \cite{Anninos:2017hhn,Anninos:2018svg} of the dS$_2$ dilaton potential, $V(\phi)=-2\phi$, that results in an asymptotically near-AdS$_2$ geometry ending at the near-Nariai black hole horizon. Our goal is to study the thermodynamic properties of such models in the presence of a Dirichlet boundary (a related discussion can be found in \cite{Gross:2019ach}). We end with a comment about the stretched dS$_2$ horizon in this framework. 

\subsection{Geometry}

\begin{figure}[H]
	\centering
	\includegraphics[width=10cm]{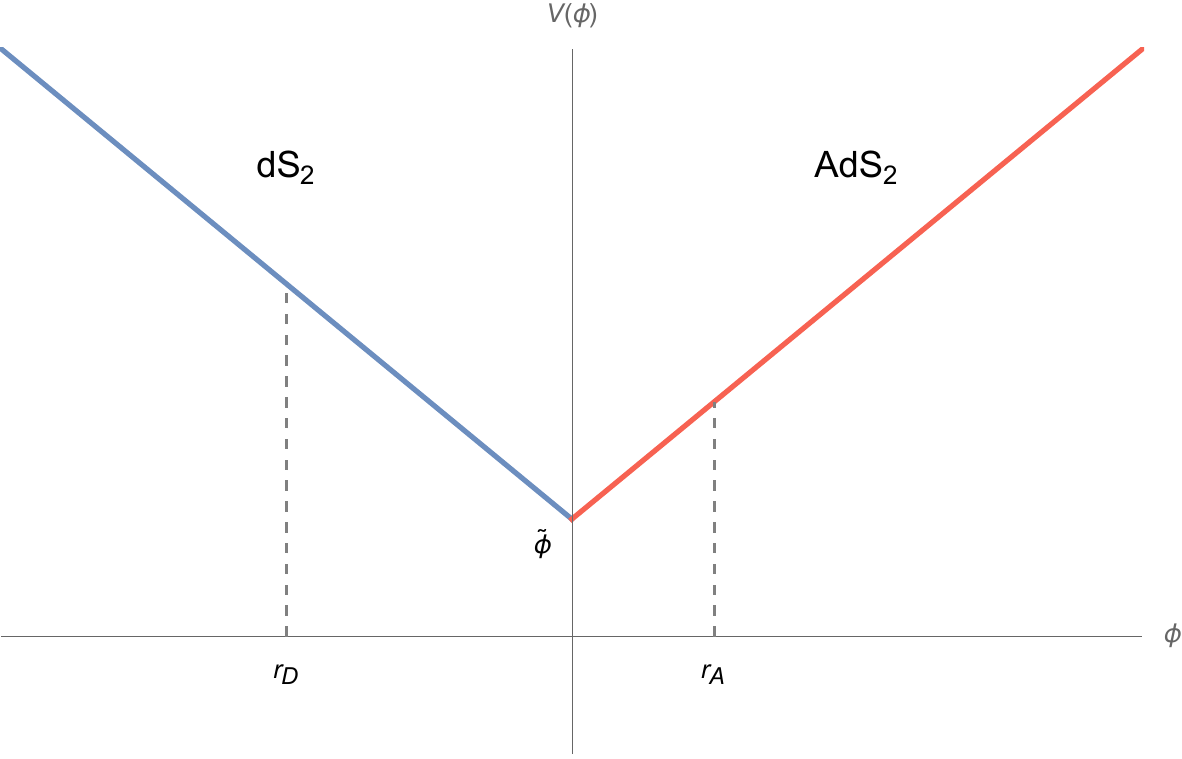}
	\caption{The dilaton potential for the interpolating geometry (\ref{centaurV}) with $\tilde{\phi} >0$. The blue slope indicates where the potential describes a dS$_2$ geometry and the red slope shows where the geometry is Euclidean AdS$_2$. At a given temperature, there can be two saddles, indicated here at $r_D$ and $r_A$.}
	\label{fig:CentaurPotential}
\end{figure}

For the sake of concreteness, we focus on the dilaton potential
\begin{equation}\label{centaurV}
    V(\phi) = 2  |\phi  | +  \tilde{\phi}\,,
\end{equation}
where $\tilde{\phi}$ is a real-valued parameter.\footnote{It is worth emphasising that any $V(\phi)$, including smooth potentials, that have regions linear in $\phi$ with oppositely signed slopes and an asymptotic growth of the form $V(\phi) = 2\phi + \mathcal{O}(\phi^{-\epsilon})$ suffices for our purposes. A simple example is $V_\varepsilon(\phi) = 2 \phi \tanh \frac{\phi}{\varepsilon}$ with $\varepsilon >0 $ small. Moreover, although we have chosen the slope in  (\ref{centaurV}) to have the same absolute value for $\phi>0$ and $\phi < 0$ one can also consider cases where the slopes differ.} The potential is shown in Figure \ref{fig:CentaurPotential}.  For a given temperature, this geometry will have up to two saddles. We shall refer to these as the interpolating geometry and the AdS$_2$ geometry. Although the potential $V(\phi)$ we study has a jump in its first derivative, the geometries it produces have continuous and differentiable metrics. We now provide the form of the metric for both $\tilde{\phi}$ positive as well as negative.

\subsubsection*{Case 1: \texorpdfstring{$\tilde{\phi} \ge 0$}{phi > 0}}

From equation (\ref{V-Areleation}) we can see that the metric for the interpolating geometry will be of the form (\ref{generalsolution}) with 
\begin{equation} \label{InterpCentaursaddle}
    N^{D_+}(r, r_D) = 
    \begin{cases} 
    (r - r_D)( \tilde{\phi} - r - r_D)\,, \qquad & r_D \leq r \leq 0 \,,\\
    r^2 + r_D^2 + \tilde{\phi}(r - r_D)\,, \qquad & 0 < r\,,
    \end{cases} 
\end{equation}
where $r_D$ defines the Euclidean dS$_2$ black hole horizon. The Euclidean time periodicity is given by
\begin{equation}\label{betaD}
\beta_{D} = \frac{4\pi}{|  \tilde{\phi} -2  r_D |}\,.
\end{equation}
We therefore have a portion of the two-sphere (Euclidean dS$_2$) for $r \leq 0$ (in particular, when $\tilde{\phi} = 0$ this is exactly the metric in (\ref{dS2metric})), and when $r > 0$ we have a Euclidean AdS$_2$ metric. Depending on the sign of $\tilde{\phi}$ the two-sphere is glued to a portion of a quotient of the hyperbolic disc ($\tilde{\phi}>0$) or a portion of the hyperbolic strip ($\tilde{\phi} \le 0$). \\
\\
The metric of the AdS$_2$ saddle is 
\begin{equation} \label{AdSCentaursaddle}
	N^{A_+}(r,r_A) = r^2 - r_A^2 + \tilde{\phi}(r - r_A)\,, \qquad 0 \leq r_A \leq r  \,,
\end{equation}
where $r_A$ defines the location of the Euclidean AdS$_2$ black hole horizon. The Euclidean time periodicity is given by
\begin{equation}\label{betaA}
\beta_{A} = \frac{4\pi}{|\tilde{\phi} + 2  r_A  |}\,.
\end{equation}

\subsubsection*{Case 2: \texorpdfstring{$\tilde{\phi} <0$}{phi <0}}

When $\tilde{\phi} < 0$, we must further ensure that the metric is everywhere positive. The AdS$_2$ saddle will remain of the same form, but the range of $r_A$ is modified as
\begin{equation} \label{AdSCentaursaddleNeg}
	N^{A_-}(r, r_A) = r^2 - r_A^2 + \tilde{\phi}(r - r_A)\,, \qquad |\tilde{\phi}| /2 \leq r_A \leq r  \,.
\end{equation}
The interpolating solution similarly restricts the range of $r_D$, such that
\begin{equation} \label{InterpCentaursaddleNeg}
    N^{D_-}(r, r_D) = 
    \begin{cases} 
    (r - r_D)( \tilde{\phi} - r - r_D)\,, \qquad &  r_D \leq r \leq 0 \,, \qquad r_D < \left(\frac{1+\sqrt{2}}{2} \right) \tilde{\phi}\,,\\
    r^2 + r_D^2 + \tilde{\phi}(r - r_D)\,, \qquad & 0 < r\,.
    \end{cases} 
\end{equation}
In the above we are assuming that the coordinate $r$ can become large compared to $\tilde{\phi}$ and $r_D$. The periodicities $\beta_D$ and $\beta_A$ will be as for $\tilde{\phi}>0$, namely (\ref{betaD}) and (\ref{betaA}) respectively. 

\subsection{Thermodynamic properties} \label{Interpolating Thermodynamic properties}

As in the previous sections, we consider Dirichlet boundary conditions whereby the proper size of the boundary circle $\beta_T$ and the boundary value of the dilaton $\phi_b$ are fixed. We take $\phi_b >0$ here since non-positive $\phi_b$ reproduces the near-Nariai black hole setup described in Section \ref{Near-Nariai geometry}. In general, there are two solutions obeying the boundary conditions which we call $r_D$ and $r_A$, as shown in Figure \ref{fig:CentaurPotential}. 
\newline\newline
The temperatures of the interpolating geometry and the AdS$_2$ geometry are 
\begin{eqnarray}
	\beta_T^D &=& \frac{4\pi }{|\tilde{\phi} - 2 \phi_D|} \sqrt{ r_b^2 + \phi_D^2 + \tilde{\phi}(r_b - \phi_D)} \,, \\
	\beta_T^A &=& \frac{4\pi }{|\tilde{\phi} + 2 \phi_A|} \sqrt{r_b^2 - \phi_A^2 + \tilde{\phi}(r_b - \phi_A)} \,,
\end{eqnarray}
where we have introduced the notation $ \phi_D = r_D$ to be the value of the dilaton at the dS$_2$ black hole horizon and $ \phi_A = r_A$ to be the value of the dilaton at the AdS$_2$ black hole horizon. We must set the temperatures equal to each other $\beta_T^D = \beta_T^A = \beta_T$ to compare their thermodynamic properties at a given temperature. 

\subsubsection*{Case 1: \texorpdfstring{$\tilde{\phi} \ge 0$}{phi > 0}}
Let us begin by taking $\phi_b$ to be large compared to $\tilde{\phi}$ and $\phi_{A,D}$. 
The thermodynamics for the Euclidean AdS$_2$ solution (\ref{AdSCentaursaddle}) with $\tilde{\phi}=0$ are reviewed in Appendix \ref{appAdS2}. For non-vanishing $\tilde{\phi}$ we find a similar result for $\phi_A>0$, namely 
\begin{equation}\label{FA}
-\beta_T F_{\text{AdS}_2} =  2\pi \left( \phi_0 - \frac{\tilde{\phi}}{2} \right) + \frac{1}{2}\left(2 \phi_b + \tilde{\phi} \right)  \sqrt{\beta_T^2+4\pi^2} \,.
\end{equation}
The specific heat follows readily
\begin{equation}\label{CA}
	C_{\text{AdS}_2} = \left(  \phi_b + \frac{\tilde{\phi}}{2}\right) \, \frac{4 \pi^2 \beta_T^2 }{\left(\beta_T^2+4 \pi ^2\right)^{3/2}}\,,
\end{equation}
which is positive  for all $\phi_b > 0$. \\
\\
The other solution has $\phi_D < 0$ such that the region of the geometry near and including the horizon has positive curvature. For this interpolating solution, we have 
\begin{equation} \label{Finterp1}
-\beta_T F_{\text{interp}} = 2\pi \left( \phi_0 + \frac{\tilde{\phi}}{2} \right)  + \left(\beta_T^2 - 4 \pi^2\right) \sqrt{\frac{(\phi_b + \chi_+)(\phi_b - \chi_- )}{\beta_T^2 - 4 \pi^2}} \,,
\end{equation} 
where for convenience we have defined $\chi_\pm \equiv \left(\tfrac{\sqrt{2} \pm 1}{2} \right)\tilde{\phi}$. In order to ensure the above expression is real we must lie in one of two regimes. The first, which connects to parametrically large $\phi_b$, is $\phi_b \ge \chi_-$ and $\tfrac{4\pi}{\tilde{\phi}} \sqrt{\phi_b( \phi_b  + \tilde{\phi})} \ge \beta_T \ge 2\pi$, where the upper bound for $\beta_T$ ensures the negativity of $\phi_D$. Given $F_{\text{interp}}$, we can compute the specific heat of the interpolating saddle for this range:
\begin{equation} \label{CInterpunstable}
      C_{\text{interp}} = - \sqrt{(\phi_b + \chi_+)(\phi_b - \chi_-  )} \times  \frac{4 \pi^2 \beta_T^2 }{( \beta_T^2 - 4 \pi^2)^{3/2}} \,. 
\end{equation}
For this case the specific heat $C_{\text{interp}}$ is negative, and the only thermodynamically stable saddle is the Euclidean AdS$_2$ black hole. We thus retrieve, within this range of boundary conditions a near-Nariai black hole horizon with negative specific heat. The second regime is $0< \phi_b < \chi_-$ and $\tfrac{4\pi}{\tilde{\phi}} \sqrt{\phi_b( \phi_b  + \tilde{\phi})} < \beta_T < 2\pi$. For this case the specific heat $C_{\text{interp}}$ is 
\begin{equation} \label{CInterpstable}
      C_{\text{interp}} =  \sqrt{(\phi_b + \chi_+)(-\phi_b +\chi_- )} \times  \frac{4 \pi^2 \beta_T^2}{(4 \pi^2 - \beta_T^2 )^{3/2}} \,,
\end{equation}
which is positive. Note that taking $\phi_b = 0$ in (\ref{CInterpstable}) recovers the result (\ref{Cnariai}) for the black hole in the Nariai geometry up to an overall constant, as expected. Upon solving $\beta_T^D = \beta_T^A = \beta_T$, we see that 
\begin{equation}
	\phi_A = \frac{1}{2} \left(\frac{(2 \phi_b + \tilde{\phi} ) (\tilde{\phi} - 2 \phi_D)}{\sqrt{(2 \phi_b + \tilde{\phi})^2+ 8 \phi_D^2 - 8 \tilde{\phi}  \phi_D}} - \tilde{\phi} \right) , \qquad \phi_b > \chi_- \,.
\end{equation}
A solution for $\phi_A$ for a given $\phi_D$ can only be found for large enough $\phi_b$. 
Therefore, in the range $0 < \phi_b < \chi_-$ the interpolating geometry is the only saddle and is thermodynamically stable. We now explore the case $\tilde{\phi}<0$. 

\subsubsection*{Case 2: \texorpdfstring{$\tilde{\phi} < 0$}{phi < 0 }}

We now consider the thermodynamic properties for $\tilde{\phi} < 0$. The Euclidean AdS$_2$ saddle will be thermodynamically stable, with the same free energy and specific heat as for $\tilde{\phi}>0$, namely (\ref{FA}) and (\ref{CA}) respectively.\\
\\
The interpolating saddle will now permit some additional properties. The free energy will still take the form (\ref{Finterp1}), with the additional restriction coming from the requirement that the metric (\ref{InterpCentaursaddleNeg}) remain positive for all $r$, namely that $\phi_D \leq \chi_+ $. This additional restriction on the value of the dilaton at the horizon modifies the reality conditions for (\ref{Finterp1}). We again find two possible regimes, the first being $\sqrt{2}\pi \left|\tfrac{2 \phi_b + \tilde{\phi}}{\tilde{\phi}}\right| \ge \beta_T \ge 2 \pi$ and $\phi_b \ge - \chi_+$. In this range the interpolating saddle has the same specific heat as in (\ref{CInterpunstable}) and hence is unstable. \\
\\
The other possibility is the regime $\sqrt{2}\pi \left|\tfrac{2 \phi_b + \tilde{\phi}}{\tilde{\phi}}\right| < \beta_T < 2 \pi$ and $0< \phi_b < - \chi_+$. In this case, the heat capacity is given by (\ref{CInterpstable}) and so is thermodynamically stable. Solving $\beta_T^D = \beta_T^A = \beta_T$, we find  
\begin{equation} \label{phiA}
	\phi_A =  \frac{1}{2} \left(  \frac{(2 \phi_b + \tilde{\phi} ) (\tilde{\phi} - 2 \phi_D)}{\sqrt{(2 \phi_b + \tilde{\phi})^2+ 8 \phi_D^2 - 8 \tilde{\phi}  \phi_D}} - \tilde{\phi}  \right) \,.
\end{equation}
Therefore, in this case for a given $\phi_D$, there is always a $\phi_A$ provided that $|\tilde{\phi}|/2 < \phi_b < - \chi_+$. The difference in free energies between the stable interpolating saddle and the AdS$_2$ saddle is given by 
\begin{multline} \label{FdiffInterp}
F_{\text{interp}} - F_{\text{AdS}_2} 
= \frac{ \sqrt{(\phi_b + \chi_+ )(-\phi_b + \chi_-) ( 4 \pi^2 - \beta_T^2  )}}{ \beta_T} 
	+ \frac{(2 \phi_b+ \tilde{\phi} ) \sqrt{\beta_T^2 + 4 \pi^2} - 4 \pi \tilde{\phi} }{2 \beta_T} \,.
\end{multline}
For $\tilde{\phi} < 0$ this expression is positive. To give a numerical example, we can take $\tilde{\phi}= -1$, $\phi_b =1$, and $\beta_T = 5$ which satisfy the conditions on $\beta_T$ and $\phi_b$. These values correspond to taking $\phi_D = - \tfrac{1}{2} - \tfrac{\pi}{\sqrt{ 4 \pi^2 - 25}} \approx -1.33$ in (\ref{phiA}) such  that $\phi_A = \tfrac{1}{2}+ \tfrac{\pi}{\sqrt{ 4 \pi^2 + 25}} \approx 0.89$ lies in the allowed range for $\phi_A$. They lead to a positive difference (\ref{FdiffInterp}).  So, the AdS$_2$ saddle will be thermodynamically favoured over the interpolating saddle. If instead $0< \phi_b < |\tilde{\phi}|/2$, then there will be no AdS$_2$ saddle since $V(\phi_A)< 0$ in this range and we will have the same situation as in (\ref{CInterpstable}), with the interpolating saddle being the only stable solution. 

\subsection{Interpolating the stretched \texorpdfstring{dS$_2$}{dS2} horizon}

\begin{figure}[H]
 	\centering
 	\includegraphics[width=16cm]{ 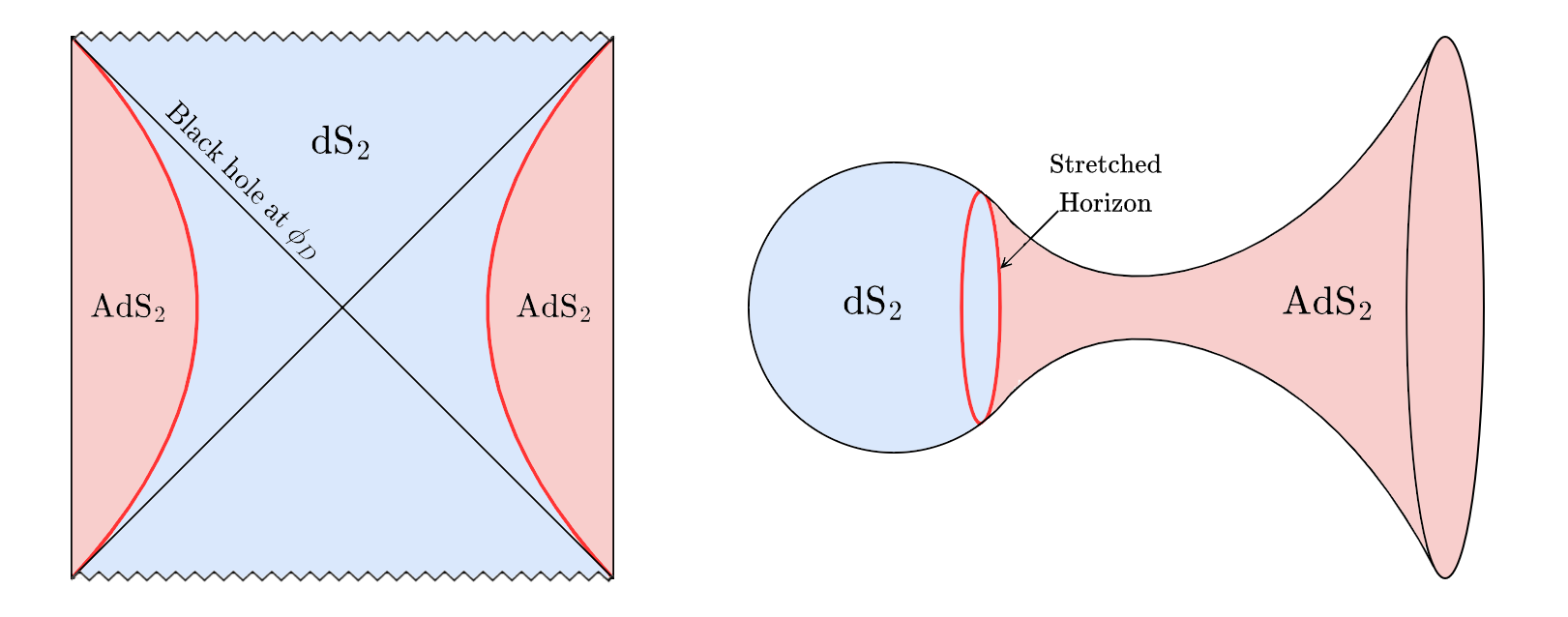}
 	\caption{Left: The Penrose diagram of the interpolating geometry that contains the black hole in a de Sitter static patch in the deep interior and interpolates to an AdS$_2$ boundary. Right: The same geometry in Euclidean signature, where the Euclidean black hole horizon now  caps off at the pole of the two-sphere. }
 	\label{fig:CentaurPenroseDiagram}
 \end{figure}

We end this section by considering the interpolating solution for $\tilde{\phi}< 0$ in some more detail.  For the purpose of our discussion, it will prove useful to  slightly generalise the dilaton potential (\ref{centaurV}) to the following
\begin{equation}
V_\delta(\phi) = 2 \phi  \left( - \Theta(-\phi) + \frac{1}{\delta}    \, \Theta(\phi)   \right) + \tilde{\phi}\,,
\end{equation} 
with $\tilde{\phi}<0$, and $\delta$ a small positive number. In order to have an interpolating solution with an asymptotic AdS$_2$ region, the horizon $r_h$ must lie below the critical value $r_\delta \equiv \tilde{\phi} (1 + \sqrt{1+\delta} \, )/2$. At precisely $r_h = r_\delta$ the geometry caps off for a second time at $r = -\tilde{\phi}\, \delta/2$, creating a closed Euclidean universe.
\newline\newline
If we now tune $r_h$ to be parametrically close to and below $r_\delta$ and take $\phi_b \gg 1$, we find an asymptotically near-AdS$_2$ geometry which includes a significant portion of the two-sphere. The interpolating saddle stemming from $V_\delta(\phi)$ admits a portion of the two-sphere for which the excised region is a disc of area $\sim \delta$. The excised region that would have contained the cosmological dS$_2$ horizon is replaced by a region of negative curvature that reaches all the way out to the AdS$_2$ boundary. The geometry and corresponding Penrose diagram is shown in Figure \ref{fig:CentaurPenroseDiagram}. 
An excision of this type is often considered when placing a stretched surface \cite{Susskind:1993if} a small region away from the dS$_2$ horizon and has been explored as a potential holographic screen of de Sitter  in various works including \cite{Banks:2006rx,Bousso:1999cb,Shaghoulian:2022fop,Dong:2018cuv}. Though physically appealing, one of the challenges in making the stretched horizon picture precise is that the surface lies in the midst of a gravitating spacetime and it is difficult to obtain concrete observable quantities analogous to those at the boundary of AdS or the flat space $S$-matrix.\footnote{One approach is to set up a type of timelike Dirichlet boundary near the cosmological horizon, as explored in \cite{Hayward:1990zm,Wang:2001gt,Anninos:2011zn,Banihashemi:2022jys,Banihashemi:2022htw}. However, at least in four and higher spacetime dimensions, caution must be exercised since this leads to various instabilities and issues with well-posedness of the Dirichlet problem \cite{Anderson:2007jpe,An:2021fcq}.}  Instead of placing a holographic theory at the stretched horizon, we might then view the interpolating geometry as an ultraviolet completion of the stretched dS$_2$ horizon with a near-AdS$_2$ boundary. 
\newline\newline
As it stands, the interpolating geometries we have discussed so far have negative specific heat whenever they are endowed with an asymptotically AdS$_2$ boundary. This can be ameliorated by bringing the near-AdS$_2$ boundary into the interior, by introducing a Dirichlet wall \cite{Gross:2019ach} along the lines we have discussed in the previous sections. This comes at the cost of sharp AdS/CFT type observables. In the next section we discuss a simple generalisation that admits a stretched dS$_2$ horizon that is capped off by a horizon with positive specific heat in the deep interior while preserving the asymptotic near-AdS$_2$ boundary. 

\section{Double interpolating geometries}\label{disec}
In this section, we will propose a geometry that has some of the benefits of the  interpolating geometry discussed in the previous section while also having the virtue of a positive specific heat, ensuring that the solution is locally thermodynamically stable. For the sake of concreteness, we will consider a theory with the following dilaton potential
\begin{equation} \label{SandPot}
    V(\phi) = \begin{cases}
  2 \phi + \tilde{\phi} - 4 x\,, \quad\quad & \phi \leq x\,, \\
 2 | \phi | + \tilde{\phi}\,, \quad\quad & x< \phi \,, \\
\end{cases}
\end{equation}
where $x < 0$. As in the previous section, the potential can be viewed as an idealisation of a smooth potential where we have only retained the piecewise linear pieces. 
\newline\newline
We now discuss the asymptotically AdS$_2$ Euclidean saddles and their thermodynamic properties, while relegating the discussion with finite boundary to Appendix \ref{Double Interpolating Geometry finite}. Here, we keep the slopes of each linear piece of $V(\phi)$ to have the same magnitude for the sake of simplicity. More generally, they can be taken to be different. 

\subsection{Geometry}

\begin{figure}[H]
	\centering
	\includegraphics[width=10cm]{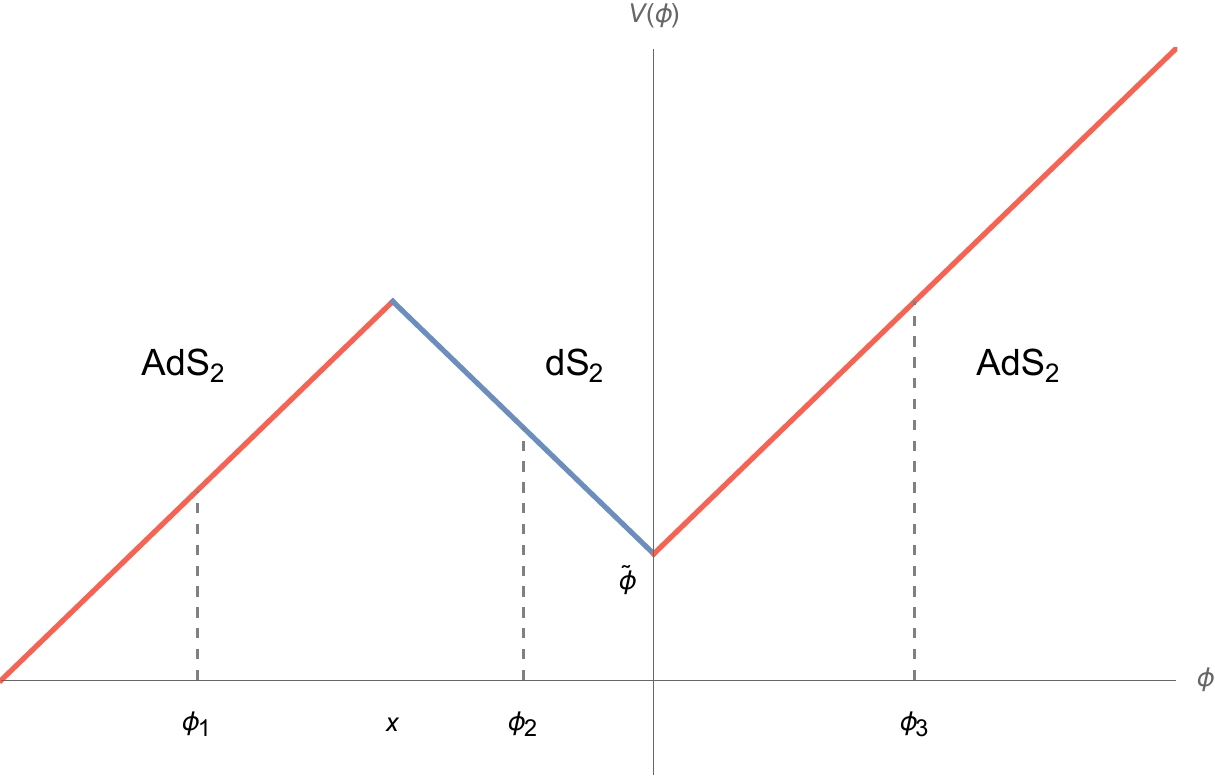}
	\caption{Plot showing the dilaton potential (\ref{SandPot}) with $\tilde{\phi} > 0$. This potential gives an AdS$_2$ geometry for $\phi \leq x$ and $\phi \geq 0$ but a dS$_2$ static patch in the region $x \leq \phi \leq 0$. At a given temperature, there may be up to three candidate black hole solutions. }
	\label{fig:SandwichPotential}
\end{figure}

The geometry is given by merging the interpolating geometry (\ref{InterpCentaursaddle}) to a second AdS$_2$ region in the deep interior at the distance $r = x$, as is shown in Figure \ref{fig:SandwichPotential}. In the range  $\tilde{\phi} < V(\phi) < \tilde{\phi} - 2x$, we therefore have three possible saddles which we label $\phi_1 < \phi_2 <0 < \phi_3$. Outside of this range, the solution will only contain a stable AdS$_2$ black hole saddle if $\tilde{\phi} - 2x <V(\phi)$, or only the stable double interpolating geometry if $V(\phi) < \tilde{\phi}$.
\subsubsection*{Case 1: \texorpdfstring{$\tilde{\phi} \ge 0$}{phi >0}}
The first saddle is for $r_1 < x$ and the metric for this system will have coefficients given by 
\begin{equation}  \label{Doubleintepsaddle1}
	N(r, r_1) = 
	\begin{cases} 
		(r - r_1)(r + r_1 + \tilde{\phi} - 4 x)\,, &  \tfrac{4x - \tilde{\phi}}{2} \le r_1 \leq r \leq x \,,\\
		-r^2 + \tilde{\phi} \,r - 2 x^2 + r_1 ( 4x - r_1 - \tilde{\phi}) \,, & x < r \leq 0 \,,\\
		r^2 + \tilde{\phi} \, r - 2x^2 + r_1 (4x - r_1 - \tilde{\phi})\,, & 0 < r \,,
	\end{cases} 
\end{equation}
where the first condition on $r_1$ ensures that $V(r_1) \ge 0$. This describes a geometry which is AdS$_2$ in the deep interior, flowing to a piece of the Euclidean dS$_2$ static patch, and then flowing to another AdS$_2$ region near the boundary. The size of the static patch region is controlled by the two free parameters $x$ and $\tilde{\phi}$.\footnote{By further adjusting the slopes of the linear pieces in $V(\phi)$, the AdS$_2$ in the deep interior can be made parametrically small.}\\
\\
The second saddle has a Euclidean horizon located in the range $x \le r_2 <  0$. This is the interpolating geometry described in Section \ref{(A)dS$_2$ Interpolating geometries} where the metric is precisely (\ref{InterpCentaursaddle}) with $r_D = r_2$. The final saddle has a Euclidean horizon located at $0 \leq r_3 \leq r$. For this range, the metric is (\ref{AdSCentaursaddle}) with $r_A = r_3$, which again is the AdS$_2$ saddle described in the previous section. 
 
\subsubsection*{Case 2: \texorpdfstring{$\tilde{\phi} < 0$}{phi < 0}}
\begin{figure}[H]
	\centering
	\includegraphics[width=15cm]{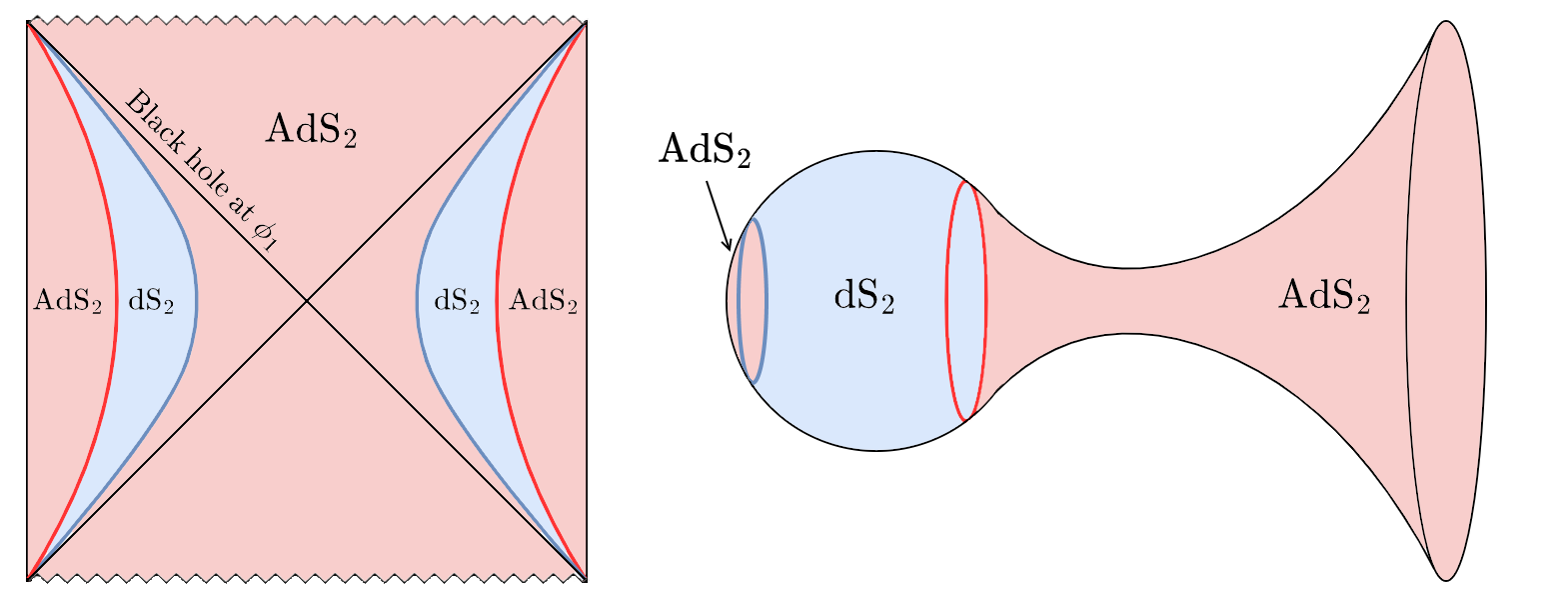}
	\caption{Left: Penrose diagram of the double interpolating geometry. Right: corresponding Euclidean geometry.}
	\label{fig:doubleinterpolating}
\end{figure}

In this case, the geometries are as above, but each saddle has an added restriction. Firstly, in order to have values of $r_h = r_1, r_2$ with $V(r_h)>0$, we require $V(x)>0$ such that $\tilde{\phi} > 2x$. This ensures that the lower bound on $r_1$ in (\ref{Doubleintepsaddle1}) remains less than $x$ when $\tilde{\phi}<0$. To ensure the positivity of the metric (\ref{Doubleintepsaddle1}) we must further have $4 r_1( 4x - \tilde{\phi} - r_1) > 8 x^2 + \tilde{\phi}^2$. The restriction for the second saddle at $r_2$ is the same as in (\ref{InterpCentaursaddleNeg}), namely $x \leq r_2 \leq \chi_+$ and for the third saddle at $r_3$ we have the same added restriction as in (\ref{AdSCentaursaddleNeg}) that $|\tilde{\phi}|/ 2 \le r_3$.
\newline\newline
As for the interpolating solution with $\tilde{\phi}<0$ discussed in the previous section, we can view the completion of the excised two-sphere as a way to push the stretched dS$_2$ horizon all the way to the near-AdS$_2$ boundary. This is shown in Figure \ref{fig:doubleinterpolating}. As we now explore, in the current setup the thermal stability properties are improved as compared to the previous case. 

\subsection{Thermodynamic properties}

Let us introduce the notation $\phi_1 = r_1$, $\phi_2 = r_2$ and $\phi_3 = r_3$ as the values of the dilaton at each horizon. In what follows we will take $\phi_b \gg 1$ so that the boundary is asymptotically AdS$_2$. The finite boundary case is treated in Appendix \ref{Double Interpolating Geometry finite}. In this limit we can employ the asymptotic formulas (\ref{asympESC}). Along with these, in the large $\phi_b$ limit the difference in free energy of any two saddles $\phi_1$ and $\phi_2$ is given by \cite{Witten:2020ert}
\begin{equation} \label{Fdiffasympt}
	\Delta F = F_2 - F_1 = \frac{1}{2  \phi_b} \left[  N(\phi_2, \phi_1) + V(\phi_1) ( \phi_1 - \phi_2 )  \right] \,.
\end{equation}
As $\phi_b \rightarrow  \infty$, we have that $\beta_T  \approx  \beta \phi_b$, and so from the definition (\ref{beta}) we must also have $V(\phi_1) = V(\phi_2)$. 
\subsubsection*{Case 1: \texorpdfstring{$\tilde{\phi} \ge 0$}{phi > 0}}
For $\phi_b \gg 1$ the first saddle at $\phi_1$ will have heat capacity 
\begin{equation} \label{asymC1}
	C_1 = \pi ( 2 \phi_1 + \tilde{\phi} - 4x ) \,,
\end{equation}
which is positive due to the lower bound on $r_1$ in (\ref{Doubleintepsaddle1}). As promised, this geometry has the benefit of being thermodynamically stable while also possessing an asymptotically AdS$_2$ boundary. The ground state with vanishing temperature is given by the configuration starting at the point $\phi_1 = 2x-\tilde{\phi}/2$ where the dilaton potential crosses the $\phi$-axis. 
\newline\newline
The thermodynamic properties of the saddles at $\phi_2$ and $\phi_3$ were described in Section \ref{Interpolating Thermodynamic properties}. For large $\phi_b$, the saddle at $\phi_2$ will have negative heat capacity (\ref{CInterpunstable}) and hence will be thermodynamically unstable. The saddle at $\phi_3$ will have heat capacity given by (\ref{CA}) and so will be thermodynamically stable. To see which saddle is thermodynamically favoured, we can use (\ref{Fdiffasympt}) with $V(\phi_1) = V(\phi_3)$, which leads to 
\begin{equation} \label{asympFdiff}
	\Delta F = F_3 - F_1 = \frac{x}{\phi_b} \left( 2\phi_1 - 3 x \right) \,. 
\end{equation}
This will be positive if $2 \phi_1 < 3x $ and negative otherwise. 
\newline\newline
Thus, for  temperatures satisfying $\beta > {4\pi}/({\tilde{\phi}-x})$, the double interpolating solution containing a large portion described by Euclidean dS$_2$ dominates the thermodynamics in this model. Once $\beta$ reaches the critical value $\beta = {4\pi}/({\tilde{\phi}-x})$ we have a first order phase transition to the Euclidean AdS$_2$ black hole. In this case, the interpolating solution is  metastable.

\subsubsection*{Case 2: \texorpdfstring{$\tilde{\phi} < 0$}{phi < 0}}

The model has two vanishing temperature configurations, one at $\phi_1 = 2x-\tilde{\phi}/2$, with $x$ restricted to yield a positive metric,  and the other at $\phi_3 = -\tilde{\phi}/2$. At finite temperature,  the specific heat  of the $\phi_1$ saddle is  (\ref{asymC1}) which remains positive. For $\phi_b \gg 1$, the saddle at $\phi_2$ is again unstable and hence we again only have thermally stable saddles at $\phi_1$ and $\phi_3$ with their difference in free energies again given by (\ref{asympFdiff}). 
\newline\newline
Given that $\Delta F$ does not depend on $\tilde{\phi}$, the sign of $\tilde{\phi}$ will not affect which saddle dominates. Indeed, the model exhibits a first order phase transition at the critical temperature $\beta= {4\pi}/({\tilde{\phi}-x})$ between a low temperature phase dominated by the double interpolating saddle, and a high temperature phase dominated by the Euclidean AdS$_2$ black hole. 
\newline\newline
To summarise, we have established the existence of dilaton-gravity models permitting locally (and even globally) thermally stable asymptotically near-AdS$_2$ geometries that encode a significant portion of the Euclidean dS$_2$ static patch in their interior. Slightly generalising the potential (\ref{SandPot}) to have different slopes in each linear regime, one can have asymptotically near-AdS$_2$ geometries that contain a stretched de Sitter horizon parametrically close to the actual cosmological dS$_2$ horizon. The geometries end at an AdS$_2$ black hole type horizon with positive specific heat. Another way to stabilise the interpolating geometries in the deep interior might be to include an end-of-the-world brane, which would be interesting to explore.

\section{Dual theory proposal}\label{outlooksec}

We have explored a variety of dilaton-gravity models whose solution space includes interpolating solutions containing a portion of the static patch of two-dimensional de Sitter space. In judiciously chosen circumstances, the stability properties of these geometries can be locally and even globally stable. In particular, the double interpolating solutions discussed in Section \ref{disec} have asymptotically near-AdS$_2$ solutions which contain a portion of the dS$_2$ static patch. Moreover, these solutions have positive specific heat and can be thermodynamically dominant over the AdS$_2$ black hole.
\newline\newline
Given the presence of a near-AdS$_2$ boundary, one is tempted to investigate whether these models permit a microphysical realisation in terms of an AdS$_2$/CFT$_1$ type picture, or some other ultraviolet completion of dilaton-gravity. One such approach might be to consider the more general relation between dilaton-gravity models with general dilaton potential and matrix models \cite{Saad:2019lba,Witten:2020wvy,Maxfield:2020ale,Eberhardt:2023rzz}. The challenge here is that the type of dilaton-potentials discussed in those works take a somewhat specific form
\begin{equation}\label{deformedV}
V_f(\phi) = 2\phi + \int_0^{2\pi} d\alpha f(\alpha) e^{-\alpha\phi}\,,
\end{equation}
with $f(\alpha)$ small.  Although it is suggested that the class of perturbative deformations in (\ref{deformedV}) might span to the larger class of holomorphic functions, $V_f(\phi)$ is subject to non-perturbative corrections. Whenever $x$ is small and $\tilde{\phi}$ is positive, the potential (\ref{SandPot}), or a smoothed-out version, can be viewed as a perturbation $\delta V(\phi)$ of $V(\phi)= 2\phi$. So, provided that the deformed potentials (\ref{deformedV}) span a sufficiently large space of deformations, one might identify a corresponding matrix model. Employing the results of \cite{Witten:2020wvy,Maxfield:2020ale}, the corresponding eigenvalue distribution to linear order in the deformation (upon shifting $\phi$ such that $V_f(0)=0$) reads
\begin{equation}\label{rhocompact}
\rho(\lambda) = e^{S_0}\left( \frac{\sinh 2\pi\sqrt{\lambda}}{4\pi^2} + \frac{e^{2\pi\sqrt{\lambda}}\delta V(\sqrt{\lambda}) +e^{-2\pi\sqrt{\lambda}}\delta V(-\sqrt{\lambda})}{8\pi\sqrt{\lambda}} + \ldots   \right)\,.
\end{equation}
For the models we consider $\rho(\lambda)$ increases indefinitely, or at least up to some large cutoff, reflecting the near-AdS$_2$ boundary. For a closed Euclidean universe, as suggested by (\ref{rhocompact}), perhaps one should take $\rho(\lambda)$ to fall back to a vanishing value \cite{Anninos:2021ydw,Anninos:2020geh,Anninos:2022ujl}, reflecting the presence of two horizons in dS$_2$ \cite{Anninos:2017hhn,Dong:2018cuv}. It would be interesting to explore such matrix models further. \\
\\
More generally, one could ask whether the SYK model \cite{Sachdev:1992fk,Kitaev:2017awl,Maldacena:2016hyu} permits deformations leading to theories such as (\ref{SandPot}). Since the vacuum is no longer pure AdS$_2$ in the interior, we expect that the SYK model is deformed by some relevant deformation. Relevant deformations are indeed permitted in SYK, and were explored in \cite{Anninos:2020cwo,Anninos:2022qgy} where flows between near-CFTs were identified. In particular, this was shown for the sum of two SYK Hamiltonians, $\hat{H}_{\text{tot}} = \hat{H}_q+ s \, \hat{H}_{q/2}$, where 
\begin{equation}
\hat{H}_q = {i^{q/2}} \sum_{i_1\leq \ldots\leq i_q \leq N} J_{i_1,\ldots,i_q} \psi_{i_1} \ldots \psi_{i_q}\,,
\end{equation}
with $J_{i_1,\ldots,i_q}$ independently drawn from a suitable Gaussian ensemble, and $\psi_i$ with $i=1,\ldots,N$ being Majorana fermions. Provided $s$ is sufficiently small, $\hat{H}_{\text{tot}}$ flows between two near-CFT regions. 
For the theory  (\ref{SandPot}) we have an intermediate region of near-dS$_2$ between the two near-AdS$_2$ regions close to the boundary and in the deep interior. It is likely that this will require adding an additional term to $\hat{H}_{\text{tot}}$, with a coupling that may have to take complex values. We leave the exploration of such constructions for future work. 

\chapter{Correlation functions in de Sitter space}
\label{ComplexGeodesics}
In this chapter, we study scalar correlation functions in a fixed de Sitter background in order to characterise the behaviour of such functions in the presence of the cosmological horizon. In doing so, we resolve a puzzle stemming from the fact that such correlation functions can be expressed as a sum over geodesics lengths, and yet real geodesics do not exist between arbitrary points in de Sitter. 
\section{The scalar two-point function in de Sitter}
\label{sec_twopoint}
Consider the action for a free massive scalar field in dS$_d$,
\begin{equation}
S_\phi = -\frac{1}{2} \int d^dx \sqrt{-g} \left[ g^{\mu\nu} \partial_\mu \phi \partial_\nu \phi  +m^2 \phi^2 \right] \,,
\end{equation}
where $m$ is the mass of the scalar field and $g_{\mu \nu}$ is the $d$-dimensional de Sitter metric. We will study the Wightman two-point function in the Bunch-Davies (or Euclidean) vacuum state $| E \rangle$ \cite{Spradlin:2001pw, Anninos:2012qw}. This is denoted $G(X,Y) = \langle E | \phi(X) \phi(Y) |  E \rangle$, where $X$ and $Y$ are arbitrary points on global de Sitter. This two-point function is a solution to the Klein-Gordon equation in dS$_d$,
\begin{equation}
(\square - m^2) G(X,Y) = 0 \,.
\end{equation}
Given that the Euclidean state is invariant under the de Sitter isometries, the two-point function can only depend on $X$ and $Y$ through their de Sitter invariant length $P_{X,Y}$. This quantity is defined in embedding space as
\begin{equation}
	P_{X,Y} \equiv \frac{1}{\ell^2} \eta_{IJ} X^I Y^J , \quad \eta_{IJ} = \text{diag} \overbrace{(-1, 1, \ldots, 1)}^{d + 1}  \,. \label{PXY}
\end{equation} 
Note that in this expression $X$ and $Y$ are the coordinates describing the embedding of the $d$-dimensional de Sitter hyperboloid inside $(d+1)$-dimensional Minkowski space, as in \eqref{ds embedding}. For instance, if we choose to parameterise the hyperboloid with the global coordinates in equation \eqref{globalembedding}, the de Sitter invariant length is given by
\begin{equation} \label{P_global12}
	P_{X,Y} = - \sinh T_X \sinh T_Y + \cosh T_X \cosh T_Y \sum_{i = 1}^d \omega^i_X {\omega}^i_Y \,,
\end{equation}
where $\omega^i$ are coordinates on the $(d-1)$-sphere and obey $\sum_{i=1}^{d} \left(\omega^i \right)^2 = 1$. It is useful to note that \cite{Spradlin:2001pw}
\begin{equation}
P_{X,Y} \quad
\begin{cases}
\, > 1 \, ,  &\text{for timelike separated points} \,; \\
\, =1  \, ,  &\text{for coincident or null separated points} \,; \\
\, <1  \, ,  &\text{for spacelike points} \,. \\
\end{cases} \nonumber
\end{equation}
Additionally, for spacelike points 
\begin{equation}
P_{X,Y} \quad
\begin{cases}
 >-1 \, ,  &\text{when $X$ is spacelike separated from the antipodal point of $Y$} \,; \\
 =-1 \, ,  &\text{when $X$ is null separated from the antipodal point of $Y$} \,; \\
 <-1 \,  ,  &\text{when $X$ is timelike separated from the antipodal point of $Y$} \,.
\end{cases} \nonumber
\end{equation}
Going back to the two-point function, if we write $G(X,Y) = G(P_{X,Y})$, then the Klein-Gordon equation becomes
\begin{equation}
(1-P_{X,Y}^2) \partial_{P_{X,Y}}^2 G(P_{X,Y}) - d P_{X,Y} \partial_{P_{X,Y}} G(P_{X,Y}) - m^2 \ell^2  G(P_{X,Y}) = 0 \,. \label{kg_P}
\end{equation}
This hypergeometric equation has two linearly independent solutions:
\begin{equation} \label{GenSol}
    G(P_{X,Y}) = A \,  _2F_1 \left( h_+, h_-; \frac{d}{2}; \frac{1+P_{X,Y}}{2} \right) + B\, _2F_1 \left( h_+, h_-; \frac{d}{2}; \frac{1 - P_{X,Y}}{2} \right) \, ,
\end{equation}
where 
\begin{equation}
 h_\pm = \frac{(d-1)}{2} \pm \sqrt{\left(\frac{d-1}{2}\right)^2-m^2 \ell^2} \,.
\end{equation}
The second of these solutions is singular when $P_{X,Y} = -1$, corresponding to antipodal points. We do not expect to see this behaviour in the two-point function, and therefore choose to set $B=0$. The first solution, on the other hand, diverges at $P_{X,Y} = 1$, which mimics the expected short-distance behaviour of the two-point function. In order to fix the overall coefficient, we must match this solution to the short distance singularity in flat space \cite{Spradlin:2001pw,Sleight:2019mgd,Castro:2020smu}, which for $d>2$ takes the form 
  \begin{equation} 
G (P_{X,Y} \sim 1)  \approx   \frac{\Gamma \left(\frac{d}{2}\right)}{(2 \pi )^{\frac{d}{2}} \ell^{d-2}(d-2)} \frac{1}{(1-P_{X,Y})^{\frac{d}{2}-1}}  \,, \label{P1}
\end{equation}
and is independent of $m$. Note that for $d=2$ the correlator diverges logarithmically as 
 \begin{equation}
   G (P_{X,Y} \sim 1)  \approx - \frac{1}{4\pi}\log \left(1-P_{X,Y}\right) \, , \label{d=2flat} 
 \end{equation}
which is consistent with the expected QFT behaviour. Near $P_{X,Y} = 1$, the hypergeometric function behaves as
\begin{equation}
    _2F_1 \left( h_+, h_-;  \frac{d}{2}; \frac{1+P_{X,Y}}{2} \right) \approx \left(\frac{2}{1-P_{X,Y}} \right)^{\frac{d}{2}-1} \frac{\Gamma(\frac{d}{2})\Gamma(\frac{d-2}{2})}{\Gamma(h_+)\Gamma(h_-)} \, . 
\end{equation}
Matching the expansion of \eqref{GenSol} to equation \eqref{P1} thus shows that the solution that correctly reproduces the expected short distance behaviour of the two-point function (and does not have singularities at antipodal points) is given by \cite{Mottola:1984ar, Allen:1985ux, Allen:1985wd}
\begin{equation}
G(P_{X,Y}) = \frac{\Gamma(h_+) \Gamma(h_-)}{\ell^{d-2}(4\pi)^{d/2} \Gamma \left( \frac{d}{2}\right)} \, _2F_1 \left( h_+, h_-; \frac{d}{2}; \frac{1+P_{X,Y}}{2} \right)\, .  \label{exact_2pt}
\end{equation}
Note that, by definition, this is also the two-point function obtained from analytically continuing the two-point function on $S^d$ to Lorentzian de Sitter spacetime. We are interested in the large mass expansion of the above correlator. The asymptotic form of the hypergeometric function when some of the parameters are large is non-trivial. The expansion of interest in our case is the one where the first two parameters become large in the imaginary direction (and with opposite signs). Asymptotic expressions in this limit were first found in \cite{watson}. See also \cite{cvitkovic2017asymptotic}  for the latest state of affairs. The form of the correlator in the large mass limit is thus given by
\begin{equation}
\begin{split}
G(P_{X,Y}) & \approx   \frac{m^{\frac{d-3}{2}}}{2(2\pi \ell) ^{\frac{d-1}{2}}} \left[ \frac{   e^{-m \ell \arccos P_{X,Y}}}{\left(1-P_{X,Y}^2\right)^{\frac{d-1}{4}}}  +  \frac{ \left(-1\right)^{\frac{d-1}{2}} e^{-m \ell (2\pi - \arccos P_{X,Y})}}{(1-P_{X,Y}^2)^{\frac{d-1}{4}}} \right] \,,\label{Pminus1} \\
\end{split}
\end{equation}
where this asymptotic form is valid for all points satisfying 
  \begin{equation}\label{p condition}
   |1-P_{X,Y}^2| \gtrsim (m \ell) ^{-2}\;.
   \end{equation}
For $P_{X,Y} < -1$, the correlator is manifestly real, consistent with the fact that the points are spacelike separated, and in our conventions spacelike separated points have real length between them:
\begin{equation} \label{P_less}
    G(P_{X,Y}) \approx  \frac{ m^{\frac{d-3}{2}}}{(2\pi \ell) ^{\frac{d-1}{2}} } \, \Re \left[ \frac{e^{-m \ell \arccos P_{X,Y} }}{(1-P_{X,Y}^2)^{\frac{d-1}{4}}} \right] \,, \qquad P_{X,Y} < -1 \,. 
\end{equation}
When $P_{X,Y} > -1$, the second term in (\ref{Pminus1}) is always exponentially suppressed in the large mass limit, so the correlator takes the form
\begin{equation}
\begin{split}
G(P_{X,Y}) \approx \frac{m^{\frac{d-3}{2}}}{2(2\pi \ell) ^{\frac{d-1}{2}}} \frac{ \exp  \left( -m \ell \arccos P_{X,Y}\right)}{\left(1-P_{X,Y}^2\right)^{\frac{d-1}{4}}}  \, , \qquad P_{X,Y} >-1 \,. 
\end{split}
\label{P_greater}
\end{equation}
For $-1<P_{X,Y}<1$, it is straightforward to check that the correlator is real. When points become timelike separated, then $P_{X,Y}>1$ and the expression becomes manifestly complex, consistent with the fact that, in our conventions, timelike separated points have an imaginary length between them. \\
\\
Both (\ref{P_less}) and (\ref{P_greater}), can be obtained by solving the Klein-Gordon equation (\ref{kg_P}) in a WKB expansion, which provides an independent check of these asymptotic expansions; see Appendix \ref{app_WKB}. It is clear that both approximations break down when $P_{X,Y} \sim \pm 1$. As previously stated, when $P_{X,Y}$ is close to one, the correlator is chosen to mimic the short distance singularity in flat space \eqref{P1}. When $P_{X,Y}=-1$, the last argument in the hypergeometric function is zero, so in the large mass limit we obtain
 \begin{equation}
G (P_{X,Y}=-1)  \approx   \frac{m^{d-2}}{\Gamma(\frac{d}{2}) 2^{d-1} \pi^{\frac{d}{2}-1}} e^{-m \ell \pi} \,. \label{t_zero_d}
\end{equation}
This limit does not commute with taking the large mass limit first.

\section{The geodesic approximation} \label{sec:geo}
In the worldline formalism \cite{Schubert:2001he, Bastianelli:2005rc}, one can schematically compute $G(X, Y)$, the correlator of a free massive scalar field between two points $X$ and $Y$, as a path integral,
\begin{equation}\label{geoworld}
G(X,Y) = \int D{\mathcal{P}} \, e^{-m L[\mathcal{P}]} \,,
\end{equation}
where $m$ is the mass of the scalar field, $\mathcal{P}$ is a path connecting $X$ and $Y$, and $L[\mathcal{P}]$ is the length of that path. In the large mass limit, it is possible to take a saddle point approximation that reduces the path integral to a sum over geodesic lengths $L_g$,
\begin{equation}\label{geo_approx_eq}
G(X,Y)  \approx \sum_{g \, \in\, {\text{geodesics}}} e^{-m L_g}\, . 
\end{equation}
If there is more than one geodesic, we need to sum over them appropriately. These discussions go back to the seminal work by Bekenstein and Parker \cite{Bekenstein:1981xe}. See also \cite{Parker:1979mf, Parker:2009uva}. So far, we have obtained the two-point correlator in the large mass limit by finding the asymptotic form of the relevant hypergeometric function. The main purpose of this chapter is to reconcile these results with the expression coming from the geodesic approximation \eqref{geo_approx_eq},
even when real geodesics do not exist. We start by reviewing how to compute geodesics in de Sitter in cases where real geodesics do exist. For simplicity, we demonstrate this in $d=2$, but the results can be generalised to higher dimensions.

\subsection{Review of geodesics in \texorpdfstring{dS$_2$}{dS2}} \label{sec_geo}
The global metric of dS$_2$ is given by,
\begin{equation}
\frac{ds^2}{\ell^2} = -dT^2 + \cosh^2 T \, d\varphi^2 \,,
\end{equation}
with $\varphi \sim \varphi + 2\pi $ and $T \in \mathbb{R}$. 
The length functional is given by
\begin{equation} \label{LorentzianLength}
	L = \int ds = \int d\lambda \, \mathcal{L} (T, \dot{T}, \varphi, \dot{\varphi}, \lambda) = \ell \int d \lambda \sqrt{ (- \dot{T}^2 + \cosh^2 T \dot{\varphi}^2)} \,  ,
\end{equation}
where the dots represent derivatives with respect to the parameter $\lambda$ along the geodesic. We are working in a slightly unusual convention where timelike geodesics will have a complex length. Often the length functional is redefined to be real for timelike geodesics, but we wish to use the same definition for both timelike and spacelike separated points. The Lagrangian $\mathcal{L}$ does not depend explicitly on $\varphi$, so we can define the following conserved quantity
\begin{equation} \label{phieqn}
	\frac{\partial \mathcal{L}}{\partial \dot{\varphi}} \equiv Q =  \frac{\ell \, \dot{\varphi}  \cosh^2 T  }{\sqrt{ -\dot{T}^2 + \cosh^2 T \dot{\varphi}^2}} \,.
\end{equation} 
Since the length functional is invariant under reparametrisation, we may select $\lambda$ such that it is an affine parameter, 
\begin{equation}  \label{lagrangian}
	\mathcal{L}^2=\left( \frac{ds}{d\lambda} \right)^2 = \pm 1 = \ell^2(- \dot{T}^2 + \cosh^2 T \dot{\varphi}^2 )\, ,
\end{equation}
where the $\pm$ depends on whether the geodesic is spacelike or timelike, respectively. 
The first order equations (\ref{phieqn}) and (\ref{lagrangian}) can be integrated to find the trajectories of the geodesics, which read 
\begin{eqnarray}
		\tan ( \varphi + \tilde{\varphi} ) &=& \pm \frac{Q \sinh T }{\sqrt{Q^2 - \ell^2 \cosh^2 T}} \, , \label{lorenztiansol} 
\end{eqnarray}
where $\tilde{\varphi}$ is a constant of integration which, as well as $Q$, can be determined by the choice of the endpoints of the geodesic. The $\pm$ in \eqref{lorenztiansol} depends on whether $T(\phi)$ is taken to be an increasing or decreasing function. Choosing the first endpoint of the geodesic to be at $(T_1, \varphi_0)$ and taking $T(\phi)$ to be increasing at this point, we obtain
\begin{equation}
	\tilde{\varphi} = \arctan \left(\frac{Q \sinh T_1}{\sqrt{Q^2 - \ell^2 \cosh^2 T_1}} \right)- \varphi_0 \, .
\end{equation}

\paragraph{Timelike geodesics.} First, we consider timelike separated points at a fixed angle. Requiring that the geodesic passes through the point $(T_2, \varphi_0)$ implies $Q=0$. The result is the equation of motion $\varphi = \varphi_0$, and using (\ref{LorentzianLength}) we find that these geodesics have length
\begin{equation}
L_g = i \ell |T_2 - T_1| \,. \label{timelike_length}
\end{equation}

\begin{figure}[H]
        \centering
         \subfigure[$T_1=0$]{
                \includegraphics[height=4.5cm]{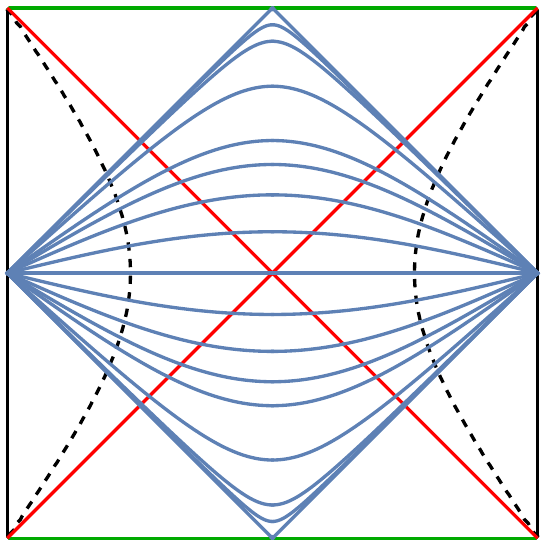}\label{fig:ds0}}  \qquad \qquad
        \subfigure[$T_1=1$]{
                \includegraphics[height=4.5cm]{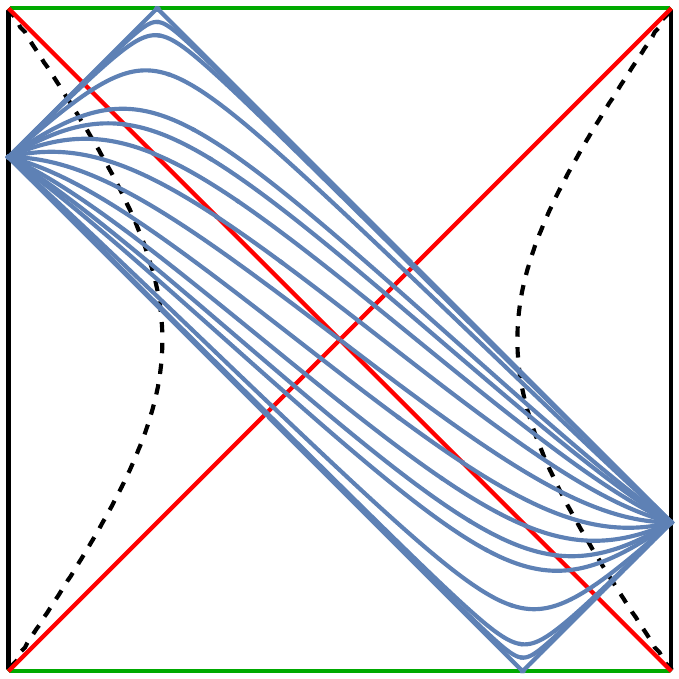} \label{fig:ds1}} 
                 \caption{Penrose diagram for (half of) dS$_2$ with geodesics in blue, horizons in red, and past and future infinity in green. In dashed black, we also plot the position of stretched horizons.}\label{fig:penroseperlim}
\end{figure}

\noindent \textbf{Spacelike geodesics.}
We will consider points at opposite sides of the spatial circle by taking the second endpoint to be $(T_2, \varphi_0 + \pi)$. Plugging this into \eqref{lorenztiansol}, we find that for any $Q$ we have $T_1 = \pm T_2$. However, only one of these solutions is smooth. To see this, it is useful to invert the geodesic trajectory (\ref{lorenztiansol}) to find the two possibilities
\begin{equation} \label{SpacelikeLorentzian}
	T(\phi) = \pm \text{ arccosh} \sqrt{ \frac{Q^2 \sec^2 ( \varphi + \tilde{\varphi})}{Q^2 + \ell^2 \tan^2 ( \varphi + \tilde{\varphi})}}  \, . 
\end{equation}
Choosing the positive sign for the entire geodesic gives the corresponding trajectory for $T_1 = T_2 > 0$, and the negative sign gives the trajectory $T_1 = T_2 < 0$, but neither of these solutions are continuous in their first derivatives, and so we discount them. Requiring a smooth trajectory necessitates gluing the two trajectories at the turning point of the function, leading to $T_1 = - T_2$. We therefore recover the result that spacelike geodesics between the poles of de Sitter only exist between times $T_1$ and $- T_1$ as shown in figures \ref{fig:ds0} and \ref{fig:ds1}. This was explored in the context of holographic complexity in \cite{Chapman:2021eyy}. There is a one parameter family of such geodesics labelled by $|Q| > \ell^2 \cosh T$, and using equation (\ref{LorentzianLength}), we find that all of these geodesics have the same length, $L_g = \ell \pi$ \cite{Chapman:2021eyy, Jorstad:2022mls}. Note that this result is completely different from the AdS$_2$ black hole case, where geodesics exist for arbitrarily long times and their length grows linearly with time \cite{Brown:2018bms}.

\subsection{From geodesics to two-point correlators}  \label{sec:tension}
One would like to use the results obtained for the real geodesics in de Sitter to reproduce the form of the correlators studied in Section \ref{sec_twopoint}. However, it is clear from the result just shown for points on opposite sides of the spatial circle that real geodesics are not enough to reproduce the correlator for points such as $T_1 = T_2 \neq 0$, for example.\\
\\
On the sphere geodesic distance between two points is given by $\ell \arccos P_{X,Y}$ with $P_{X,Y}$ defined similarly to \eqref{PXY} but with the metric replaced with the Euclidean one $\delta_{IJ}$. Drawing inspiration from this, one can formally define $L_g = \ell \arccos P_{X,Y}$ to be the geodesic distance between the points $X$ and $Y$ in de Sitter. Timelike separated points have $P_{X,Y}>1$, and for these points $L_g$ reproduces the complex length \eqref{timelike_length}. On the other hand, spacelike separated points have $P_{X,Y}<1$, but real geodesics only exist for $-1 \leq P_{X,Y} < 1$. As an example, consider points on opposite sides of the spatial circle, as in the previous section, for which $P_{X,Y}\leq -1$. Requiring $L_g$ to be real then imposes that $P_{X,Y} = -1$. In this case, $L_g = \ell \pi$ in agreement with the result of the last section. \\
\\
In terms of $L_g$, (\ref{Pminus1}) becomes
\begin{equation}
G(L_g) \approx \frac{m^{\frac{d-3}{2}}}{2 (2\pi \ell) ^{\frac{d-1}{2}}} \left[ \frac{   e^{-m \, L_g}}{\left(\sin L_g \right)^{\frac{d-1}{2}}}  +  \frac{  e^{-m (2\pi\ell - L_g)}}{\left(\sin (2\pi\ell - L_g) \right)^{\frac{d-1}{2}}} \right] \,. \label{GofL}
 \end{equation} 
For any $P_{X,Y}>-1$, the second term is always exponentially suppressed, and so the propagator can be written in the form of the geodesic approximation  (\ref{geo_approx_eq}). Moreover, we have seen that in those cases, the geodesic length is actually $L_g = \ell \arccos P_{X,Y}$, both for spacelike and timelike separated points. The only term that needs to be explained is the denominator, that will come from perturbations around the geodesic length.\\
\\
If $P_{X,Y}<-1$, there seems to be more tension. In this case, we derived that there are no real geodesics connecting these points in de Sitter, and the naive continuation of $L_g$ gives a complex geodesic length. Moreover, from the form of (\ref{GofL}) it seems that there are two geodesics contributing to the correlator. As in the previous case, the denominator in each term needs to be explained. In the next section, we will show that the tension for $P_{X,Y}<-1$ can be cured by looking at geodesics on the sphere.  In all cases, denominators will appear as one-loop corrections to the geodesic length.\\
\\
Before moving on, let us comment on the special degenerate case of $P_{X,Y} = -1$. In this case, we showed that $L_g = \ell \pi$, which is consistent with $P_{X,Y}=-1$. But we found that there are infinitely many geodesics, which would naively yield an infinite correlator, unless properly regulated. Note also that equation (\ref{P_greater}) diverges in this limit, but that the correct large mass correlator is (\ref{t_zero_d}). In what follows, we will restrict to $P_{X,Y} \neq -1$.

\section{Sphere two-point functions} \label{sec:sphere}
In this section, we compute the two-point correlator on the sphere, where we know that any two points are connected by a geodesic with real (Euclidean) length. Considering the sphere is natural since, as we have seen previously, it is the Euclidean continuation of Lorentzian de Sitter spacetime. For simplicity, we will focus for now on two dimensions. Repeated here for convenience, the round sphere metric \eqref{d-dimsphere} for $d=2$ is 
\begin{equation}
\frac{ds^2}{\ell^2} = d\theta^2 + \cos^2 \theta \, d\varphi^2 \, ,  \label{line_element}
\end{equation}
where $\theta \in [-\pi/2,\pi/2]$ and $\varphi \sim \varphi + 2 \pi$. It is known that, given two points on the sphere, there is always a great circle that passes through them. The great circle is defined by the intersection of the plane containing the two points and the origin with the sphere. The two segments of the great circle are geodesics connecting the two points.\\
\\
It will be convenient to use a new set of coordinates $\{ \Theta, \Phi \}$, where the $\Phi$ angle moves around the great circle between the two points, with $\Phi \in [0, 2\pi )$ and $\Theta \in [-\pi/2, \pi/2]$. The great circle lies at $\Theta = 0$. Note that you can always move to this coordinate frame for any two points $X$ and $Y$ on the sphere. The metric on this coordinate system is given by
\begin{equation} \label{new_metric}
\frac{ds^2}{\ell^2} = d\Theta^2 + \cos^2 \Theta \, d\Phi^2 \,,
\end{equation}
so the length functional becomes
\begin{equation} \label{length_funct_2}
\tilde{L} [\Phi_X, \Phi_Y, \Theta] = \ell \int^{\Phi_Y}_{\Phi_X} d\Phi \, \sqrt{\dot{\Theta}^2 (\Phi) + \cos^2 \Theta (\Phi)}  \,.
\end{equation}
In this section, we use tildes to denote Euclidean quantities. We would like to compute the Euclidean two-point function on the sphere, 
\begin{equation} \label{euc_path_int}
\tilde{\mathcal{G}}(\Phi_X, \Phi_Y) = \int_{\Theta(\Phi_X)=0}^{\Theta(\Phi_Y)=0} D\Theta(\Phi) \exp (-m \tilde{L}(\Phi_X, \Phi_Y, \Theta)) \,, 
\end{equation}
using the geodesic approximation. We start by considering geodesics in $d=2$ in Section \ref{sec:euclidean_saddles}. We then consider perturbations to the geodesic length in Section \ref{sec:pert} and finally, we generalise our results to higher dimensions in Section \ref{sec_higher_d}.

\subsection{Euclidean saddle points} \label{sec:euclidean_saddles}
In the large mass limit, (\ref{euc_path_int}) is dominated by its saddle points which are the geodesics connecting $X$ and $Y$. To find these, we extremise the length (\ref{length_funct_2}). The equation of motion stemming from this length functional is
\begin{equation}
	\ddot{\Theta} + 2 \dot{\Theta}^2  \tan \Theta +\sin \Theta \cos \Theta = 0 \,, 
\end{equation}
which is solved by
\begin{equation} \label{theta of phi}
	\Theta ( \Phi) = \pm \arcsin \left[\frac{\sqrt{c_1-1 } \tan (\Phi + c_2)}{\sqrt{1+c_1 \tan^2(\Phi +c_2)}}\right]\,,
\end{equation}
where $c_1$ and $c_2$ are constants of integration. Boundary conditions at the endpoints of integration set $\Theta(\Phi_X) = \Theta(\Phi_Y) = 0$. A generic solution obeying the boundary conditions has $c_1=1$, reducing the solution to $\Theta_{\text{geodesic}} = 0$, {\it i.e.,} the geodesic goes through the great circle, as expected.  Evaluating the action on-shell gives the length of the shorter geodesic,
\begin{equation} \label{L-}
\tilde{L}_{g} =  \ell \int^{\Phi_Y}_{\Phi_X} d\Phi  \sqrt{\dot{\Theta}_{\text{geodesic}}^2 + \cos^2 \Theta_{\text{geodesic}}} =  (\Phi_Y - \Phi_X ) \ell \,,
\end{equation}
assuming, without loss of generality, that $0 < \Phi_Y - \Phi_X \leq \pi$. There is also another geodesic that goes around the other side of the great circle and has length
\begin{equation} \label{L+}
\tilde{L}_+ = \ell \int^{2\pi + \Phi_X}_{\Phi_Y} d\Phi \sqrt{\dot{\Theta}_{\text{geodesic}}^2 + \cos^2 \Theta_{\text{geodesic}}} = 2\pi\ell - (\Phi_Y - \Phi_X)\ell  = 2\pi\ell  - \tilde{L}_g \,.
\end{equation}
Note that, generically, $\tilde{L}_g \leq \tilde{L}_+$, so only $\tilde{L}_g$ will contribute to the Euclidean correlator in the large mass limit. However, as we will see in Section \ref{lorentzian_props}, we require both geodesics in certain cases to reproduce the correct Lorentzian correlator from analytic continuation of the Euclidean result. \\
\\
There also exists another (infinite) set of geodesics that wrap multiple times around the great circle. Their lengths are given by  $\pm\tilde{L}_g+ 2\pi \ell n$ with $n \in \mathbb{N}$, and their contribution to the correlator will always be exponentially suppressed in the large mass expansion, even after analytic continuation to Lorentzian spacetime.\footnote{With the exception of $-\tilde{L}_g + 2\pi\ell$, which is actually $\tilde{L}_+$.}\\
\\
On the sphere, the only case where both $\tilde{L}_g$ and $\tilde{L}_+$ will contribute corresponds to having $\Phi_Y - \Phi_X = \pi$. In this case, there is another set of solutions to (\ref{theta of phi}) that satisfy  the boundary conditions. These are a one-parameter family labelled by $c_1 \in \mathbb{R}^{\geq 1}$ and obtained by setting $c_2 = -\Phi_X$ in (\ref{theta of phi}). These geodesics correspond to rotating the great circle around the sphere, while keeping $\Phi_Y$ and $\Phi_X$ fixed. Note that you can only do this when $\Phi_Y - \Phi_X = \pi$. The length of all these geodesics is the same and reads
\begin{equation}
\tilde{L}_g= \pi \ell \,,
\end{equation}
independently of the choice of $c_1$. In what follows, we will assume that $\Phi_Y - \Phi_X < \pi$.
    
\subsection{Quadratic perturbations} \label{sec:pert}
We have found geodesics that are saddle points of the Euclidean propagator (\ref{euc_path_int}). Now we can compute the corrections to the propagator stemming from quadratic perturbations to the geodesic length on the sphere. One-loop path integrals on the sphere have been computed in, for instance, \cite{Anninos:2020hfj, Law:2020cpj}. Recall that we are parameterising our paths as $\Theta(\Phi)$, so we want to consider perturbations to the geodesics of the form
\begin{equation}
\Theta (\Phi) = \Theta_{\text{geodesic}} (\Phi) + \delta \Theta (\Phi) \,,
\end{equation}
where the geodesic equation just gives $
\Theta_{\text{geodesic}} (\Phi) = 0 \,. $  The variation of the Euclidean two-point function (\ref{euc_path_int}) is given by\footnote{In principle, the measure in this path integral should include a factor of $\cos(\Theta)$ that comes from the determinant of the metric \eqref{new_metric}. The inclusion of this factor is needed for the path integral to be diffeomorphism invariant \cite{Bekenstein:1981xe}. Note, however, that this term will not contribute to the path integral in the large mass expansion.}
\begin{equation}
\tilde{\mathcal{G}} (\Phi_X, \Phi_Y) \approx  \sum_{*=g,+} \left( e^{-m \tilde{L}_* (\Phi_X, \Phi_Y) }\int D \delta \Theta (\Phi) \exp \left( - m\, \delta\tilde{L}_* (\Phi_X, \Phi_Y, \delta\Theta, \delta\dot{\Theta}) \right) \right) \,. \label{GE}
\end{equation}
Evaluating the length functional (\ref{length_funct_2}) to second order around each geodesic, we obtain that,
\begin{eqnarray}
\tilde{L} (\Phi_X,\Phi_Y, \delta \Theta)  =  \tilde{L}_{*} (\Phi_X, \Phi_Y) + \delta \tilde{L}_{*} (\Phi_X, \Phi_Y, \delta\Theta, \delta\dot{\Theta})  \,, 
\end{eqnarray}
with
\begin{eqnarray}
\delta \tilde{L}_{*} (\Phi_X, \Phi_Y, \delta\Theta, \delta\dot{\Theta})  \equiv  \frac{\ell}{2} \int d\Phi \left(\delta \dot{ \Theta} (\Phi)^2 - \delta \Theta (\Phi)^2 \right)\,,
\end{eqnarray}
where the integration limits in this integral depend on which geodesic we are expanding around and are the same as in equations  (\ref{L-}) and (\ref{L+}). Given that $\delta\tilde{L}_{*}$ is quadratic in $\delta \Theta$, this path integral can be computed exactly. In general, consider the following quantum mechanical path integral,
\begin{equation}
Z (\Phi_0, \Phi_N) \equiv \int_{\delta \Theta(\Phi_0) = 0}^{\delta \Theta(\Phi_N)=0   } D\delta\Theta (\Phi)  \exp \left( - \frac{m \ell}{2} \int_{\Phi_0  }^{ \Phi_N} d\Phi \left( \delta \dot{ \Theta}^2 - \delta\Theta^2 \right) \right) \,,
\end{equation}
where the generic endpoints of the path integral are named $\Phi_0$ and $\Phi_N$, we assume $\Phi_N > \Phi_0$, and we require a vanishing variation of the trajectory at these points. In Appendix \ref{path_integral_app}, we show how to compute a more general class of quadratic path integrals in quantum mechanics, including this one. The final result is
\begin{equation} \label{one_loop_path_integral}
Z (\Phi_0, \Phi_N)= \sqrt{\frac{m \ell}{2\pi \sin (\Phi_N - \Phi_0)}} \,.
\end{equation}
Inserting this back into equation (\ref{GE}), we can write the correlator between any two points on the sphere, in the large mass limit, as a function of the geodesic length $\tilde{L}_g$ between the two points. This yields, 
\begin{equation} \label{euclidean_one_loop}
\tilde{\mathcal{G}} (\tilde{L}_g) = \sqrt{\frac{m \ell}{2\pi \sin \tilde{L}_g}} e^{-m \tilde{L}_g} + \sqrt{\frac{m \ell}{2\pi \sin(2\pi\ell-\tilde{L}_g)}} e^{-m (2\pi\ell - \tilde{L}_g)} \,,
\end{equation}
which looks suggestively similar to (\ref{GofL}) with $d=2$. We stress that this result is valid for any arbitrary two points on the sphere, as long as $\tilde{L}_g \gtrsim m^{-1}$ and $\pi \ell - \tilde{L}_g \gtrsim m^{-1}$. As previously mentioned, in Euclidean signature the second term in (\ref{euclidean_one_loop}) will always be exponentially suppressed in the large mass limit. However, we will keep both saddle points, because interestingly, in some cases, after doing the analytic continuation back to Lorentzian signature, they will both contribute to the Lorentzian, large mass two-point correlator.

\subsection{Generalising to higher dimensions} \label{sec_higher_d}
It is possible to extend the previous calculation on $S^2$ to higher dimensions. In this case, the analytic continuation of the global de Sitter metric is given by the round metric on $S^d$, equation \eqref{d-dimsphere}. In any dimension, it is again true that geodesics between any two points are sections of the great circle connecting those two points. As in the case of two dimensions, it is convenient to rotate the coordinates to a frame where the $\Phi$ coordinate goes around the great circle between the two endpoints. We will call these coordinates $\{\Theta_1, \cdots, \Theta_{d-1}, \Phi\}$. In this frame, the metric on $S^d$ is given by
\begin{equation}
\frac{ds^2}{\ell^2}  = d\Theta_1^2 + \cos^2 \Theta_1 d\Theta_2^2 + \cos^2 \Theta_1 \cos^2 \Theta_2 d\Theta^2_3 + \cdots + \cos^2 \Theta_1 \cos^2 \Theta_2 \cdots \cos^2 \Theta_{d-1} d\Phi^2  \,,
\end{equation}
so that the great circle lies at $\Theta_i = 0$. 
As in the two dimensional case, we can use $\Phi$ to parameterise the geodesic, which will follow a path $(\Theta_1(\Phi), \cdots, \Theta_{d-1} (\Phi))$, that extremises the length functional,
\begin{equation} 
		\tilde{L} = \ell \int d\Phi \, \sqrt{\dot{\Theta}^2_1 + \cos^2 \Theta_1 \left( \dot{\Theta}_2^2 + \cos^2 \Theta_2 \dot{\Theta}_3^2+   \cdots + \cos^2 \Theta_2 \cdots \cos^2 \Theta_{d-1} \right)} \, . 
\end{equation}
In Appendix \ref{app_ddim}, it is shown that the equations of motion imply that the saddle point is given by
\begin{equation}
\Theta_i^{(\text{geodesic})} = 0  \, , \qquad \text{for} \,\, 1 \leq i \leq d-1 \,.
\end{equation}
As in the case of two dimensions, this implies that there will be two geodesics leading the saddle point approximation. The one with minimal length is
\begin{equation} \label{L-d}
\tilde{L}_{g} = \ell \int^{\Phi_Y}_{\Phi_X} d\Phi  = (\Phi_Y - \Phi_X) \ell \,,
\end{equation}
where again we assume that $0 < \Phi_Y - \Phi_X < \pi$. The other geodesic goes around the remainder of the great circle and has length
\begin{equation} \label{L+d}
\tilde{L}_{+} = \ell \int^{2\pi + \Phi_X}_{\Phi_Y} d\Phi  = 2\pi\ell - (\Phi_Y - \Phi_X)\ell = 2\pi\ell - \tilde{L}_g \,.
\end{equation}
The contributions from other geodesics that wrap around the great circle multiple times will be exponentially suppressed in the two-point function, so we neglect them. We can expand the length functional around each geodesic trajectory, $\Theta_{i} = 0 + \delta \Theta_i$, and this gives 
\begin{equation}
\tilde{L} = \ell \int d\Phi \,  \left(1 + \frac{1}{2} \sum_{i=1}^{d-1} \left( \delta \dot{\Theta}_i^2 - \delta\Theta_i^2 \right)  \right) \,,
\end{equation}
where each of the $(d-1)$ terms in the sum give the same contribution to the path integral, and this is exactly the same contribution as in the two-dimensional case. So, finally, we get that for most\footnote{The same restrictions as in the $d=2$ case apply, {\it i.e.,} $\tilde{L}_g \gtrsim m^{-1}$ and $\pi \ell - \tilde{L}_g \gtrsim m^{-1}$.} two points on the higher dimensional sphere, the Euclidean correlator in the large mass limit can be written as a function of the geodesic length between the two points. The Euclidean correlator, to this order, is given by
\begin{equation} \label{euclidean_higher_d}
\tilde{\mathcal{G}} (\tilde{L}_g) = \left(\frac{m \ell}{2\pi \sin \tilde{L}_g}\right)^{\frac{d-1}{2}} e^{-m \tilde{L}_g} + \left(\frac{m \ell}{2\pi \sin (2\pi \ell  - \tilde{L}_g)}\right)^{\frac{d-1}{2}} e^{-m (2\pi\ell - \tilde{L}_g)} \,.
\end{equation}

\section{Lorentzian two-point functions} \label{lorentzian_props}

We will now use the results of the last section to reproduce the Lorentzian two-point function for a free massive scalar field in the Euclidean state $|E\rangle$. For simplicity, let us consider $d=2$. On the Euclidean sphere, the geodesic distance between any two points $(\theta_1, \varphi_1)$ and $(\theta_2, \varphi_2)$ is given by 
\begin{equation} \label{spheregeodesicdist}
	\tilde{L}_g = \ell \arccos \left[\cos \theta_1 \cos \theta_2 \cos (\varphi_2 - \varphi_1) + \sin \theta_1 \sin \theta_2 \right] \, . 
\end{equation}
Analytically continuing back to Lorentzian de Sitter space by taking $\theta \to i T$, we recover $L_g = \ell \arccos P_{X,Y}$, with $P_{X,Y}$ as in equation \eqref{P_global12}, even in the regime where Lorentzian geodesics do not exist. So it is straightforward to verify that Wick rotating \eqref{euclidean_one_loop} gives the large-mass expansion of the hypergeometric, up to a factor of $2m$. In fact, using (\ref{GofL}) and (\ref{euclidean_higher_d}), one can also verify that in higher $d$,
\begin{equation} \label{lorentzian_euclidean}
	\left. \tilde{\mathcal{G}} (\tilde{L}_g) \right|_{\theta \to i T} \, = 2 m \ell^{d-1}\, G(L_g) \, . 
\end{equation}
The apparent tension in Section \ref{sec:tension} is now resolved. For $P_{X,Y}<-1$, the complex nature of the Lorentzian geodesic length comes from analytic continuation of Euclidean geodesic lengths on the sphere. In this particular case, the Lorentzian geodesic length can be written as $L_g = \pi \ell - i \ell \,\text{arccosh} |P_{X,Y}|$. The second saddle (corresponding to the Euclidean geodesic encircling the sphere from the opposite side) has length $2\pi\ell - L_g$ and so, in Lorentzian signature, they both have the same real part. Thus, neither of them can be neglected in the large mass limit. This explains the need for both terms in (\ref{GofL}). \\
\\
When $P_{X,Y}>-1$, it is always true that the second saddle is exponentially suppressed, as it will always have a larger real part than $L_g$, after analytically continuing back the Euclidean answer. For any $P_{X,Y}$, it is important to keep the next order correction to the saddle point answer in order to reproduce the large mass correlator. In the remainder of this section, we explore several choices of points that illustrate the different features of the correlator and the geodesics in the different regimes.

\subsection{Timelike separated points}
\begin{figure}[H]
        \centering
         \subfigure[Timelike separated points]{
                \includegraphics[height=7cm]{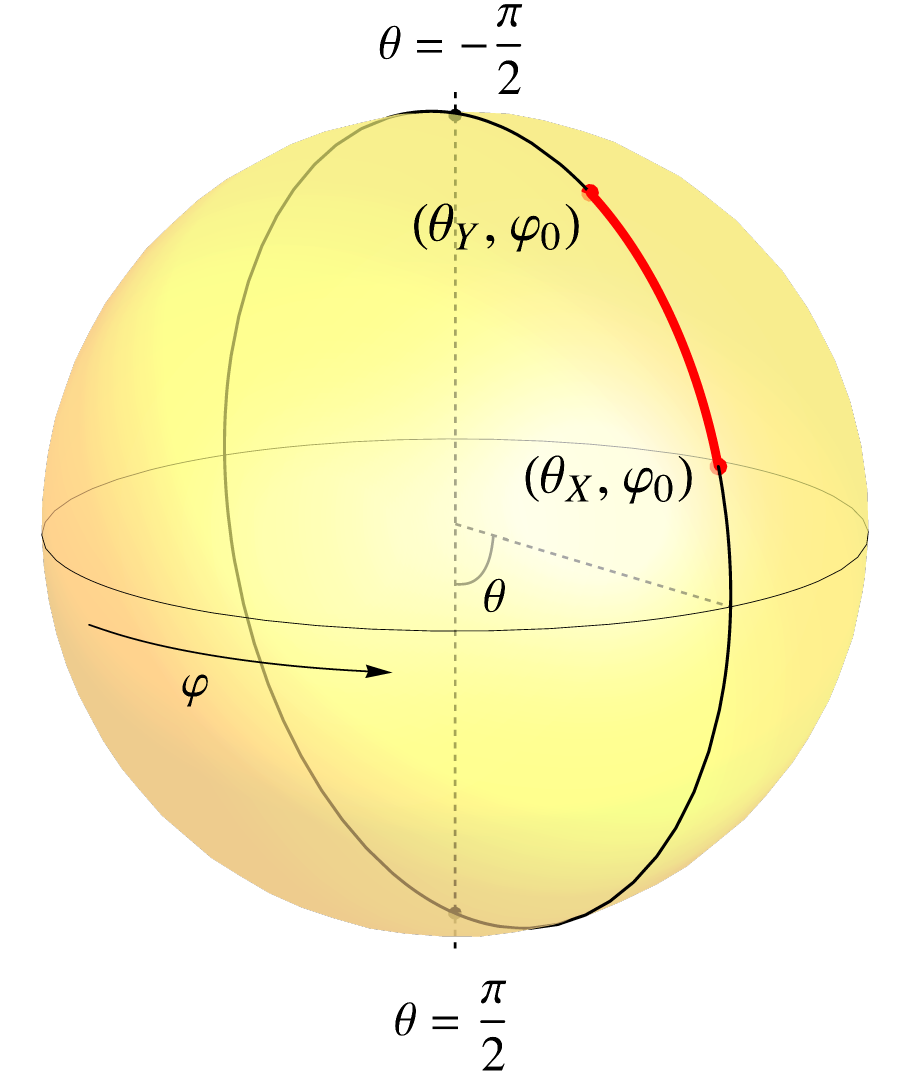}\label{fig:Timelike}}  \qquad \qquad 
        \subfigure[Spacelike separated points]{
                \includegraphics[height=7cm]{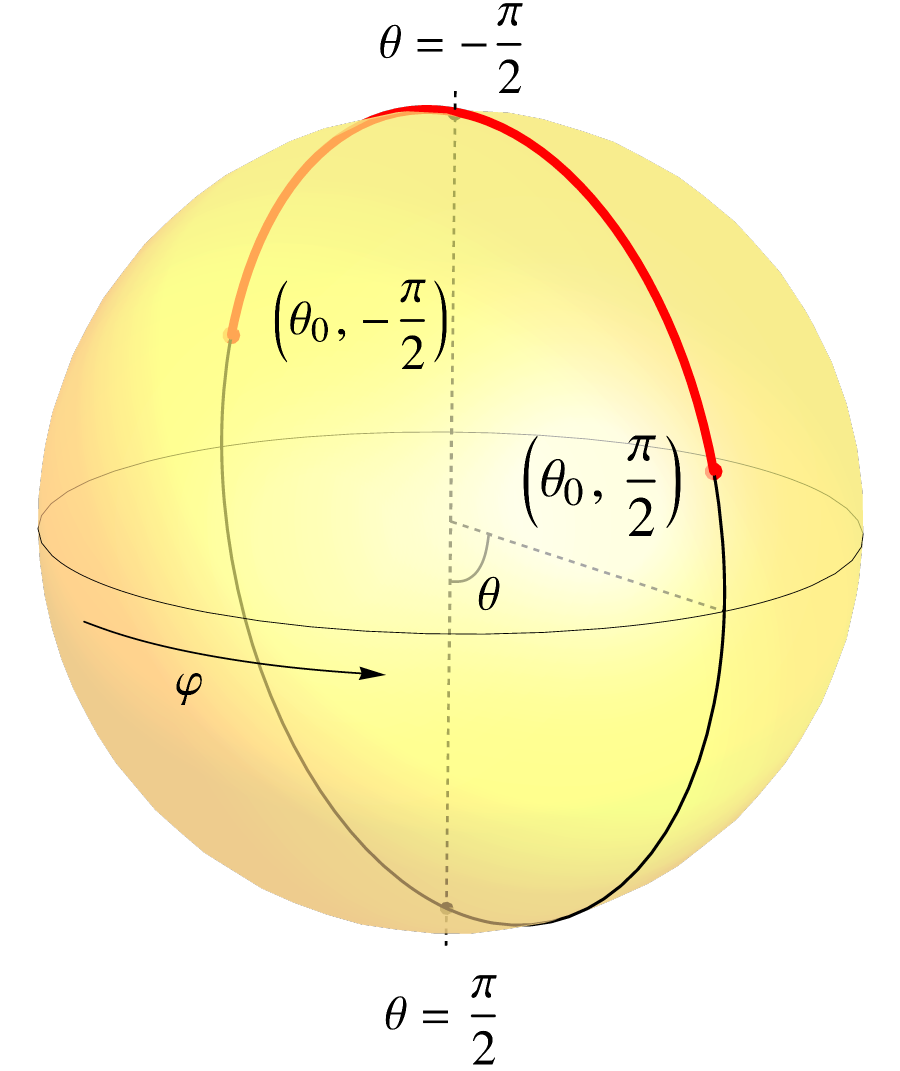} \label{fig:Spacelike}}
                                \caption{Upon analytic continuation to the Euclidean sphere, we look for two different types of geodesics, depending on the type of correlator under consideration. The geodesic with shorter length is shown in red, while the one going around the other side of the great circle is shown in black. }\label{fig:spheres}
\end{figure}

As a first example, consider a fixed point on the spatial $S^{d-1}$, at two different global times $T_1$ and $T_2$. In this case, $P_{X,Y} = \cosh (T_2 - T_1)>1$ and the relevant geodesics have a complex length $L_g = i \ell \, \text{arccosh} |P_{X,Y}|$.\\
\\
It follows from (\ref{P_greater}) that in the large mass limit the correlator is given by
\begin{equation} \label{timelike_d}
	G (P_{X,Y})  \approx 
	\frac{m^{\frac{d-3}{2}}}{2^{\frac{d+1}{2}}  (\pi \ell)^{\frac{d-1}{2}}  \sinh^{\frac{d-1}{2}} \left| T_2-T_1 \right|} e^{-i m \left| T_2-T_1 \right|} e^{-i \frac{\pi}{4} (d-1)} \, . 
	\end{equation}
This correlator can be obtained from the geodesic approximation both in Lorentzian and in Euclidean signature. Given that $L_g$ is purely imaginary, it is clear that the contribution coming from the geodesic with length $\tilde{L}_+ = 2\pi\ell - \tilde{L}_g$  will always have a larger real part when taken to Lorentzian signature, so it will be exponentially suppressed. To illustrate this case, consider the sphere for $d=2$. We choose points with the same spatial angle $\varphi = \varphi_0$. On the sphere they will look as in Figure \ref{fig:Timelike}. The Euclidean geodesic length will be given by $\theta_Y - \theta_X$, with $\theta_Y > \theta_X$, and this is enough to reproduce (\ref{timelike_d}) for $d=2$.

\subsection{Spacelike separated points}
Next, consider opposite points on the $S^{d-1}$ at a given global time $T$. 
In this case, $P_{X,Y} = -\cosh 2T < -1$. 
In the large mass limit, it follows from equation (\ref{P_less}) that,
\begin{equation}
G (P_{X,Y})  \approx \frac{m^{\frac{d-3}{2}}}{2^{\frac{d}{2}}  (\pi \ell)^{\frac{d-1}{2}}} \frac{ e^{-m \ell \pi}}{\sinh^{\frac{d-1}{2}}2 |T|} \left(\cos \left(2 m \ell |T| - \frac{\pi  d}{4}\right) - \sin \left(2 m \ell |T| - \frac{\pi  d}{4}\right)\right) \,. \label{two_point_d}
\end{equation}
The correlator is real, but it oscillates with a frequency of $2m \ell$. Furthermore, it exponentially decays as a function of time and the decay rate does not depend on the mass. There do not exist real geodesics to account for this behaviour. This result for the two-point function was first found for $d=2$ and $d=4$ in \cite{Galante:2022nhj} and \cite{Anninos:2022ujl}, respectively.\\
\\
Again, to illustrate this behaviour we focus on $d=2$. On the sphere, it is natural to choose points opposite to each other at a given latitude $\theta = \theta_0$ (one at $\varphi = -\tfrac{\pi}{2}$ and the other one at $\varphi = \tfrac{\pi}{2}$), as in Figure \ref{fig:Spacelike}. Then, the Euclidean geodesic lengths are given by
\begin{equation} \label{l- spacelike}
\tilde{L}_g (\theta_0) = (\pi  - 2 \theta_0 )\ell \,, \qquad \tilde{L}_+ (\theta_0)  = (\pi  + 2 \theta_0 )\ell \,.
\end{equation}
Note that, in this case, both will have the same real part when we analytically continue back to Lorentzian spacetime, and so neither can be neglected. Plugging this $\tilde{L}_g$ into (\ref{euclidean_one_loop}), it is straightforward to verify (\ref{lorentzian_euclidean}).

\subsection{Spacelike separated points between stretched horizons}
\begin{figure}[h!]
        \centering 
        \subfigure[Penrose diagram]{
                \includegraphics[height=5.85cm]{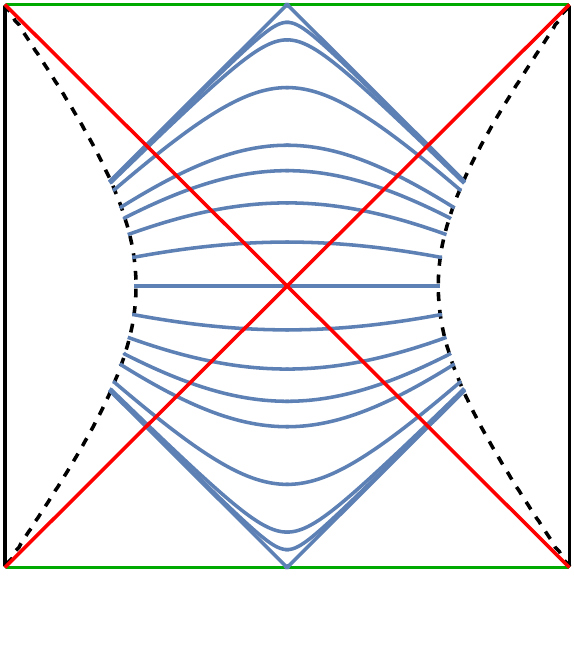} \label{fig:stretched}} \qquad
                  \subfigure[Correlator]{
                \includegraphics[height=5.8cm]{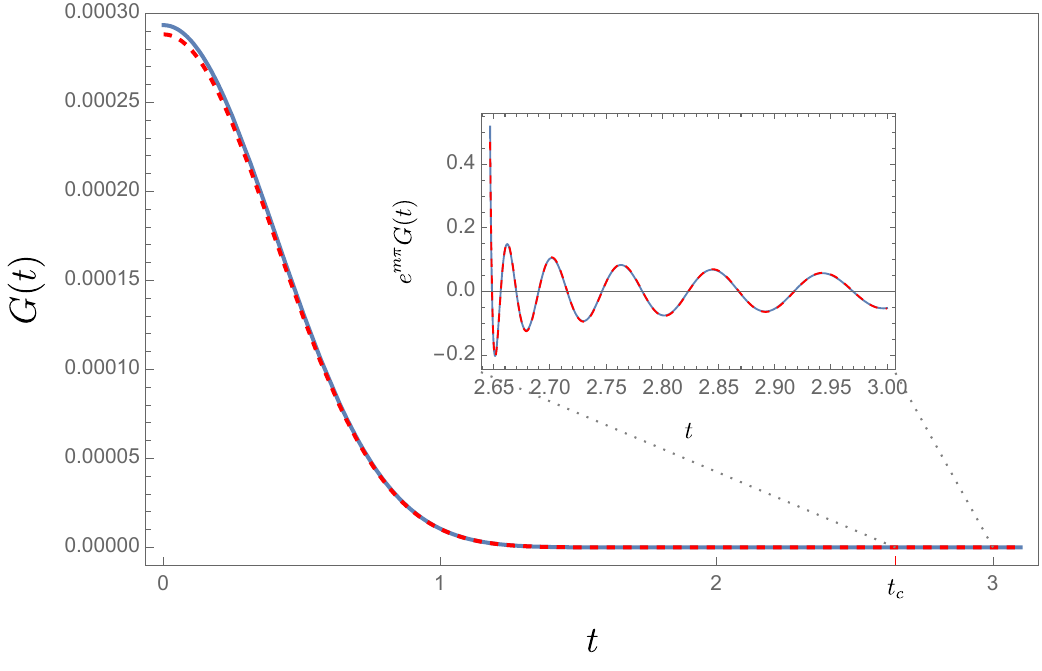}\label{fig:stretchedP}}  
                               
\caption{(a) Penrose diagram showing the geodesics (in blue) between two stretched horizons (in dashed black). The position of the stretched horizon is set to $r_{st}=2/3$. (b) Exact (dashed red) and large mass (solid blue) correlator for two symmetric points on opposite stretched horizons as a function of the static patch time. At early times  we observe a decay of the correlation function while at later times we observe an oscillatory behaviour. We multiply by $e^{m\pi}$ in the inset to make the oscillations apparent. In the plot, $m=20$ and $r_{st}=0.99$, so $t_c \sim 2.65$. Given the stretched horizon is very close to the actual horizon, at $t\sim0$, the two points become very close to each other, and so the approximation breaks down.}
\end{figure}

So far we have only considered points for which either $P_{X,Y} >1 $ or $P_{X,Y}  < -1$. It is interesting to consider a case where $P_{X,Y}$ goes through the transition point $P_{X,Y} = -1$. A concrete example of this involves studying the form of the correlator  between points anchored at opposite stretched horizons as a function of the static time; see Figure \ref{fig:stretched}.  For simplicity, we again restrict to $d=2$. As discussed, all we need in order to find the correlator is $P_{X,Y}$ between these points. The stretched horizon  $r_{st}$ is defined as a constant $r$ surface in the static patch metric (\ref{ddimStatic}). In order to get the position of the opposite stretched horizon, we need to relate static patch coordinates to global coordinates. Explicit expressions are shown in Appendix \ref{app:details}.

\begin{figure}[H]
        \centering
         \subfigure[$d=2$]{
                \includegraphics[height=4.5cm]{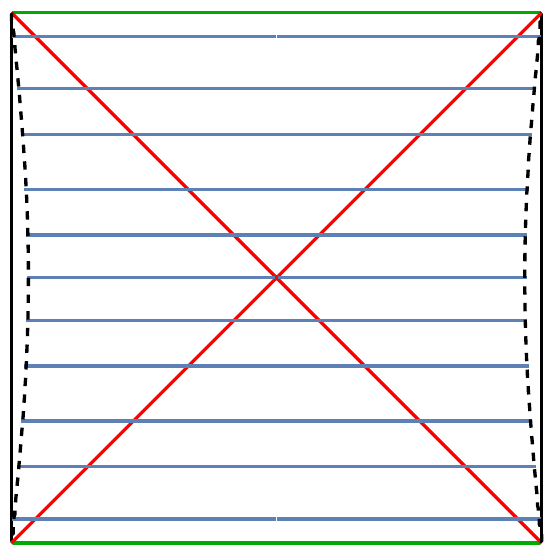}\label{fig:ads2}}  \qquad \qquad
        \subfigure[$d\geq 3$]{
                \includegraphics[height=4.5cm,trim={3.1cm 3.1cm 3.1cm 3.1cm},clip]{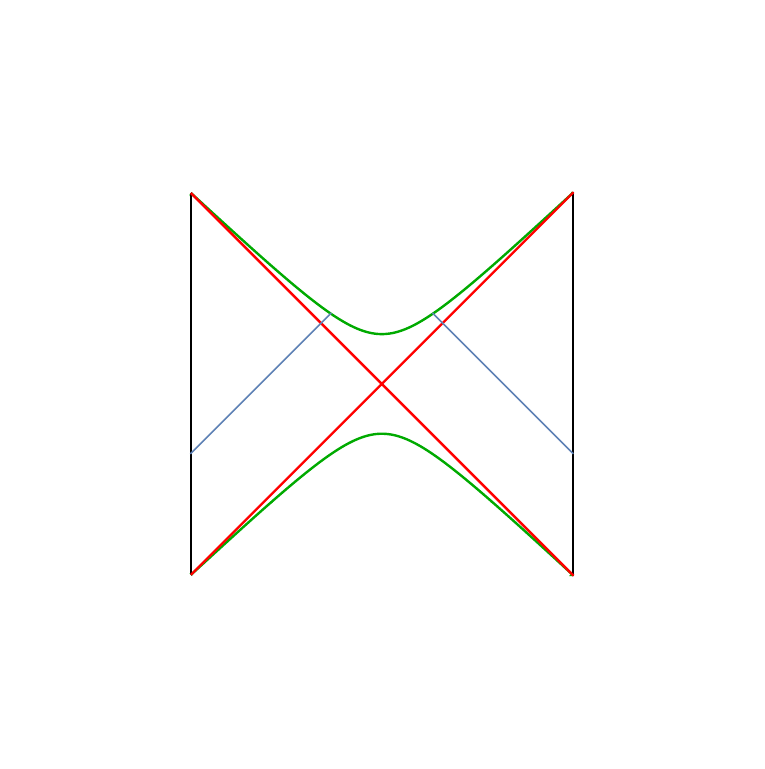} \label{ads_bhd}} 
                 \caption{Penrose diagrams for an anti-de Sitter black hole in (a) $d=2$ and in (b) $d\geq 3$. Geodesics are indicated in blue, horizons in red, and the singularities are in green. Figures adapted from those in \cite{Chapman:2021eyy}.}
                 \label{fig:penroseadsbh}
\end{figure}

It is clear that opposite points on the Penrose diagram with a fixed $r = r_{st}$ will have the same $T_0$ and their angles will be at $\varphi_0$ and $- \varphi_0$. 
In this setup then, one point is at embedding coordinates $X$ and the second one, $Y$, is at the same coordinates except for $Y_2 = - X_2$. This gives,
\begin{equation} 
    P_{X,Y}^{st} = \frac{1}{\ell^2} \left[-(X_0)^2+(X_1)^2-(X_2)^2 \right] = \left(r_{st}^2-1\right) \cosh 2 t+r_{st}^2 \,. \label{StretchedP}
\end{equation} 
Note that $P_{X,Y}^{st} = -1$ at the critical time
\begin{equation}
t_c \equiv \frac{1}{2} \text{arccosh}\left(\frac{1+r_{st}^2}{1-r_{st}^2}\right) = \frac{1}{2} \log \left(\frac{1+r_{st}}{1-r_{st}}\right) \,. \label{critical_t}
\end{equation} 
For times between $0 < |t| < t_c$, the de Sitter invariant length has range $1 > P_{X, Y}^{st} > -1$, but for $|t| > t_c$, the range becomes $P_{X,Y}^{st} < -1$. The large mass correlator will then be given by (\ref{P_greater}) with $P_{X,Y}^{st}$ before $t_c$, and by (\ref{P_less}), after. We plot the correlator in Figure \ref{fig:stretchedP}. Before $t_c$ the correlator decays monotonically, while afterwards it starts oscillating with frequency proportional to the mass. This differs from what happens to the correlator for the black hole in AdS$_2$, for which real geodesics exist at all times, and hence there is no oscillatory behaviour \cite{Chapman:2021eyy, Galante:2022nhj}; see Figure \ref{fig:ads2}. \\
\\
We can easily obtain geodesics anchored at the stretched horizon by simply cutting parts of the geodesics obtained in Section \ref{sec_geo}. See Figure \ref{fig:stretched}. Note that this breaks the degeneracy of geodesics at $T=0$, and now there is at most one geodesic at each static time $t$. Their length is simply given by $L_g = \ell \arccos P_{X,Y}^{st}$, up to $|t| = t_c$. At these times, the last geodesics have length $L_g = \ell \pi$, as we are just removing two null pieces from the geodesics from Section \ref{sec_geo} that have that same length.\\
\\
After $t_c$, there are no more real geodesics. To find the complex geodesics, on the sphere we look for points at $(\theta_0,\varphi_0)$ and $(\theta_0, - \varphi_0)$. Using the geodesic length formula (\ref{spheregeodesicdist}), we find that
\begin{equation}
	\tilde{L}_g = \ell \arccos \left[\cos^2 \theta_0 \cos ( 2\varphi_0) + \sin^2 \theta_0 \right] \,  , 
\end{equation}
and for the longer geodesic, we have $\tilde{L}_+ = 2 \pi\ell - \tilde{L}_g $. After analytic continuation, we can transform the global coordinates into static patch ones, and this Euclidean geodesic length recovers the Lorentzian two-point function, as in (\ref{lorentzian_euclidean}). For $|t|>t_c$, we do need the contributions from both geodesics. It is interesting to note that the Euclidean length above works for both $P_{X,Y}^{st}$ smaller and larger than $-1$, so it seems that the Euclidean computation does not know about $t_c$.\\
\\
It is interesting to view these results from the perspective of the interpolating geometries that were introduced in Chapter \ref{Interpolating}. For such geometries, the presence of an asymptotic, timelike boundary permits us to interpret bulk correlators in terms of correlation functions of the boundary quantum mechanics. For heavy fields, we can employ a geodesic approximation to compute such correlators. In the two-sided geometries, the result is that geodesics between opposite boundaries only exist for a short period of time (of order of the inverse temperature), after which there are no more real geodesics \cite{Chapman:2021eyy, Galante:2022nhj}. The last geodesic is almost null and goes all the way to the future (or past) infinity. This is reminiscent of what happens for the AdS double-sided black hole in dimensions higher than three. In that case, the singularity in the Penrose diagram bends inwards, and spacelike geodesics from the boundary are able to reach it \cite{Fidkowski:2003nf, Festuccia:2005pi}. This is shown in Figure \ref{ads_bhd}. However, the boundary correlator is not expected to have singularities of this type. The resolution is that there exist complex geodesics, whose contributions to the correlator are necessary to reproduce the correct answer for the boundary correlator. A similar story might hold for interpolating geometries where both the correlator and the geodesics can be computed in the bulk.

\begin{figure}[H]
        \centering
        \includegraphics[height=4cm]{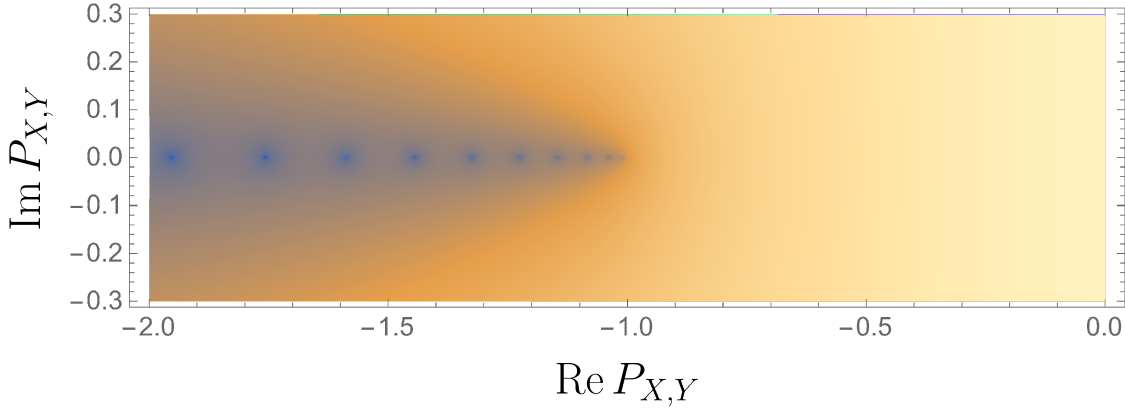}
\caption{Density plot of $|G(P_{X,Y})|$ for $d=4$ and $m=25$. In the large-$m$ limit zeroes (blue dots) accumulate near $P_{X,Y}=-1$. }\label{fig:zeroes}
\end{figure}

One might worry that some of the effects shown here could be hard to observe, due to the exponential suppression in the large mass limit. However, the characteristic oscillations\footnote{As in the case of the AdS black hole \cite{Fidkowski:2003nf, Amado:2008hw}, these oscillations are probably related to the quasinormal mode frequencies of the cosmological horizon \cite{Lopez-Ortega:2006aal}.} of the two-point function will already be present for large separations in the exact correlator as soon as  $m>(d-1)/2$. A harder question is whether there are signatures of the breakdown of the approximation in the exact correlator at $P_{X,Y}=-1$ in the large mass limit. As in \cite{Fidkowski:2003nf}, we see \textit{subtle but distinct} signatures of this breakdown in the exact correlator that show up as an accumulation of zeroes near $P_{X,Y}=-1$ as we increase the mass; see Figure \ref{fig:zeroes}.

\part{Three spacetime dimensions}
\label{Part2}
\chapter*{Introduction}
Uncovering the microscopic origin of the cosmological horizon entropy is an important and intriguing pursuit. Essentially, one wishes to test the proposal of Gibbons and Hawking that the entropy is given by the logarithm of the Euclidean gravitational path integral, as in equation \eqref{GibHawk}. The Euclidean path integral is somewhat of a black box - it seems to produce the correct answer for the horizon entropy, but how it does so is a bit mysterious given that we do not live in a Euclidean universe, but rather a Lorentzian one. Verifying the proposal, therefore, comes down to a search for an interpretation of this entropy in terms of a Hilbert space where we can count microstates explicitly, as this would constitute a more `Lorentzian' picture of the thermodynamics. We have already noted that the sphere topology is the sensible saddle to compute the Euclidean gravitational path integral around, since both the global spacetime and the static patch metric Wick rotate to the round sphere metric. A more Lorentzian picture could be to consider quantum fields in a thermal density matrix within a single static patch. This can be achieved by placing them in a pure Hartle-Hawking state across the full $S^{d-1}$ spatial slice at $T=0$ and tracing out one of the two static patch regions. The Lorentzian topology that we will consider is therefore that of a single static patch, which is $D \times \mathbb{R}$ where $D$ is a disc of appropriate dimension. This can be seen by taking constant time slices in the static patch metric \eqref{ddimStatic}. Thus, $S^{d-1}$ timeslices in the global spacetime  \eqref{ddimGlobal} get split into two static patch discs with a common $S^{d-2}$ cosmological horizon. \\
\\
In this part of the thesis, we conduct this search for a microscopic cosmological entropy in the context of three-dimensional gravity. The main reason we have chosen this setting is that three-dimensional gravity is semi-classically equivalent to a Chern-Simons gauge theory, where the choice of the gauge group determines the sign of the cosmological constant and the signature of the metric \cite{Achucarro:1987vz,Witten:1988hc}. This equivalence is beneficial as there already exists a plethora of results for Chern-Simons theory on both compact and bounded manifolds that will be useful for our purposes. For example, placing Chern-Simons theory on a manifold with a boundary, such as the $D \times \mathbb{R}$ topology of the static patch, induces an edge-mode theory at that boundary. One can also draw inspiration from literature on topological entanglement entropy \cite{Kitaev:2005dm,Levin:2006zz,Fendley:2006gr}, which argues that upon quantising Chern-Simons theory on a spatial $S^2$, dividing the space into two discs produces an entanglement entropy between the two regions. This entropy is given by the ${\mathcal{S}_0}^0$ component of the modular $S$-matrix ${\mathcal{S}_m}^n$ of the edge-mode theory. Furthermore, as was shown in the seminal work \cite{Witten:1988hf}, ${\mathcal{S}_0}^0$ also computes the partition function of Chern-Simons theory on an $S^3$ topology. One might therefore expect to see a match between the partition functions of the Euclidean theory evaluated on the three-sphere saddle and that of the edge-mode theory. However, the gauge group required to produce de Sitter in Lorentzian signature is $SL(2,\mathbb{C})$. Such complexified Chern-Simons theories have been the subject of various interesting works including \cite{Witten:1989ip,Witten:2010cx,Gukov:2016njj,Dimofte:2009yn,Dimofte:2016pua,Vafa:2015euh}. The non-compact nature of this group makes it difficult to calculate the spectrum of the theory explicitly. We therefore tackle the problem by first considering the contributions to the thermodynamics of the cosmological horizon of massless Chern-Simons fields on a fixed de Sitter background. We proceed by studying analogous Abelian Chern-Simons theories that have a gauge group structure similar to that of the $3$D gravity theory. For instance, for the Lorentzian theory we will take the gauge group to be $U_{\mathbb{C}}(1)$ to mimic the $SU_{\mathbb{C}}(2) \cong SL(2,\mathbb{C})$ of gravity. The possibility of using Chern-Simons theory to study three-dimensional gravity also appears in \cite{Smolin:1995vq, Soo:2001qf, Smolin:1994qb, Hikida:2022ltr, Castro:2023dxp}.\\
\\
In Chapter \ref{SoftDeSitter}, we explore contributions from massless fields to the thermodynamic properties of the de Sitter horizon in three dimensions. In Section \ref{css3} we review and place several relevant results from the Chern-Simons literature into the context of three-dimensional de Sitter space. In Section \ref{csab} we describe in detail the relation between the edge-mode theory residing at the boundary of a spatial disc and the three-sphere partition function for Abelian Chern-Simons theory. The edge-mode theory introduced in this section serves as the basic template for the latter sections. In Section \ref{complexCSL} we discuss an Abelian Chern-Simons theory with complexified $U(1)$ gauge group. This is meant to serve as an Abelian toy model for Lorentzian three-dimensional gravity expressed as a Chern-Simons theory with a complexified $SU(2)$ gauge group. In \ref{complexCS} we introduce an Abelian Chern-Simons theory with complexified level that is meant to act as a simplified toy model of Euclidean gravity expressed as a Chern-Simons theory with a complexified level. In Section \ref{oneloop3dgrav} we briefly discuss how to extend the results of sections \ref{complexCSL} and \ref{complexCS} to the non-Abelian case relevant to general relativity. In Section \ref{adscft} we make some comments on AdS$_4$/CFT$_3$ in the case where AdS$_4$ has a three-dimensional de Sitter boundary. In Chapter \ref{GeneralBCs}, we consider a more general edge-mode theory to see whether specific boundary conditions may be able to tame the unboundedness that we saw in the original problem. Appendices \ref{S3det}, \ref{Zthermal}, and \ref{LedgeApp} provide further details on various computations presented in the main text of this section.

\chapter{Topological contributions to the de Sitter entropy}
\label{SoftDeSitter}
In three dimensions, for a theory of pure Einstein gravity with a positive cosmological constant $\Lambda = 1/\ell^2$, one has\cite{Anninos:2020hfj,Carlip:1992wg,Guadagnini:1995wv,Castro:2011xb}
\begin{equation}\label{logZgrav}
\log \mathcal{Z}_{\text{grav}} =   S_{\text{dS}} - 3 \log  S_{\text{dS}} + 5\log 2\pi  + i \varphi_{\text{grav}} + \ldots ,
\end{equation}
where $S_{\text{dS}} = \pi \ell/2G \gg 1$ is the tree-level Gibbons-Hawking entropy of the de Sitter horizon, which is parametrically large in the semiclassical limit. Unlike the leading term, the subleading terms in (\ref{logZgrav}) stem from perturbative quantum corrections of the gravitational fluctuations. The logarithmic term is related to the residual $SO(4)$ subgroup of the diffeomorphism group preserved by the three-sphere saddle. The constant part is unambiguous in odd spacetime dimensions as it cannot be absorbed into a local counterterm, while the phase $\varphi_{\text{grav}}$ can arise due to the unboundedness of the conformal mode. In the treatment of \cite{Polchinski:1988ua} one finds that $\varphi_{\text{grav}} = - {5 \pi  }/{2}$ to leading order at large $S_{\text{dS}}$. Although three-dimensional gravity carries no propagating degrees of freedom, the structure (\ref{logZgrav}) is non-trivial, and a microscopic understanding is so far lacking (attempts include \cite{Park:1998qk, Maldacena:1998ih,Banados:1998tb,Govindarajan:2002ry,Dong:2010pm}). \\
\\
Further corrections to (\ref{logZgrav}) will appear as even inverse powers of $S_{\text{dS}}$. Upon coupling the theory to matter fields, the expansion (\ref{logZgrav}) may receive additional contributions. Consider first matter fields that are parametrically heavy with respect to the de Sitter length $\ell$. Integrating them out results in an effective gravitational theory that can be organised as a power series in the inverse mass, with coefficients that are functions of the metric \cite{Vassilevich:2003xt}. In three dimensions these coefficients can only be built out of powers of the Ricci tensor. Since the mass is large, this theory will be perturbed, to high accuracy, by a local functional of the metric containing higher derivative terms. Provided parity is preserved, due to the absence of local degrees of freedom in three-dimensional gravity, such a theory can always be brought back to an Einstein theory with a cosmological constant through local field redefinitions of the metric \cite{Anninos:2020hfj}.\footnote{Note that if the number of massive fields becomes large the coefficient of each term in the power series may be enhanced, and therefore this is only strictly true for a finite number of fields.} From this perspective, integrating out massless or light fields, or some more general conformal matter theory, is interesting in that one can affect the details of (\ref{logZgrav}) which are of a more non-local nature. The two-dimensional version of this problem was recently explored in \cite{Anninos:2021ene,Muhlmann:2021clm}. In what follows we focus on the contribution to (\ref{logZgrav}) from Chern-Simons gauge fields. These are massless gauge theories which are under solid theoretical control while producing various interesting modifications to (\ref{logZgrav}). We will be interested in investigating the hypothesis that
\begin{equation}\label{hypothesis}
\mathcal{Z}_{\text{grav}} \overset{?}{=} \lim_{\beta_T\to0^+}  Z_{\text{edge}}[\beta_T]\,,
\end{equation}
where $\beta_T$ is the inverse Tolman temperature as defined in equation \eqref{boundarymetric} and $Z_{\text{edge}}$ is the thermal partition function of the proposed edge-mode theory, which is evaluated at high temperature due to the parametric proximity of the edge-modes to the de Sitter horizon. In this part of the thesis, we provide evidence that the subleading terms in equation \eqref{logZgrav} can be interpreted in terms of such edge-modes, but the interpretation of the leading term in terms of entanglement entropy of an edge-mode theory is left open.  \\
\\
As already mentioned, in much of what follows we will consider contributions to the thermal properties of a single static patch region from Chern-Simons and gravitational fields. With this in mind, we will repeat here the static patch metric. Here, it will be useful to make the coordinate transformation $r = \sin \rho$ in the static patch metric \eqref{ddimStatic}, which for $d=3$ reads
\begin{equation}\label{sp}
\frac{ds^2}{\ell^2} = -\cos^2 \rho \,  dt^2 + d\rho^2 + \sin^2\rho \, d\varphi^2\,, 
\end{equation}
where $t \in\mathbb{R}$ and $0 \leq \rho \leq \pi/2$. The de Sitter horizon now resides at $\rho = \pi/2$ and has size $2\pi \ell$ in these coordinates. We note that the original global Cauchy surface at $T=0$, which takes the form of a spatial $S^2$, has been split into two discs whose common $S^1$ boundary comprises the dS$_3$ horizon. The induced metric on each disc at a constant $t$ slice is given by
\begin{equation}\label{diskmetric}
\frac{ds^2}{\ell^2} = d\rho^2 + \sin^2\rho \, d\varphi^2\,.
\end{equation}
In three dimensions, the hypersurface \eqref{ds embedding} preserves an $SO(3,1) \cong SL(2,\mathbb{C})/\mathbb{Z}_2$ subgroup of the Poincar{\'e} symmetries of $\mathbb{R}^{1,3}$, which constitutes the isometry group of dS$_3$. As discussed in the introduction, analytically continuing equation (\ref{ds embedding}) to Euclidean signature results in an $S^3$ embedded in $\mathbb{R}^4$, and the isometries become $SO(4)\cong SU(2)\times SU(2)/\mathbb{Z}_2$. This reflects the fact that three-dimensional gravity with positive cosmological constant can be written as a combination of Chern-Simons theories with gauge group $SL(2,\mathbb{C})$ for Lorentzian signature and $SU(2) \times SU(2)$ for Euclidean signature. 

\section*{Entanglement \texorpdfstring{$\&$}{and} gravitational entropy}\label{graventropy}
It is generally a complicated problem to separate two regions of a gauge theory (including gravitational theories) in a strictly local way. In discussions of entanglement entropy for ordinary gauge theories there are various proposals for how to deal with the problem of the entangling surface \cite{Buividovich:2008gq,Donnelly:2011hn,Donnelly:2014gva,Casini:2013rba,Ghosh:2015iwa,Lin:2018bud}. For standard Yang-Mills theories, the possibilities are often labelled by a choice of `centre'  of the gauge invariant operator algebra. In Chern-Simons theories, which we will proceed to study, electric and magnetic fields do not commute. Consequently, it is not reasonable to classify choices in terms of electric and magnetic centres. Alternatively, \cite{Buividovich:2008gq,Donnelly:2014gva} one might extend the gauge invariant Hilbert space to admit non-trivial charges at the entangling surface. As overviewed in \cite{Lin:2018bud}, the extended Hilbert space picture matches calculations of entanglement entropy employing Euclidean path integral techniques. Our considerations will be in line with this approach since we are interested in making contact with the manifestly gauge-invariant three-sphere partition function. \\
\\
Consider Chern-Simons theory with gauge group $\mathcal{G}$ and level $k$ and whose three-sphere partition function we denote by $Z_{\mathcal{G}_k}[S^3]$.  Although $Z_{\mathcal{G}_k}[S^3]$ is a manifestly gauge invariant object, it will be related to a Lorentzian edge-mode calculation, which we describe in sections \ref{css3} and \ref{csab}, for which gauge invariance must be manifestly broken. Both the entanglement entropy of the edge-mode theory as well as $Z_{\mathcal{G}_k}[S^3]$ are ultraviolet divergent quantities. We will make sense \cite{Anninos:2020hfj} of the divergences of $Z_{\mathcal{G}_k}[S^3]$ by coupling the theory to three-dimensional gravity with cosmological constant $\Lambda = 1/\ell^2>0$, as in equation \eqref{GibHawk+matter}. The low energy effective theory in Euclidean signature is thus given by\footnote{The Chern-Simons action (being insensitive to the metric) is insensitive to the signature and hence appears in the same form for both the Lorentzian and Euclidean path integral.}
\begin{equation}
-S_E[g_{\mu\nu},A_\mu] = \frac{1}{16\pi G} \int d^3x \sqrt{g} \left( R-\frac{2}{\ell^2} \right) + i S_{\text{CS}}[A_\mu]\,,
\end{equation}
where $S_{\text{CS}}[A_\mu]$ is the Chern-Simons action. Unlike pure Chern-Simons theory, the above theory is no longer topological due to the explicit metric dependence in the gravitational sector. Here, we are distinguishing between a theory being topological, and a theory having no propagating degrees of freedom. These phrases are often used interchangeably but, specifically, we refer to a theory with no metric dependence as topological. When three-dimensional gravity is written as a Chern-Simons theory, the action depends explicitly on the metric through the gauge field. In this sense, the theory is no longer metric independent. The classical solution space for the metric remains that of three-dimensional gravity due to the metric independence of Chern-Simons theory. In three dimensions, this is nothing more than the round metric on the three-sphere or quotients thereof. The path integral of interest now becomes
\begin{equation}
\mathcal{Z}_{\text{grav} + \text{CS}} = Z^{\text{reg}}_{\mathcal{G}_k}[S^3] \times \mathcal{Z}_{\text{grav}}\,.
\end{equation}
The pure gravitational path integral $\mathcal{Z}_{\text{grav}}$ is given by (\ref{logZgrav}). Any ultraviolet divergences stemming from quantum fluctuations of the Chern-Simons sector can be absorbed into the cosmological constant $\Lambda$ and Newton constant $G$, such that $Z^{\text{reg}}_{\mathcal{G}_k}[S^3]$ is the finite ultraviolet regularised part of the Chern-Simons partition function. The overall phase of $Z^{\text{reg}}_{\mathcal{G}_k}[S^3]$ is affected by the choice of framing and is given by $\varphi_{\text{CS}} = 2 \pi n c_k/24$ where $c_k$ is the left-moving central charge of the WZW model with current algebra $\mathcal{G}_k$ \cite{Witten:1988hf} --- for $SU(2)_k$ this would be $c_k =3k/(k+2)$ --- and $n\in\mathbb{Z}$. We note that the total phase of $\mathcal{Z}_{\text{grav} + \text{CS}}$ is then given by
\begin{equation}
\varphi_{\text{grav} + \text{CS}} = \varphi_{\text{grav}} + \frac{2 \pi n c_k }{24} + \ldots 
\end{equation}
For the gravitational phase $\varphi_{\text{grav}} = -5\pi/2$ proposed in \cite{Polchinski:1988ua}, special choices of $n$ and $k$ can lead to a vanishing $\varphi_{\text{grav} + \text{CS}}$ modulo $2\pi$, at least to leading order in the large $S_{\text{dS}}$ expansion. \\
\\
At least for vanishing phase, $\log Z^{\text{reg}}_{\mathcal{G}_k}[S^3]$ can be viewed as a physical, ultraviolet finite correction to the tree-level Gibbons-Hawking de Sitter horizon entropy $S_{\text{dS}} = \pi\ell/2G$ \cite{Gibbons:1976ue}. So long as it is small compared to $S_{\text{dS}}$, there is no issue with the contribution to the entanglement entropy from the Chern-Simons sector being negative.

\section{Chern-Simons theory on \texorpdfstring{$S^3$ and dS$_3$}{S3 and dS3}}\label{css3}

In this section, we consider Chern-Simons theory on the manifolds $S^3$, $S^2\times \mathbb{R}$, and $D\times \mathbb{R}$, where $D$ denotes the two-dimensional disc. We frame the discussion in the language of Lorentzian and Euclidean three-dimensional de Sitter space. Chern-Simons theory with gauge group $\mathcal{G}$ and level $k \in \mathbb{Z}^+$ is described by the action
\begin{equation}\label{CSaction}
S_{\text{CS}}[A_\mu]  
= \frac{k}{4 \pi}   \int_M d^3 x \, \varepsilon^{\mu \nu \rho} \, \text{Tr} \,\left(A_{\mu}\partial_{\nu}A_{\rho}  + \frac{2}{3}A_{\mu}A_{\nu} A_{\rho}\right)\,,
\end{equation}
with $A_\mu = A_{\mu}^a T^a$, where $T^a$ are the anti-Hermitian generators of $\mathcal{G}$ satisfying the normalisation $\text{Tr}(T^a T^b) = - \frac{1}{2} \delta^{ab}$, and $a = 1,..., \text{dim}\,\mathcal{G}$, while $ \varepsilon^{\mu \nu \rho}$ is the Levi-Civita symbol. The non-Abelian field strength tensor is $F_{\mu\nu} = \partial_{\mu}A_{\nu} - \partial_{\nu}A_{\mu} + [A_{\mu},A_{\nu}]$. \\
\\
Our reason for studying Chern-Simons theories is threefold. Firstly, the topological nature of Chern-Simons theory ensures that all excitations have vanishing energy -- all physics is `soft physics'. In particular, it contributes to the non-local structure of the de Sitter entropy (\ref{logZgrav}). Secondly, we can draw from a host of exact results for Chern-Simons theory on compact manifolds, as well as on manifolds with boundaries. Both these features will prove to be particularly relevant to questions regarding de Sitter space. Thirdly, general relativity in three dimensions is equivalent, at least at the semiclassical level, to a Chern-Simons theory exhibiting certain unusual features such as a non-compact gauge group and/or a complexified value for the level \cite{Achucarro:1987vz,Witten:1988hc}. For now, we will steer clear of the third point, which we return to in Section \ref{oneloop3dgrav}, and focus on the first two.

\subsection{Chern-Simons theory on \texorpdfstring{$S^3$}{S3}}

We now consider the behaviour of Chern-Simons gauge fields on a fixed dS$_3$ or $S^3$ background. Pure Chern-Simons theory on a three-manifold $\mathcal{M}$ will only perceive topological features, so one might wonder if there is any lesson to be learned specifically about de Sitter. Here, it is helpful to consider the more general setup whereby we couple Chern-Simons theory to three-dimensional gravity. The Lorentzian/Euclidean theory admits dS$_3$/$S^3$ solutions. Given that the metric is semiclassically fixed, the theory is no longer topological. In this context we can ask how the presence of a Chern-Simons gauge field contributes to the thermodynamic properties of the dS$_3$ horizon. \\
\\
In the remainder of this section and the next our focus will be entirely on the case of a fixed background. We take the gauge group $\mathcal{G}$ to be compact and semi-simple, and the level $k \in \mathbb{Z}^+$. Recall that Chern-Simons theory is topological, and hence only sensitive to the type of manifold on which it resides. From our discussion of the dS$_3$ geometry one is led to consider various three manifolds. These are $S^3$, $S^2\times \mathbb{R}$, and $D \times \mathbb{R}$. 
\newline\newline\newline
\textbf{Chern-Simons theory on $S^3$.} The first object we consider is the partition function of Chern-Simons theory on $S^3$. Although na{\"i}vely this is nothing more than a number, and hence as good as our choice of normalisation, it turns out that Chern-Simons theory is sufficiently structured that one can make sense of this number in an essentially unambiguous way.\footnote{There is an ambiguity in the overall phase of the $S^3$ partition function, which is fixed by a choice of framing. Unless otherwise stated, we will select a framing for which the overall phase vanishes.} The result  \cite{Witten:1988hf} can be stated concisely in terms of the modular $S$-matrix ${\mathcal{S}_m}^n$ associated to the WZW CFT with gauge group $\mathcal{G}$ and level $k$, namely
\begin{equation}
Z_{\mathcal{G}_k}[S^3] = {\mathcal{S}_0}^0\,.
\end{equation}  
(An account of modular $S$-matrices and their properties can be found in the latter chapters of \cite{DiFrancesco:1997nk}. They encode the transformation of extended characters under inversion.) It should be emphasised that $Z_{\mathcal{G}_k}[S^3]$ is a gauge invariant object. As a concrete example, we can consider $\mathcal{G} = SU(5)$ for which
\begin{equation}
Z_{SU(5)_k}[S^3] = \frac{1}{\sqrt{5}(k+5)^2} \prod_{j=1}^{4} \left[2\sin \frac{\pi j}{k+5}\right]^{(5-j)}\,.
\end{equation}  
Notice that in the perturbative limit, where $k \to \infty$, the above expression is approximated by 
\begin{equation}
\lim_{k\to\infty} Z_{SU(5)_k}[S^3] \approx \frac{9}{\pi ^{14} \, 512\sqrt{5}}   \left(\frac{2\pi}{\sqrt{k+5}}\right)^{24}\,.
\end{equation}
The above expression has some identifiable features. For instance the power of $\sqrt{k+5}$ corresponds to $\text{dim} \,SU(5)=24$. Whereas the numerical pre-factor is the reciprocal of the canonical group volume of $SU(5)$, given by (see for example \cite{Ooguri:2002gx,Anninos:2020ccj}) 
\begin{equation}
\text{vol} \, SU(N) = \frac{\sqrt{N}}{2\pi} \frac{(2\pi)^{N(N+1)/2}}{G(N+1)}\,,
\end{equation}
where $G(z)$ is the Barnes $G$-function. More generally, the large $k$ limit of the sphere partition function of Chern-Simons theory with gauge group $\mathcal{G}$ is given by 
\begin{equation}\label{CSpert}
\lim_{k\to\infty} Z_{\mathcal{G}_k}[S^3] \approx \frac{1}{\text{vol} \, \mathcal{G}} \left( \frac{2\pi}{\sqrt{k+h}}\right)^{\text{dim}\, \mathcal{G}}\,,
\end{equation} 
where $h$ is the dual Coxeter number of $\mathcal{G}$. Schematically, we may view the above structure as stemming from the fact that the path integral requires division by the space of gauge transformations, of which all but the constant part are cancelled by redundancies in the local description of the theory. When a quantum field theory lives on a compact space, the constant part of a compact gauge group is finite and must be taken into account. Indeed, its contribution is responsible for the group theoretic factors exhibited by (\ref{CSpert}). The dependence on the level $k$ is also sensible. The volume of the constant part of the gauge group must be normalised with respect to the size of the fluctuations of the theory, which are controlled by $1/\sqrt{k}$ in the perturbative limit. Such group theoretical factors have been recently explored in \cite{Anninos:2020hfj}. 

\subsection{Chern-Simons on \texorpdfstring{$S^2\times \mathbb{R}$ and $D \times \mathbb{R}$}{S2 X R and D X R}}
\begin{figure}[H]
	\centering
	\includegraphics[width=0.4\linewidth]{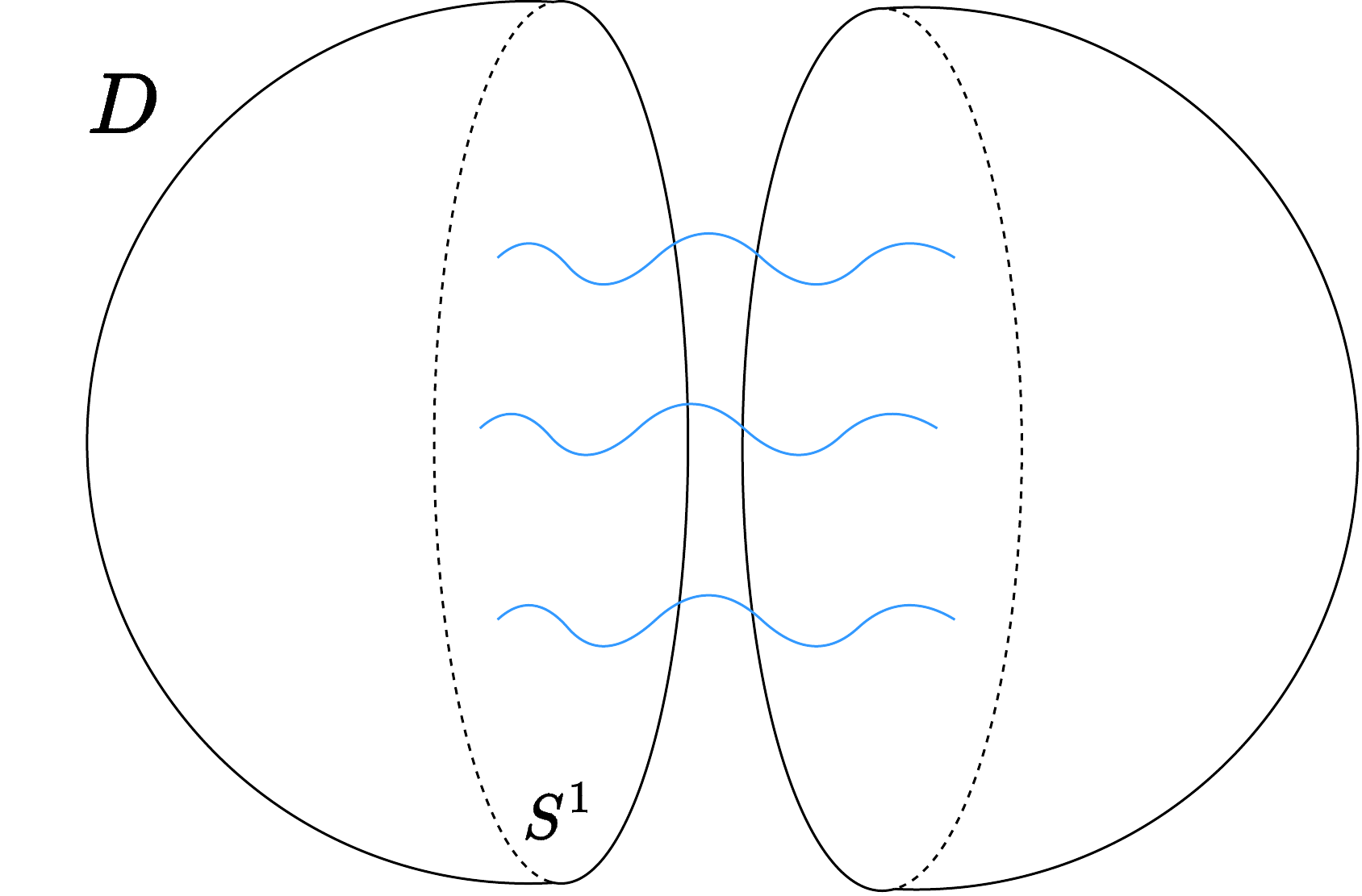}
	\caption{The $T=0$ slice of global de Sitter split into two static patch hemispheres at the $S^1$ horizon. These hemispheres are topologically discs, and we can imagine `tracing out' the degrees of freedom in one of the static patches that is behind the horizon of the other.}
	\label{fig:S2tracingoutEE}
\end{figure}

We now consider Chern-Simons theory on $S^2\times \mathbb{R}$ and $D \times \mathbb{R}$. The first is topologically equivalent to the global geometry (\ref{ddimGlobal}) of dS$_3$, while the second is topologically equivalent to the static patch (\ref{sp}). \\
\\
\textbf{{Chern-Simons theory on $S^2\times \mathbb{R}$}.} We now proceed to consider Chern-Simons theory on $S^2\times \mathbb{R}$. Here, we view $\mathbb{R}$ as the global dS$_3$ time coordinate $T$, and the $S^2$ as the spatial Cauchy slice of global dS$_3$. It is convenient to choose the following metric on $S^2$
\begin{equation}
ds^2 = \frac{4 d \mathbf{x}^2}{(1+\mathbf{x}^2)^2}\,, \quad\quad \mathbf{x} = \{x,y\} \in \mathbb{R}^2\,.
\end{equation}
We begin with a description of the Hilbert space. This is given by quantising the space of flat $\mathcal{G}$-connections on $S^2$, modulo gauge transformations. But on $S^2\times \mathbb{R}$, all flat connections are trivial. The Hilbert space is one-dimensional. The unique state of this peculiar world can be explicitly described in the Schr{\"o}dinger picture. To describe it, it is convenient to work in the gauge $A_T=0$. The gauge constraints that we must solve are then given by
\begin{equation}
F_{ij} = 0\,, \quad\quad i,j \in \{x,y\}\,.
\end{equation}
The following quantum state can be shown to solve the above equation \cite{Dunne:1989cz}
\begin{equation} \label{CSnon-abelianWavefucntion}
\Psi_{S^2}[A_x(\mathbf{x})] = \mathcal{N} \exp{\left[\frac{ik}{4\pi}  \text{Tr} \int_{S^2} d^2 {x}   \left[ (g^{-1}\partial_x g) ( g^{-1} \partial_y g) \right] - 2 \pi i k \int_{S^2} \, d^2 {x} \, w^0 (g) \right]}\,,
\end{equation}
where $\mathcal{N}$ is a normalisation constant, $g(\mathbf{x}) \in \mathcal{G}$, and $w^0$ is the time-component of the three-vector whose divergence gives the winding number density:
\begin{equation}
W(g) = \partial_{\mu} w^{\mu} = \tfrac{1}{24 \pi^2} \text{Tr} \, \varepsilon^{\alpha \beta \gamma} (g^{-1} \partial_{\alpha}g g^{-1} \partial_{\beta}g g^{-1} \partial_{\gamma}g)\,.
\end{equation}
The gauge field is non-locally related to $g(\mathbf{x})$ as $A_x(\mathbf{x}) = g^{-1} \partial_x g$. 
Unlike what happens in standard Yang-Mills theory, the Chern-Simons wavefunctional $\Psi_{S^2}[A_x(\mathbf{x})]$ is not gauge invariant under the residual gauge freedom 
\begin{equation}
A^h_i(\mathbf{x}) = h^{-1}\left( \partial_i + A_i(\mathbf{x})\right) h\,, \quad\quad i \in \{x,y\}\,,
\end{equation}
with $h(\mathbf{x}) \in \mathcal{G}$. Rather, under the residual gauge freedom
$\Psi_{S^2}[A_x(\mathbf{x})]$ transforms as 
\begin{equation}
\Psi_{S^2}[A_x(\mathbf{x})] = e^{2\pi i \alpha(A_x(\mathbf{x}); \,h)} \Psi_{S^2}[A_x(\mathbf{x})]\,,
\end{equation} 
where
\begin{equation}
\alpha (A_x(\mathbf{x}); h) \equiv \frac{k}{8\pi^2}  \text{Tr}  \int d^2 x \left[2A_x(\mathbf{x}) \, \partial_y h \, h^{-1} + h^{-1}\partial_x h \, h^{-1} \, \partial_y h \right] - k \int d^2 x \, w^0 (h). 
\end{equation}
As a function of the gauge field, the ground state $\Psi_{S^2}[A_x(\mathbf{x})]$ encodes non-local quantum correlations. The non-local correlations play an important role, for instance, if we are to trace out those degrees of freedom living in a certain spatial region. Our considerations of dS$_3$ suggest performing precisely such a trace if we are to consider the physics contained within a single dS$_3$ horizon, as depicted in Figure \ref{fig:S2tracingoutEE}. 
\noindent {\textbf{Chern-Simons theory on $D \times \mathbb{R}$.}} As already noted, the spatial section of a single static patch is given by the hemisphere metric on a two-dimensional disc $D$ whose coordinates we denote by $\{\rho,\varphi\}$ with $\varphi\sim\varphi+2\pi$. From this perspective, we are encouraged to consider the problem of Chern-Simons theory on $D\times \mathbb{R}$, where now $\mathbb{R}$ is associated with the static patch time $t$. In this case, the Hilbert space of Chern-Simons theory is far richer \cite{Elitzur:1989nr}  due to the presence of the spatial $S^1$ boundary. \\
\\
We can construct the edge-mode theory as follows (see \cite{Tong:2016kpv} for an overview). First, we must ensure that we have a well-defined variational problem in the presence of the $S^1$ boundary. We will not add further terms localised on the boundary $\partial\mathcal{M}$ of the disc. As such, the variational problem enforces us to impose boundary conditions on $\partial \mathcal{M}$:
\begin{equation}\label{gauge}
\left( A_t - \upsilon A_\varphi \right)|_{\partial \mathcal{M}} = 0\,.
\end{equation}
The parameter $\upsilon$ is a real number which we take to be positive. We must further ensure that the gauge transformations do not disturb our boundary condition. This can be achieved by forcing the gauge parameter $\lambda$ to vanish on the $S^1$ boundary. We can further impose (at least locally) that the gauge condition (\ref{gauge}) holds away from the boundary of the disc as well. For that to be the case the bulk gauge parameter must  satisfy
\begin{equation}
\left(\partial_t -\upsilon \partial_\varphi \right) \lambda =  0\,, \quad\quad \lambda|_{S^1} = 0\,.
\end{equation} 
The above is solved by $\lambda = \bar{\lambda}(\varphi + \upsilon t,\rho)$ with $\bar{\lambda}|_{S^1} = 0$, which constitutes a residual gauge freedom. We can further specify a gauge fixing condition throughout the interior of the disc. The constraint arising from extending the gauge condition (\ref{gauge}) into the disc interior is solved by the following configurations
\begin{equation}
A_\rho  =  U^{-1} \partial_\rho U\,, \quad\quad   A_\varphi  =  U^{-1} \partial_\varphi U\,,
\end{equation}
with $U \in \mathcal{G}$. The residual gauge freedom transforms the group valued function $U \to e^{i \varepsilon \bar{\lambda}} U$, where $\varepsilon$ is a small parameter. Consequently, the residual gauge parameter can be entirely fixed upon fixing the form of $U$. Unless otherwise specified, we impose that $A_\mu$ remains smooth throughout the interior of the disc. To do so, we must ensure that the holonomy of the gauge field vanishes around any closed loop in the interior of the disc. Non-vanishing holonomy would indicate the presence of external charge puncturing the disc. The resulting theory at the boundary of the disc, upon taking everything into account, becomes a chiral WZW model with group $\mathcal{G}$ and level $k$ built from the boundary values of $U$ \cite{Elitzur:1989nr}. Upon quantisation, this theory has an infinite number of states. We now study the Abelian case in detail.
\newpage
\section{Abelian example}\label{csab}
In this section, we construct and explore the edge-mode theory associated to Abelian Chern-Simons theory with gauge group $U(1)$ at level $k$ quantised on a spatial disc $D$. The purpose of the section is to provide a detailed discussion of this edge-mode theory and its relation to the three-sphere partition function $Z_{U(1)_k}[S^3]$. This is meant to serve as the basic template for the discussion of edge-mode theories in the latter sections. The Abelian theory on a three-manifold $\mathcal{M}$ is governed by the action
\begin{equation}
S_{U(1)_k}[A_\mu] = \frac{k}{4\pi} \int_{\mathcal{M}} d^3 x \, \varepsilon^{\mu \nu \rho} A_\mu \partial_\nu A_\rho\,,
\end{equation}
where $A_\mu$ is a real-valued Abelian gauge field. So long as the three-manifold $\mathcal{M}$ has no boundary, the above is gauge invariant under $A_\mu \to A_\mu - \partial_\mu \lambda$, where $\lambda$ is a smooth function on $\mathcal{M}$. If it is further assumed that $U(1)$ is compact, and moreover that $S_{U(1)_k}$ is the low energy limit of a theory containing both electric and magnetic monopoles, then the standard Dirac argument enforces $k\in\mathbb{Z}$. For the sake of simplicity, we  take $k \in 2  \times \mathbb{Z}^+$ in what follows.\footnote{This is, in part, to avoid subtleties associated to fermionic states in the edge-mode theory.}

\subsection{Three-sphere partition function} 

We begin by computing the $S^3$ partition function. We have 
\begin{equation}
Z_{U(1)_k}[S^3] =  \left( {\text{vol} \, \mathcal{G}}\right)^{-1} \, {\int \mathcal{D}A_\mu \, e^{i S_{U(1)_k}[A_\mu]}}\,,
\end{equation}
where we have manifestly divided by the volume of the gauge group, which is generated by the space of smooth real functions on $S^3$. Given that $S^3$ is compact, we must also consider the constant part of the gauge group which is generated by the constant function $\lambda_c$ on $S^3$. We take the group elements generated by $\lambda_c$ to be given by
\begin{equation}
U(\lambda_c) = e^{i \lambda_c}\,,
\end{equation}
such that $\lambda_c$ is compact with radius $2\pi$. It is convenient to rescale the gauge field by $\sqrt{2\pi/k}$, to remove the $k$ dependence from the action. In doing so, we change the volume of the constant part of the gauge group to $\sqrt{2\pi k}$. Whenever we evaluate path integrals, we will assume this normalisation for the action. To get a feel for the structure of the result \cite{Giombi:2015haa}, we can parameterise $A_\mu$ as
\begin{equation}
A_\mu = {A}^T_\mu + \partial_\mu \mathcal{B}\,, \quad\quad \nabla^\mu {A}^T_\mu= 0\,,
\end{equation} 
where we exclude the constant part of the scalar field $\mathcal{B}$ as it does not contribute to the configuration space of the $A_\mu$. We thus find
\begin{equation}\label{unfixed}
Z_{U(1)_k}[S^3] =  \left( {\text{vol} \, \mathcal{G}}\right)^{-1} \times \sqrt{{\det}' ( -\nabla^2)} \times \int \mathcal{D}'\mathcal{B}  \int \mathcal{D}{A}^T_\mu \, e^{i S_{U(1)_k}[{A}^T_\mu]}\,.
\end{equation}
The prime indicates we are dropping the zero-mode. Although the above expression is still rather schematic, one can already note that the path-integral over $\mathcal{B}$ will mostly cancel the volume of $\mathcal{G}$. The difference lies in the zero-mode sector which is absent in the space of functions $\mathcal{B}$. Since no other term in the partition function depends on $k$, we can already conclude that
\begin{equation}
Z_{U(1)_k}[S^3]  \propto \sqrt{\frac{1}{{k}}}\,.
\end{equation}
We now fix the remaining proportionality constant. One way to do so involves calculating and regularising the divergent functional determinants stemming from the Gaussian path integral. Generally speaking, a local quantum field theory in three dimensions on a compact space with metric $g_{\mu\nu}$ will have a partition function of the form
\begin{equation}\label{divergence}
\log Z[g_{\mu\nu}] =  c_0 \, \ell_{\text{uv}}^{-3} \int d^3x \sqrt{g} + c_1 \, \ell_{\text{uv}}^{-1} \int d^3x \sqrt{g} R + \text{finite}\,,
\end{equation}
where $\ell_{\text{uv}}$ is an ultraviolet length scale. The finite piece will encode some non-local functional of $g_{\mu\nu}$.  If the theory is not parity invariant, as is the case for Chern-Simons theory, one will generally have a local contribution proportional to the gravitational Chern-Simons term contributing to the $\Lambda$-independent part. In certain circumstances, due to the presence of additional symmetries or structures, such as supersymmetry, some of the divergences can be argued to be absent. As we discuss in Appendix \ref{S3det}, applying the Fadeev-Popov procedure in the Lorenz gauge to the Abelian Chern-Simons theory yields the following contribution to the sphere partition function
\begin{equation}\label{Zdet}
Z_{U(1)_k}[S_3] = e^{i \varphi_{\text{CS}}} \left({\text{vol} \,S^3\, \ell_{\text{uv}}^{-3}}\right)^{-1/2}  \, \sqrt{\frac{1}{k}} \times \frac{\left({\det' \left(- \nabla^2 /\ell_{\text{uv}}^{-2}\right)}\right)^{1/2}}{\left(\det L L^\dag /\ell_{\text{uv}}^{-2} \right)^{1/4}}\,,
\end{equation}
where $L$ is the operator $i\varepsilon^{\mu\nu\rho}\partial_\rho$ acting on the space of transverse vector fields, and the numerator stems from path integration over the ghost fields. The prime in the determinant means we are omitting any zero modes, which must be dealt with separately. Any of the local ultraviolet divergences in (\ref{divergence}) come from evaluating the functional determinants in (\ref{Zdet}). \\
\\
To assess the nature to the leading cubic divergence, it is sufficient to consider the problem on $\mathbb{R}^3$ (or a large enough box). In this case the spectra are straightforward to obtain, and one finds that the eigenvalues of both $L L^\dag$ and $-\nabla^2$ are given by $\mathbf{k}\cdot\mathbf{k}$ with $L L^\dag$ having twice the multiplicity due to the two polarisations of $A^T_\mu$. Consequently, at least the absolute value of $Z_{U(1)_k}[S_3]$ is free of cubic ultraviolet divergences. This resonates well with the absence of local degrees of freedom in Chern-Simons theory. So far, our arguments do not suffice to conclude anything about the linear divergence of $Z_{U(1)_k}[S_3]$.  As we discuss in Appendix \ref{S3det}, a linear divergence is indeed present in the heat kernel regularisation scheme. We find
\begin{equation}\label{AbelianZ}
|Z_{U(1)_k}[S^3]| = \sqrt{\frac{1}{k}} e^{-\frac{3\pi}{4\varepsilon}}\,, \quad\quad \varepsilon \equiv \frac{2 e^{-\gamma}\ell_{\text{uv}}}{\ell}\,.
\end{equation}
The phase $\varphi_{\text{CS}}$ of $Z_{U(1)_k}[S_3]$ requires a careful treatment. We can heuristically argue that it will not contribute to the cubic divergence either. Indeed, given that the phase is associated to the parity non-invariance of the theory, we might expect any associated divergence to also be parity non-invariant such as the gravitational Chern-Simons term.\footnote{Alternatively, we could consider adding two Chern-Simons terms with equal and opposite level and regularise in such a way that the phase of the sphere partition function vanishes.}\\
\\
A more sophisticated approach to compute the sphere partition function, and its phase, follows \cite{Witten:1988hf}. Since this approach will be of use more generally, we now discuss it. Consider two solid tori, $\mathbf{T}_L^2 = D_L\times S^1$ and $\mathbf{T}_R^2 = D_R\times S^1$, with boundary tori $T^2_L$ and $T^2_R$. If we glue these together by identifying the points on $T_L^2$ with those on an oppositely oriented $T_R^2$,  we obtain an $S^2 \times S^1$. The partition function $Z_{\mathcal{G}_k}[S^2\times S^1]$ counts the number of states of Chern-Simons theory with gauge group $\mathcal{G}$ at level $k$ quantised on a spatial $S^2$, that is to say
\begin{equation}
Z_{\mathcal{G}_k}[S^2\times S^1] \equiv 1\,.
\end{equation} 
That the above equation holds, regardless of the gauge group and level (so long as there are no punctures on the $S^2$) is simply the statement that Chen-Simons theory on a spatial $S^2$ has a unique state in its Hilbert space. What is perhaps less immediate is that by identifying the points on $T^2_L$ with a particular modular transformation $\tau_R$ of the points on $T^2_R$, we instead get an $S^3$. This is true for a variety of $\tau_R$. For our immediate purpose we will consider an inversion $\tilde{\tau}_R  = -1/\tau_R$. Finally, we must now take into account the general relation 
\begin{equation}\label{ZSmod}
Z_{\mathcal{G}_k} [\tilde{\mathbf{T}}^2] = {\mathcal{S}_0}^0 \times Z_{\mathcal{G}_k} [\mathbf{T}^2 ]\,,
\end{equation}
where $\tilde{\mathbf{T}}^2$ is a solid torus, whose boundary is given by that of $\mathbf{T}^2$ upon performing the inversion $\tilde{\tau}  = -1/\tau$. The above formula follows from a careful consideration of the solid torus Hilbert space \cite{Witten:1988hf}, but we will not attempt to derive it here. If we can view the partition functions on the left- and right-hand sides of (\ref{ZSmod}) as quantum wavefunctions in the Hilbert space of the theory quantised on a spatial $T^2$, we can take an inner product of both sides with the state on the right hand side to get
\begin{equation}
{Z_{U(1)_k} [S^3]}  = {\mathcal{S}_0}^0 \times Z_{U(1)_k}[S^2 \times S^1]\,.
\end{equation}
Using the modular $S$-matrix for $U(1)_k$, it is concluded that
\begin{equation}\label{Zu1}
{Z_{U(1)_k} [S^3]} = \sqrt{\frac{1}{k}}\,.
\end{equation}
In such a way, we can fix the overall constant and phase in (\ref{unfixed}). We now proceed to consider the above expressions from the perspective of the theory quantised on a spatial disc. 

\subsection{Classical edge-mode theory}\label{edgemodeu1}

Quantisation of Chern-Simons theory on a spatial disc requires a careful consideration of boundary conditions on the $S^1$ boundary. As already mentioned, a well-posed variational problem with Dirichlet conditions on the gauge field enforces the boundary condition (\ref{gauge}). Imposing this condition throughout the remainder of space fixes our gauge entirely. The resulting constraints impose that the spatial components of the gauge field take the following form
\begin{equation}\label{constrain}
A_\rho  = i \, e^{i \, \Xi(t,\varphi,\rho)} \, \partial_\rho \, e^{-i \, \Xi(t,\varphi,\rho)}\,, \quad\quad {{A_\varphi   = i \, e^{i \, \Xi(t,\varphi,\rho)} \partial_\varphi e^{-i \, \Xi(t,\varphi,\rho)}}}\,,
\end{equation}
where we recall that $\rho\in[0,\pi/2]$ and $\varphi \sim \varphi + 2\pi$ are coordinates on the disc (\ref{diskmetric}). The edge-mode theory residing at the boundary of the disc where $\rho=\pi/2$ in the case of $\mathcal{G} = U(1)$ is described by the Floreanini-Jackiw action \cite{Floreanini:1987as}
\begin{equation}\label{edge}
S_{\text{edge}} = \frac{k}{4\pi} \int d t d\varphi \, \partial_\varphi \zeta  \left( \partial_t  - \upsilon \partial_\varphi  \right) \zeta\,,
\end{equation}
where the field $\zeta(t,\varphi) = \Xi(t,\varphi,\rho)|_{\rho=\pi/2}$ is a compact scalar field with periodicity $\zeta \sim \zeta + 2\pi$ mapping $\mathbb{R} \times S^1 \to S^1$. The above theory is that of a compact chiral boson. Classically, the solution space is given by
\begin{equation}\label{solns}
\zeta(t,\varphi) = g(\varphi + \upsilon t) +  m\, \varphi + f(t)\,.
\end{equation}
The first term corresponds to a chiral excitation moving at angular velocity $\upsilon$ \cite{Wen:1990se}. The second term, which arises due to the compactness of $\zeta$, provides a winding number $m\in\mathbb{Z}$ counting the number of times $\zeta$ wraps around the spatial $S^1$. 
The third term does not contribute to the classical energy of the solution, as can be seen from the classical Hamiltonian
\begin{equation}
H = \frac{k\upsilon}{4\pi}\int d\varphi \, (\partial_{\varphi}\zeta)^2\,.
\end{equation}
As can be seen from (\ref{constrain}), $f(t)$ carries no physical information and can be dropped all together. The global $U(1)$ symmetry of the edge-mode theory corresponds to shifts of $\zeta$ (modulo integer multiples of $2\pi$), and is generated by the following charge
\begin{equation}\label{U1charge}
\mathcal{Q} = \int \frac{d\varphi}{2\pi} \partial_\varphi \zeta\,.
\end{equation}
It is convenient to expand the non-winding mode sector in a Fourier expansion
\begin{equation}
\zeta(t,\varphi) = \frac{1}{\sqrt{2\pi}} \sum_{n \in\mathbb{Z}^+} \left(\alpha_n(t) e^{-i n \varphi} + \bar{\alpha}_n(t) e^{i n \varphi} \right)\,.
\end{equation}

\subsection{Quantum edge-mode theory}

\begin{figure}[H]
	\centering
\includegraphics[width=0.9\linewidth]{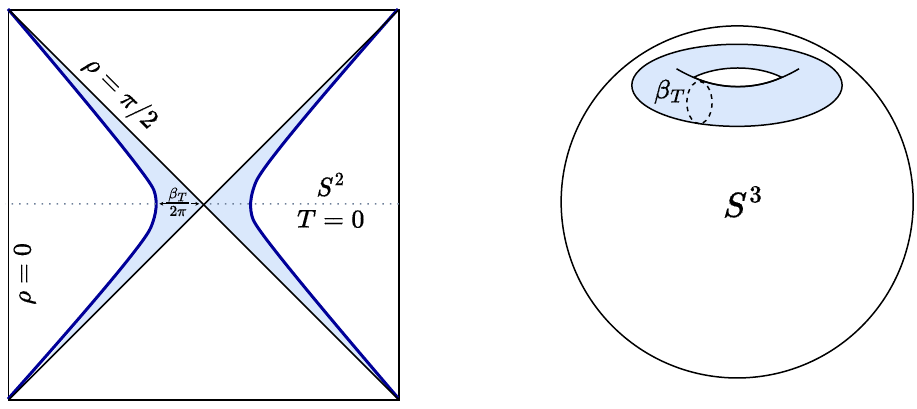}
	\caption{The Penrose diagram on the left shows the Lorentzian picture of the edge-modes, which live on the thick, blue line at a distance $\beta_T/2\pi \ll 1$ from the horizon. On the right is the Euclidean picture, where the edge-modes live on the surface of a solid torus with thermal cycle $\beta_T$. Shrinking the thermal cycle is the same as taking the high temperature limit, or taking the edge-modes to live very close to the horizon.}
	\label{edgemodes}
\end{figure}

Upon quantisation, the complex functions $\alpha_n(t)$ are promoted to operators satisfying the equal-time commutation relations
\begin{equation}\label{commutators}
[\hat{\alpha}_n,\hat{\alpha}_m]  = [\hat{\alpha}_n^\dag,\hat{\alpha}_m^\dag]  =0\,, \quad\quad\quad [\hat{\alpha}_n,\hat{\alpha}_m^\dag] =  \frac{2\pi}{k n}  \, \delta_{m,n}\,.
\end{equation}
To derive the above, we should keep in mind that this problem involves a constrained phase space. It follows from the commutation relations (\ref{commutators})  that the edge-mode theory contains a level $k$ $\mathfrak{u}(1)$ Kac-Moody algebra. The periodicity $\zeta \sim \zeta  + 2\pi$ leads us to consider the vertex operators
\begin{equation}\label{Psi}
\hat{\mathcal{O}}_n = \, : e^{i n \hat{\zeta}} :\,, \quad\quad n = 1,2,\ldots,k-1\,.
\end{equation}
Acting with the $\hat{\mathcal{O}}_n$ inserts a timelike Wilson line piercing the interior of the disc. The operator carries fractional charge $\mathcal{Q}_n = n/k$ under the $U(1)$ shift symmetry (\ref{U1charge}) of the edge-mode theory. We view such anyonic insertions as external data to the edge-mode theory, and consequently not part of the edge-mode Hilbert space. For $n=m k$, with $m\in\mathbb{Z}$, the above operators are no longer singular in the disc interior. Indeed, the closed-loop integral $\oint A_\varphi d\varphi$ from a single winding mode evaluates to $2\pi m$. Instead, for  $n=m k$ these correspond to winding mode excitations which indeed reside in the edge-mode Hilbert space.\\
\\
The quantum Hamiltonian for a given winding sector $m$ is given by 
\begin{equation}\label{eHam}
\hat{H}_m = k \, \upsilon \frac{m^2}{2} \hat{\mathbb{I}}_m  +\frac{\upsilon k}{2\pi}  \sum_{n \in\mathbb{Z}^+} n^2 \, \hat{\alpha}^\dag_n \hat{\alpha}_n\,,
\end{equation}
where $\hat{\mathbb{I}}_m$ is the identify operator in a  given winding sector. 
The ground state is given by the state carrying vanishing winding and annihilated by all the $\hat{\alpha}_n$. In a given winding sector the eigenstates of the Hamiltonian are given by
\begin{equation}
|n_1,d_1;n_2,d_2;\ldots,n_p,d_p \rangle_m = \prod_{i=1}^p  \sqrt{\frac{1}{d_i !} \, \left(\frac{kn_i}{2\pi} \right)^{d_i}}\, \left( \hat{\alpha}_{n_i}^\dag \right)^{d_{i}} | 0\rangle\,, \quad\quad d_i , p \in \mathbb{N}\,, 
\end{equation}
and their corresponding energy is
\begin{equation}
E_{\{n_i,d_i\};m}  = k \, \upsilon \frac{m^2}{2} - \upsilon\left( \epsilon_0 + \frac{1}{24} \right) +  \upsilon \sum_{i=1}^p d_i n_i\,.
\end{equation}
The degeneracy of those states for fixed $\sum_i d_i n_i = N$ is given by the integer partition number $p(N)$. We have allowed for an overall shift in the energy $\upsilon \epsilon_0$ to account for normal ordering ambiguities. The thermal partition function of the edge-mode theory is thus given by
\begin{equation}\label{zedge}
Z_{\text{edge}}[\beta_T] = \text{tr} \, e^{-\beta_T \hat{H}} =   {q^{-\epsilon_0}} \times \frac{\vartheta_3(0,q^{k/2})}{{\eta(q)}}\,,
\end{equation}
where $q = e^{-\upsilon \beta_T}$, and $\vartheta_3(0,q)$ is the elliptic theta function
\begin{equation}
\vartheta_3 (z, q ) \equiv \sum_{n \in \mathbb{Z}}  q^{n^2} e^{2ni z} \,.
\end{equation}
In the coordinates \eqref{sp}, the inverse Tolamn temperature is $\beta_T = 2\pi \cos \rho$. Equation \eqref{zedge} is reproduced from the perspective of a Euclidean partition function in Appendix \ref{Zthermal}. \\
\\
We now imagine that our edge-modes are located parametrically close to the dS$_3$ horizon. One might argue, in such a case, that the edge-mode theory should be placed at a parametrically high temperature (in units measured by the inertial clock at $\rho=0$). Let us discuss this from a Euclidean perspective. Recall that  Euclidean continuation of the de Sitter horizon becomes a circle in $S^3$. Removing a small region near the horizon corresponds to excising a thin solid torus $\mathbf{T}^2$ from the $S^3$, as shown in Figure \ref{edgemodes}. The two cycles of the boundary of $\mathbf{T}^2$ correspond to a spatial cycle and a thermal cycle. As we take the region to vanishing size, the size of the thermal cycle shrinks to zero, which is the Euclidean picture of a high temperature limit. This is, to some extent, similar to the brick-wall regularisation considered by `t Hooft \cite{tHooft:1984kcu} (see also \cite{Das:2015oha,Geiller:2017xad,Wong:2017pdm}). Thus, to make contact with the dS$_3$ picture, we would like to study the edge-mode theory (\ref{edge}) at high temperature. 
\noindent In the high temperature limit, we can exploit the modular properties of $\vartheta(0,q^{k/2})$ and $\eta(q)$ to find
\begin{equation}\label{edgefinal}
\lim_{\beta_T\to0^+} \log Z_{\text{edge}}[\beta_T] = \frac{\pi^2}{6\upsilon\beta_T} -\frac{1}{2}  \log k \ldots
\end{equation}
The first term encodes the contribution from the high energy sector of the theory and is proportional to the temperature. The finite term is temperature independent, and moreover it is independent of our choice of $\upsilon$. In fact, it is equal to the three-sphere partition function (\ref{Zu1}). The entropy $S_{\text{e}}$ of the edge-modes at high temperature can similarly be computed and reads
\begin{equation}\label{Se}
S_{\text{e}} = \frac{\pi^2}{3\upsilon\beta_T} -\frac{1}{2}  \log k \ldots 
\end{equation}
Thus, we can relate the regularised expression for $Z_{U(1)_k}[S_3]$ (\ref{Zu1}) to the finite part of the entropy of the edge-mode theory in the high temperature limit. The divergent high temperature contribution is most naturally accommodated by a linearly divergent ultraviolet piece of the three-sphere partition function, indicating that
\begin{equation}
Z_{U(1)_k}[S_3] = \lim_{\beta_T\to0^+}  Z_{\text{edge}}[\beta_T]\,,
\end{equation}
in a regularisation scheme where the linear divergence of $Z_{U(1)_k}[S_3]$ in (\ref{AbelianZ}) is tuned accordingly. We note that there are no cubic divergences appearing on either side. \\
\\
It has been argued \cite{Kitaev:2005dm,Levin:2006zz} that the temperature independent term in (\ref{edgefinal}) is a universal contribution to the entanglement entropy, independent of any of the detailed features associated to the cutoff surface. Here, in view of our discussion in Section \ref{graventropy}, we can interpret $-\log \sqrt{k}$ as a contribution to the dS$_3$ horizon entropy. From this perspective, the fact that this is negative is immaterial. 

\subsection{Edge-mode symmetries.} 
It is worth pausing momentarily here and noting that the edge-mode theory, which is nothing more than a chiral compact boson theory, has a large symmetry group containing an affine extension of the Virasoro algebra.\footnote{The dS$_3$/CFT$_2$ literature discusses a Virasoro symmetry associated to the future/past boundary \cite{Strominger:2001pn}. Here we see the appearance of a Virasoro symmetry within a single static patch.} The Virasoro generators are constructed in the usual way
\begin{equation}\label{virasoro}
\hat{L}_0 =  \frac{k}{2\pi}\sum_{n\in \mathbb{Z}^+} n^2 \hat{\alpha}^\dag_{n} \hat{\alpha}_{n}\,, \quad\quad   \hat{L}_{m} = \frac{k}{2\pi}  \sum_{n=1}^\infty  (m+n)n \hat{\alpha}^\dagger_{n} \hat{\alpha}_{n+m} + \frac{k}{4\pi}\sum_{n=1}^{m-1}  (m-n)n\hat{\alpha}_{m-n} \hat{\alpha}_{n}\,,  
\end{equation}
giving rise to the corresponding Virasoro algebra
\begin{equation} 
[\hat{L}_m, \hat{L}^\dag_n] = 
\begin{cases}
(m+n) \hat{L}_{m-n} + \frac{1}{12} m(m^2-1) \delta_{n,m}, \quad m \geq n\,, \\
(m+n) \hat{L}^\dagger_{n-m} , \quad m < n \,. 
\end{cases}
\end{equation}
We notice that the edge-mode Hamiltonian in (\ref{eHam}) obeys $\hat{H}_0 = \upsilon \hat{L}_0$. We can further compute the commutation relations between the $\hat{L}_n$ and $\hat{\alpha}_n$,
\begin{equation}
[\hat{L}_n,\hat{\alpha}_m] = -(n+m) \hat{\alpha}_{n+m}\,, \quad\quad [\hat{L}_n,\hat{\alpha}^\dag_m] = \begin{cases}
		(m-n) \hat{\alpha}^\dag_{m-n}, \quad m>n\\
		(n-m) \hat{\alpha}_{n-m}, \quad m< n \,.
	\end{cases}
\end{equation}
The presence of the vertex operators (\ref{Psi}) allows the algebra to be extended beyond the Virasoro-$\mathfrak{u}(1)$ Kac-Moody algebra. The generators additionally include the vertex operators $\hat{\mathcal{J}}^{(k)}_\pm = : e^{\pm i {k} \hat{\zeta}}:$. Recalling that $k$ is even, it follows that the conformal dimension of the generators, $\Delta_\pm = k/2$, is also an integer.
The extended algebra has a finite number, $k$, of highest weight irreducible representations (see \cite{DiFrancesco:1997nk} for a pedagogical discussion). We can organise the Hilbert space of the edge-mode theory in terms of these. The different representations are labelled by a parameter $s=0,1,\ldots,k-1$, denoting the representation carried by the Wilson line inserted in the interior of the disc \cite{Witten:1988hf}. Their corresponding character is given by
\begin{equation} \label{u(1) character}
\chi_s(q)  = \frac{1}{\eta(q)} \sum_{m\in \mathbb{Z}} q^{\frac{k}{2} \left(m + \frac{s}{k}\right)^2}\,, \quad\quad s = 0,1,\ldots,k-1\,.
\end{equation}  
We observe that upon fixing $\epsilon_0 = 0$ and $q = e^{-v \beta_T}$, the above character for $s=0$ is equivalent to the thermal partition (\ref{zedge}). The modular $S$-matrix for the extended $\mathfrak{u}(1)$ character at level $k$ is given by (see for example \cite{Schellekens:1996tg,Dong:2008ft})
\begin{equation}\label{modsu1}
{\mathcal{S}_{m}}^{n} = \sqrt{\frac{1}{k}} e^{2\pi i m n/k}\,, \quad\quad m,n = 0,1,\ldots,k-1\,.
\end{equation}
Using that ${\mathcal{S}_{m}}^{n}$ relates the $\chi_s(q)$ under $\beta_T\to1/\beta_T$, we can re-derive the high temperature behaviour (\ref{edgefinal}). 

\subsubsection*{{\it Odd values of $k$ \& $\mathcal{N}=2$ supersymmetry, briefly}}
Before moving on to the non-Abelian case, we briefly comment that for odd  $k$ much of the above discussion carries through. The essential difference is that the generators $\hat{\mathcal{J}}^{(k)}_\pm = \, : e^{\pm i {k} \hat{\zeta}}:$ will carry half-integer conformal dimensions and obey fermionic anti-commutation relations. \\
\\
Let us focus on the case $k=3$, which is studied for example in \cite{Fendley:2006gr}. In this case the generators $\hat{\mathcal{J}}^{(3)}_\pm = :e^{\pm {3}i  \hat{\zeta}}:$ have weight $\Delta_\pm = 3/2$. Decomposing the generators into their respective Fourier modes $\hat{\mathcal{J}}^{(3)}_{\pm,n}$ on the cylinder,  one finds the anti-commutation relations
\begin{equation}
	\begin{split}
	&\{ \hat{\mathcal{J}}^{(3)}_{\pm,n}, \hat{\mathcal{J}}^{(3)}_{\pm,m} \} =  0\,,\\
	&\{\hat{\mathcal{J}}_{+,n}^{(3)}, \hat{\mathcal{J}}_{-,m}^{(3)} \} = \frac{1}{2} \left(n^2 - \frac{1}{4} \right) \delta_{n+m,0} + 3 \hat{L}_{n+m } + \frac{3 i}{2} \left(m^2 - n^2 \right) \hat{\alpha}_{n + m}\,.
	\end{split}
\end{equation} 
We can also compute the commutation relations between the $\hat{\mathcal{J}}^{(3)}_\pm$, the Virasoro generators (\ref{virasoro}), and the Kac-Moody generators (\ref{commutators}) leading to an $\mathcal{N}=2$ superconformal algebra.
This has been explored, for example, in \cite{Fendley:2006gr,Sagi:2016slk}.\footnote{The edge-mode theory of Chern-Simons theory with an $SU(2)$ gauge group at level $k=2$ also enjoys an $\mathcal{N}=2$ superconformal symmetry.} The operators $: e^{\pm  i  \hat{\zeta}}:$ correspond to insertions of fractional charge $\pm 1/3$ inside the spatial disc, and are primaries under the $\mathcal{N}=2$ superconformal algebra.\\
\\
Supersymmetry is not often associated to physics in de Sitter space \cite{Pilch:1984aw}, but the usual arguments do not preclude the possibility of a supersymmetric edge-mode theory (or a superconformal theory more generally \cite{Anous:2014lia}).

\subsection{Comments on the non-Abelian case} \label{Comments on the non-Abelian case}

From the modular $S$-matrix (\ref{modsu1}) we observe that inserting a single Wilson line in the interior of the disc will not affect the constant part of the partition function because ${\mathcal{S}_{m}}^{0} = {\mathcal{S}_{0}}^{0}$. This is no longer true for the non-Abelian case. For instance, if we take Chern-Simons theory with gauge group $SU(2)$ at level $k$ the relevant modular $S$-matrix is (see for example \cite{DiFrancesco:1997nk})
\begin{equation}\label{modsu2}
{\mathcal{S}_{m}}^{n} = \sqrt{\frac{2}{k+2}} \sin \left(\frac{(m+1)(n+1)\pi}{k+2} \right)\,, \quad\quad m,n = 0,1,\ldots,k\,.
\end{equation}
As mentioned earlier, ${\mathcal{S}_{0}}^{0}$ is equal to the regularised partition function $Z_{SU(2)_k}[S^3]$ for the choice of framing leading to vanishing phase. Explicitly,
\begin{equation}\label{zsu2s3}
Z_{SU(2)_k}[S^3] =  \sqrt{\frac{2}{k+2}} \sin \left(\frac{\pi}{k+2} \right)\,.
\end{equation}
When considering the theory on the disc, the thermal partition function of the edge-mode theory is again given by a character of the $\widehat{\mathfrak{su}}(2)_k$ extended algebra. These are known from the rational CFT literature. For an insertion of the spin-$s/2$ integrable highest weight representation of level $k$ in the interior of the disc, we have (see for example \cite{DiFrancesco:1997nk})
\begin{equation}\label{su2char}
Z_{\text{edge}}[\beta_T;s]  = \frac{\vartheta_{s +1}^{\, k+2} - \vartheta_{-s - 1}^{\, k+2}}{\vartheta_1^{\,2} - \vartheta_{-1}^{\,2}}\,, \quad\quad s = 0,1, \ldots,k\,,
\end{equation}
where the generalised $\vartheta$-functions are
\begin{equation}
\vartheta_s^{\, k }(q,z) \equiv \sum_{p \in \tfrac{s}{2k} + \mathbb{Z}} q^{kp^2}e^{2\pi i p k z}.
\end{equation}
As in the Abelian case, $q = e^{-\beta_T \upsilon}$ and $z=0$ in (\ref{su2char}) gives the thermal partition function of the edge-mode theory. The entropy in the high temperature limit of the edge-mode theory, which is now an $SU(2)_k$ chiral WZW model, will be given by
\begin{equation}\label{nonabelianedge}
	S^{(s)}_{\text{e}}  = \frac{3 k}{k+2}\, \frac{\pi^2}{3\beta_T v}  + \log \left(  \sqrt{\frac{2}{k+2}} \sin \left(\frac{(s+1)\pi}{k+2}\right) \right) \ldots
\end{equation}
The above is an increasing function of $s$ in the range $s=0,1,\ldots,k$.\footnote{It is also  customary to express the entropy in terms of the quantum dimensions $d^{(k)}_s$ of $SU(2)_k$. The numbers encode the multiplicity of various operators appearing upon fusion. They are given by
\begin{equation}
d^{(k)}_s = {\sin \left( \frac{\pi(s+1)}{k+2} \right)}{\arcsin \left( \frac{\pi}{k+2} \right)}\,, \quad\quad s = 0,1,2,\ldots,k\,,
\end{equation}
such that the temperature independent part of the entropy is given by $S^{(s)}_{\text{e}} = -\frac{1}{2}\log \sum_{s} d^{(k)}_s d^{(k)}_s$.
In the large $k$ limit the quantum dimensions are approximately given by the usual degeneracy formula for $SU(2)$ spin-$s/2$ irreducible representations $d^{(k)}_s \approx s + 1$\,.}
For vanishing $s$, the temperature independent piece of $S_{\text{e}}$ is the $S^3$ partition function (\ref{zsu2s3}) of Chern-Simons theory with $SU(2)$ gauge group at level $k$. For non-vanishing $s$, it is the $S^3$ partition function including the insertion of a closed unknotted Wilson loop carrying the spin-$s/2$ integrable representation of level $k$. The large $k$ expansion makes clear that in this case, horizon thermodynamics will receive contributions at all loop orders. In the weakly coupled limit, one finds
\begin{equation}
\lim_{k\to\infty} \exp {S^{(s)}_{\text{e}}}  \approx (1+s) \times \frac{1}{\text{vol} \, SU(2)} \left(\frac{4\pi^2}{k} \right)^{3/2}\times e^{\frac{\pi^2}{\beta_T v}}\,, \quad s = 0,1,\ldots,k\,.
\end{equation}
We see that the exponential of the entropy receives a multiplicative factor of $(1+s)$ in the weakly coupled limit. The factor $(1+s)$ is nothing more than the dimension of the Hilbert space attached to the puncture, and hence the entropy increases by this amount if the state is not measured with further precision.\\
\\
Finally, we can also consider the phase of $Z_{SU(2)_k}[S^3]$ from the perspective of the edge-mode theory, again following \cite{Witten:1988hf}. Usually, when we consider a two-dimensional conformal field theory on the torus, the partition function is invariant under shifts $\tau \to \tau+1$. This is simply the statement that the angular momentum is quantised. However, our edge-mode theory is chiral, and  transforms anomalously upon shifting $\tau\to\tau+1$. The integrable character of weight $h$ transforms as 
\begin{equation}
\chi_h(\tau+1) = e^{2\pi i (h-c_k/24)} \chi_h(\tau)\,,
\end{equation}
where for the $SU(2)_k$ theory under consideration the weights of the spin-$s/2$ primaries are given by $h = s(s+2)/4(k+2)$ and $c_k = 3k/(k+2)$. In particular, under $\tau \to \tau+n$ the identity character transforms by a phase given by $e^{ 2 \pi i n  c_k/24}$ where $n$ is an integer. This agrees with the set of admissible phases for $Z_{SU(2)_k}[S^3]$ stemming from the framing anomaly. 
\section{Complexified Abelian Chern-Simons: Lorentzian model}\label{complexCSL}

In this section, we proceed to consider a complexified version of the Abelian Chern-Simons theory. This theory is introduced as a simple and calculable toy model with a complexified gauge group, a property common to three-dimensional gravity with Lorentzian signature and $\Lambda > 0$ expressed as a Chern-Simons theory.

\subsection{Lorentzian model} 

The Lorentzian model is built from a complexified gauge field $\mathcal{A}_\mu = A_\mu + i B_\mu$, where $A_\mu$ and $B_\mu$ are real Abelian gauge fields. Our action is given by
\begin{equation}\label{complexSL}
S_L[\mathcal{A}_\mu] = \frac{k + i \lambda}{8\pi} \int_{\mathcal{M}} d^3 x  \, \varepsilon^{\mu\nu\rho} \mathcal{A}_\mu \partial_\nu \mathcal{A}_\rho +  \frac{k - i \lambda}{8\pi} \int_{\mathcal{M}} d^3 x  \, \varepsilon^{\mu\nu\rho} \bar{\mathcal{A}}_\mu \partial_\nu \bar{\mathcal{A}}_\rho\,,
\end{equation}
and is real-valued. The parameters $k$ and $\lambda$ are taken to be real-valued. We denote the Lie algebra of our theory by $\mathfrak{u}_{\mathbb{C}}(1)$. The gauge transformations act in the following way
\begin{equation}
\mathcal{A}_\mu \to \mathcal{A}_\mu + \partial_\mu \vartheta\,,
\end{equation}
where the gauge parameter $\vartheta$ is a complex-valued function.
In terms of $A_\mu$ and $B_\mu$ the action is given by
\begin{equation}\label{ABaction}
S_L[\mathcal{A}_\mu] = \frac{1}{4\pi} \int_{\mathcal{M}} d^3 x  \, \varepsilon^{\mu\nu\rho} \left[ k \left( A_\mu \partial_\nu A_\rho- B_\mu \partial_\nu B_\rho   \right) -  \lambda \left( A_\mu \partial_\nu B_\rho + B_\mu \partial_\nu A_\rho \right) \right]\,.
\end{equation}
At $\lambda=0$, the real and imaginary parts of the gauge field $\mathcal{A}$ decouple. Setting either $\lambda=0$ or $k=0$ reduces the model to an Abelian BF model with $U(1)$ gauge group (see \cite{Birmingham:1991ty,Cattaneo:1995tw} for an overview). In addition to being real-valued, we will impose that $k$ takes integer values upon quantisation of the theory. This follows from taking $A_\mu$ as the generator of a compact $\mathfrak{u}(1)$ gauge algebra. The parameter $\lambda$ is not constrained to be an integer. Unless otherwise specified, we will take $k \in \mathbb{Z}^+$ and $\lambda \in\mathbb{R}/\{0\}$ in what follows. \\
\\
The classical equations of motion for (\ref{ABaction}) are given by
\begin{equation}
\varepsilon^{\mu\nu\rho} \partial_\nu A_\rho =  \varepsilon^{\mu\nu\rho} \partial_\nu B_\rho= 0\,.
\end{equation}
They are satisfied when the field strengths associated to $A_\mu$ and $B_\mu$ vanish. The classical solution space is the space of flat connections modulo gauge transformations. Assuming the theory resides on a three-manifold with topology $\mathbb{R} \times S^2$, we can canonically quantise the theory in the $A_t = B_t = 0$ gauge. States must then satisfy the constraints
\begin{equation}\label{complexconstraint}
F^{A}_{ij} =   F^{B}_{ij} = 0\,, \quad\quad i,j \in S^2\,.
\end{equation}
The equal time commutation relations are given by
\begin{equation}\label{complexcomm}
[A_i(\mathbf{x}), A_j (\mathbf{y}) ] = - [B_i(\mathbf{x}), B_j (\mathbf{y}) ] = \frac{2\pi i}{k} \varepsilon_{ij} \delta (\mathbf{x} - \mathbf{y} ) \,.
\end{equation}
It follows that on a spatial $S^2$ the Hilbert space has a unique state.\\
\\
As a final remark before embarking onto the edge-mode theory, it is worth noting that the volume of $U_{\mathbb{C}}(1)$ is infinite, making it difficult to interpret the perturbative formula (\ref{CSpert}) for the $S^3$ partition function of the Lorentzian model.

\subsection{Lorentzian edge-mode theory \& quantisation}

Following the discussion in Section \ref{edgemodeu1}, we can construct an edge-mode theory at the $S^1$ boundary of the spatial disc. As before, we consider fixing the following gauge\footnote{More general boundary conditions will be considered in the next chapter.} 
\begin{equation} \label{Complexgaugechoice}
\mathcal{A}_t - \upsilon \mathcal{A}_\varphi = 0\,,
\end{equation}
where $\upsilon = \upsilon_{re} + i \upsilon_{im}$ is now a complex parameter. The gauge constraint that follows from imposing the above condition is
\begin{equation}
\mathcal{A}_i (t,\rho,\varphi) = i e^{i \, \Xi(t,\rho,\varphi)} \, \partial_i \, e^{-i \,  \Xi(t,\rho,\varphi)}\,, \quad\quad i \in \{ \rho, \varphi \}\,,
\end{equation}
where now the boundary value of $\Xi(t,\rho,\varphi)|_{\rho=\pi/2}$ is denoted by $\zeta = \zeta_{re} + i \zeta_{im}$ and has compact real part. The action governing our edge-mode theory is given by
\begin{equation}\label{SLedge}
S_{\text{edge}} = \frac{k+i \lambda}{8\pi} \int d t d\varphi  \left( \partial_t \zeta - \upsilon \partial_\varphi \zeta \right) \partial_\varphi \zeta +  \frac{k-i \lambda}{8\pi} \int d t d\varphi  \left( \partial_t \bar{\zeta}  - \bar{\upsilon} \partial_\varphi \bar{\zeta}  \right) \partial_\varphi \bar{\zeta}\,.
\end{equation}
Once again, the classical solutions are given by complexified chiral excitations of both the real and imaginary parts of $\zeta$. We find 
\begin{equation}
\zeta = f(\varphi+ \upsilon t) + m \varphi + g(t)\,,
\end{equation}
where $f(z)$ is a complex valued function. We note that $\zeta_{im}=  (\zeta-\bar{\zeta})/2i$ does not contain winding modes around the $S^1$. The classical Hamiltonian is given by
\begin{equation}
\mathcal{Q}_t = \frac{(k+i \lambda) \upsilon}{8\pi} \int d\varphi  \left( \partial_\varphi \zeta  \right)^2 +  \frac{(k- i \lambda)\bar{\upsilon}}{8\pi} \int d\varphi  \left( \partial_\varphi \bar{\zeta}  \right)^2\,,
\end{equation}
which generates time translations. We note that $\mathcal{Q}_t$ is real. The generator of $\varphi$-translations is given by
\begin{equation}
\mathcal{Q}_\varphi = \frac{(k+i \lambda) }{8\pi} \int d\varphi   \partial_\varphi \zeta  \partial_t \zeta + \frac{(k- i \lambda)}{8\pi} \int d\varphi   \partial_\varphi \bar{\zeta} \partial_t \bar{\zeta}\,.
\end{equation}
On-shell, $\mathcal{Q}_t$ and $\mathcal{Q}_\varphi$ are equivalent within the sector with vanishing winding number. The classical theory (\ref{SLedge}) is also invariant under shifts $\zeta \to \zeta + \delta$ of the field $\zeta$ by some $\delta \in \mathbb{C}$. The generators of the shift symmetries are
\begin{equation}\label{complexQ}
\mathcal{Q}_\zeta = \frac{1}{2\pi} \int d\varphi \partial_\varphi \zeta\,, \quad\quad \mathcal{Q}_{\bar{\zeta}} = \frac{1}{2\pi} \int d\varphi \partial_\varphi \bar{\zeta}\,.
\end{equation}
\textbf{Quantisation.} In order to quantise the theory, it is convenient to express $\zeta$ in terms of Fourier modes. For the non-winding mode sector, we have
\begin{equation}
\zeta(t,\varphi) = \frac{1}{\sqrt{2\pi}} \sum_{n\in\mathbb{Z}/\{0\}} \alpha_n(t) e^{i n \varphi}\,.
\end{equation}
The $\alpha_n(t) \equiv a_n(t) + i b_n(t)$ are independent complex functions of $t$ for all $n$, and we take $a_n(t)$ and $b_n(t)$ to be real-valued. To quantise the theory, it is convenient to follow the procedure outlined in \cite{Dunne:1992ew}. For a given $n$, we can define the vector with components $\xi^i_n$ by $\xi_n \equiv (a_{-n},b_{-n},a_n,b_n)$. One finds the commutator
\begin{equation}
[\hat{\xi}^i_n,\hat{\xi}^j_n] = \frac{i}{n}  M^{ij}\,, 
\end{equation} 
with
\begin{equation}
M^{-1} = \frac{2\pi}{k^2 + \lambda^2} \left(
\begin{array}{cc}
 0 & -1 \\
 1 & 0 \\
\end{array}
\right) \otimes \left(
\begin{array}{cc}
 \lambda  & k  \\
 k  & -\lambda  \\
\end{array}
\right)\,.
\end{equation}
The eigenvalues of $M^{-1}$ are $\pm 2\pi i/ \sqrt{k^2+\lambda^2}$, each doubly degenerate. In terms of the $\hat{\xi}_n$, the quantum Hamiltonian for the vanishing winding mode sector is given by
\begin{equation}
\hat{\mathcal{Q}}_t = \frac{1}{4\pi} \sum_{n\in\mathbb{Z}/\{0\}} n^2 \left(\omega \left(\hat{\xi}^3_n \hat{\xi}_n^1 - \hat{\xi}^4_n \hat{\xi}_n^2 \right) - \psi \left(\hat{\xi}^3_n \hat{\xi}^2_n + \hat{\xi}^4_n \hat{\xi}^1_n \right) \right)\,.
\end{equation}
Here we have defined $\omega \equiv k \upsilon_{re} - \lambda \upsilon_{im}$ and $\psi \equiv  \lambda \upsilon_{re} + k \upsilon_{im}$. We can  construct operators obeying the more standard creation/annihilation algebra. We find that
\begin{equation} \label{ABrels}
	\begin{split}
		\hat{A}_n^{\pm} & \equiv \hat{\xi}_n^4 \pm i \hat{\xi}_n^3 \pm \frac{(ik\mp  \lambda)  }{\sqrt{k^2 + \lambda^2 }} \left(\hat{\xi}_n^1 \pm i \hat{\xi}_n^2 \right)\,,\\   
		\hat{B}_n^{\pm} & \equiv \hat{\xi}_n^4 \mp i \hat{\xi}_n^3 \pm \frac{(ik \pm \lambda)}{\sqrt{k^2 + \lambda^2}} \left(\hat{\xi}_n^1 \mp i  \hat{\xi}_n^2\right)\,,
	\end{split}
\end{equation}
obey the commutation relations
\begin{equation}\label{complexAAdag}
	\begin{split}
		[\hat{A}_n^+, \hat{A}_m^-] = [\hat{B}_n^+, \hat{B}_m^-] = \frac{8\pi}{n\sqrt{k^2 + \lambda^2}} \delta_{n,m}\,,
	\end{split}
\end{equation}
with all others vanishing. The Hamiltonian can then be put into the following form

\begin{equation}\label{LedgeH}
\hat{\mathcal{Q}}_t = \frac{\sqrt{k^2+\lambda^2}}{16\pi} \sum_{n\in\mathbb{Z}/\{0\}} n^2 \left( \upsilon_{re} \left( \hat{A}_n^+ \hat{A}_n^- - \hat{B}_n^+ \hat{B}_n^- \right) + i  \upsilon_{im} \left( \hat{A}_n^+ \hat{B}_n^+ - \hat{A}_n^- \hat{B}_n^-  \right) \right)\,.
\end{equation}
Consequently, uncovering the spectrum of $\hat{\mathcal{Q}}_t$ reduces to a two-site hopping type problem, where the sites are labelled by ${A}$ and $B$.

\subsection{Lorentzian edge-mode spectrum}

At this stage, we are confronted with diagonalising the Lorentzian edge-mode Hamiltonian (\ref{LedgeH}). Since each Fourier mode decouples, it is sufficient to discuss the problem for a single mode number $n$. It is convenient to represent the creation and annihilation operators satisfying (\ref{complexAAdag}) in the following way
\begin{equation}
\hat{A}^\pm_n = \left( \frac{4\pi}{n\sqrt{k^2 + \lambda^2}}  \right)^{1/2} \left( x_n \pm \frac{d}{dx_n} \right)\,, \quad \hat{B}^\pm_n =  \left( \frac{4\pi}{n\sqrt{k^2 + \lambda^2}}  \right)^{1/2} \left( y_n \pm \frac{d}{dy_n} \right)\,.
\end{equation}
As detailed in Appendix \ref{LedgeApp}, it is straightforward to put the single mode Hamiltonian $\hat{\mathcal{Q}}_t^{(n)}$ problem into the Schr\"odinger problem
\begin{equation}\label{Qn}
 \frac{\text{sign} \, w_n}{2}\left(- \frac{d^2}{dw_n^2 } +w_n^2 - \frac{1/4+p_n^2}{w_n^2}\right) \psi_n = \frac{1}{\upsilon_{re} } \left( \frac{E_n}{n} +  \upsilon_{im} p_n\right) \psi_n\,,
\end{equation}
where $x_n = w_n \cosh t_n $, $y_n = w_n\sinh t_n $ and $p_n \in \mathbb{R}$ is the Fourier momentum associated to the $t_n \in \mathbb{R}$ coordinate. The $E_n$ are the eigenvalues of $\hat{\mathcal{Q}}_t^{(n)}$. The operator on the left-hand side of (\ref{Qn}) is precisely that of a conformal quantum mechanics, as studied for example in \cite{deAlfaro:1976vlx,Anous:2020nxu,Andrzejewski:2011ya, Andrzejewski:2015jya}, whose $sl(2,\mathbb{R})$ generators are given by
\begin{equation}\label{sl2rgen}
\hat{K}_n = \frac{\text{sign} \, w_n}{2} w_n^2 \, ,  \; \; \hat{D}_n = -  \frac{ i}{2} \left[ w_n \frac{d}{dw_n} + \frac{1}{2} \right]\, , \; \; \hat{H}_n = -\frac{\text{sign} \, w_n}{2} \left[\frac{d^2}{dw_n^2} + \frac{1/4+p^2}{w_n^2} \right]\,,
\end{equation}
satisfying
\begin{equation}
[\hat{D}_n,\hat{H}_n] = i\hat{H}_n\,, \quad\quad [\hat{D}_n,\hat{K}_n]=-i\hat{K}_n\,, \quad\quad [\hat{K}_n,\hat{H}_n] = 2i\hat{D}_n\,. 
\end{equation}
The generators (\ref{sl2rgen}) furnish the principal series representation with weight $\Delta = 1/2+ i p_n/2$. 
It is known that the spectrum of $\hat{H}_n+\hat{K}_n$ is given by the even integers and is thus unbounded \cite{Anous:2020nxu,Andrzejewski:2011ya, Andrzejewski:2015jya}. Consequently, the single mode energy spectrum is given by
\begin{equation} \label{Loreenergies}
E_n(m,p_n) = 2\upsilon_{{re}} n m - \upsilon_{{im}} n p_n\,, \quad\quad m \in \mathbb{Z}\,, \quad p_n \in \mathbb{R}\,.
\end{equation}
The unboundedness and continuity of the edge-mode Hamiltonian reflects the fact that the Lorentzian Chern-Simons theory has a complexified gauge group. It also implies that the thermal partition function of the edge-mode theory is no longer a sensible quantity to compute. (Sensible quantities to compute might involve \cite{Carlip:1994gy,Banados:1996ad} imposing additional constraints on the state-space or modifying the boundary conditions on the Chern-Simons gauge field at the boundary of the disc, which we will examine in Chapter \ref{GeneralBCs}.) To sharpen this issue, we consider the edge-mode theory on a torus.

\subsection{Lorentzian model in Euclidean signature?}\label{complexL}

The Euclidean continuation of the complexified edge-mode theory (\ref{SLedge}) can be achieved by taking $t = -i\xi$. If we are to place the system at a finite inverse temperature $\beta_T$, we further impose that $\xi \sim \xi + \beta_T$, such that the theory resides on a torus. Upon continuing to Euclidean time, we end up with the Euclidean edge-mode action
\begin{equation}\label{SLEedge}
S^{(E)}_{\text{edge}} = \frac{k+i \lambda}{8\pi} \int d \xi d\varphi  \left(- i \partial_\xi \zeta + \upsilon \partial_\varphi \zeta \right) \partial_\varphi \zeta +  \frac{k-i \lambda}{8\pi} \int d \xi d\varphi  \left( - i  \partial_\xi \bar{\zeta}  + \bar{\upsilon} \partial_\varphi \bar{\zeta}  \right) \partial_\varphi \bar{\zeta}\,.
\end{equation}
It is convenient to further express the action in terms of the modes on the torus
\begin{equation}\label{Fedge}
\zeta(\xi,\varphi) = \frac{1}{\sqrt{2\pi}} \sum_{(m,n) \in \mathbb{Z}^2} e^{2\pi i m  \xi/\beta_T +  i n \varphi} \zeta_{m,n}\,, \quad \bar{\zeta
}(\xi,\varphi) = \frac{1}{\sqrt{2\pi}} \sum_{(m,n) \in \mathbb{Z}^2} e^{-2\pi m i \xi/\beta_T -  i n \varphi} \bar{\zeta}_{m,n}\,,
\end{equation}
such that
\begin{multline}\label{SLEedgemodes}
S^{(E)}_{\text{edge}} =\sum_{(m,n) \in \mathbb{Z}^2}  \frac{k+i \lambda}{8\pi}\left(\beta_T \upsilon n^2 -{2\pi i m}n \right) \zeta_{m,n} \zeta_{-m,-n}
+ \frac{k-i \lambda}{8\pi} \left({\beta_T}\bar{\upsilon} n^2  - {2\pi i m} n \right) \bar{\zeta}_{m,n} \bar{\zeta}_{-m,-n}\,.
\end{multline}
Here $\zeta_{m,n} \in \mathbb{C}$ and $\bar{\zeta}_{m,n}$ is the complex conjugate of $\zeta_{m,n}$. As discussed in Appendix \ref{Zthermal}, in the Abelian case a Euclidean continuation that preserves the reality conditions of the Chern-Simons gauge field would further require the continuation $\upsilon = -i \upsilon_E$. In the complexified case, this is no longer necessitated as $\upsilon$ is allowed to take complex values. Instead, given the unboundedness of the spectrum we must confront a Gaussian unsuppressed Euclidean path integral
\begin{equation} \label{deformed}
\mathcal{Z}_{\text{edge}} [\beta_T] = \int \mathcal{D}\zeta \mathcal{D} \bar{\zeta} \, e^{-S^{(E)}_{\text{edge}}[\zeta,\bar{\zeta}]}\,.
\end{equation}
In order to render $\mathcal{Z}_{\text{edge}} [\beta_T]$ better defined, we can complexify the contour of path integration \cite{Witten:2010cx}. For simplicity, we take $\upsilon_{re}  > 0$ and $\lambda=0$ while recalling $k \in \mathbb{Z}^+$. It is then clear that taking a contour where $\zeta(\xi,\varphi)$ and $\bar{\zeta}(\xi,\varphi)$ are independent real fields will render $\mathcal{Z}_{\text{edge}} [\beta_T]$ well-defined.\footnote{Expressing $\mathcal{Z}_{\text{edge}} [\beta_T]$ in terms of the Fourier modes (\ref{Fedge}), we can view the problem as an infinite-dimensional version of the complex integral
\begin{equation}
\mathcal{I}_\sigma = \int_{\mathbb{C}\times \mathbb{C}} dz d\bar{z} dw d\bar{w} \, e^{- \sigma z w -  {\sigma} \bar{z} \bar{w}}\,,
\end{equation}
where $\sigma \in \mathbb{C}$ with positive real part. To define $\mathcal{I}_\sigma$ we first take $z$ and $\bar{z}$ as well as $w$ and $\bar{w}$ to be independent complex variables, and then integrate over the contour $w= \bar{z}$ and $\bar{w} = z$. In this sense, we have that $\mathcal{I}_\sigma = {\pi^2}/{{\sigma^2}}$.}
Upon continuing the contour of $\zeta$ and $\bar{\zeta}$, the symmetry group of the edge-mode theory becomes $U(1)\times U(1)$. As we will see in the following section, the price to pay in curing the unboundedness of the Hamiltonian is unitarity of the edge-mode theory. For $\lambda \neq 0$, we must consider a more elaborate contour to render $\mathcal{Z}_{\text{edge}} [\beta_T]$ better defined. This is discussed in the next section where we view the $\lambda \neq 0$ case as a continuation of the $\lambda=0$ case with $k \in \mathbb{Z}^+$.\\
\\
From the perspective of the original Chern-Simons gauge theory (\ref{complexSL}) with complexified gauge group $U_{\mathbb{C}}(1)$, the analytic continuation rendering $\zeta(\xi,\varphi)$ and $\bar{\zeta}(\xi,\varphi)$ as independent real fields can be achieved by continuing the complexified Chern-Simons gauge fields $\mathcal{A}_\mu$ and $\bar{\mathcal{A}}_\mu$ to two independent real-valued $U(1)$ gauge fields, whilst maintaining a complexified level.

\section{Complexified Abelian Chern-Simons: Euclidean model}\label{complexCS}
In this section, we proceed to consider a different complexified version of the Abelian Chern-Simons theory. This theory is introduced as a simple and calculable toy model with a complexified level, a property common to three-dimensional gravity with Euclidean signature and $\Lambda > 0$ viewed as a Chern-Simons theory.

\subsection{Euclidean model on \texorpdfstring{$S^3$}{S3}}
The action for the Euclidean model is given by
\begin{equation}\label{EucU1}
S_E[{A}^\pm_\mu] = \frac{\kappa + i \gamma}{4\pi} \int_{\mathcal{M}} d^3 x  \, \varepsilon^{\mu\nu\rho} {A}^+_\mu \partial_\nu {A}^+_\rho +  \frac{\kappa - i \gamma}{4\pi} \int_{\mathcal{M}} d^3 x  \, \varepsilon^{\mu\nu\rho} {A}^-_\mu \partial_\nu {A}^-_\rho\,,
\end{equation}
where now ${A}^+$ and ${A}^-$ are two real-valued $\mathfrak{u}(1)$ gauge fields. We assume that both $U(1)$ groups are compact and take the parameter $\kappa \in \mathbb{Z}^+$. There is no restriction on $\gamma$ so we take $\gamma \in \mathbb{R}^+$. 
The $S^3$ partition function must be carefully defined since $i S_E[{A}^\pm_\mu]$ is no longer purely oscillatory but rather an unbounded functional rendering the path integral over $e^{i S_E}$ problematic. To make sense of it we can again consider a complexification of the original path integration contour. On $S^3$ the only $U(1)$ flat connection (modulo gauge transformations) is the trivial one. Moreover, given that the action is quadratic in the fields the choice of contour can be reduced to a problem of Gaussian integration. The Gaussian integral we are interested in is of the form
\begin{equation}
\mathcal{I}_\sigma = \int_{\mathcal{C}} d x  \, e^{-\sigma x^2/2}\,,
\end{equation}
with $\sigma = e^{i \vartheta}|\sigma| \in \mathbb{C}$. We take $\mathcal{C}$ to be along a path $e^{i\varphi} \tilde{x}$ with $\tilde{x} \in \mathbb{R}$ and $\vartheta+2\varphi \in (-\pi/2,\pi/2)$ such that
\begin{equation}
\mathcal{I}_\sigma = \sqrt{\frac{2\pi}{|\sigma|}} e^{-i\vartheta/2}\,.
\end{equation}
We further take that $\mathcal{I}_{\bar{\sigma}} \equiv \left( \mathcal{I}_{\sigma} \right)^*$ such that the corresponding contour is $e^{-i\varphi} \tilde{x}$, and the product $\mathcal{I}_{\bar{\sigma}} \mathcal{I}_{\sigma}$ is positive and real. Applying the same reasoning to the Chern-Simons case, we arrive at
\begin{equation}\label{complexS3}
Z_{\kappa,\gamma}[S^3] = \left(\frac{1}{\text{vol} \, U(1)} \right)^2 \times \left| \frac{4\pi^2}{\kappa+i\gamma} \right| = \frac{1}{\sqrt{\kappa^2+\gamma^2}}\,.
\end{equation}
The above result is invariant under complex conjugation, in other words $\gamma\to-\gamma$. For $\gamma=0$ we retrieve the standard result (\ref{Zu1}) with a choice of framing leading to a vanishing phase. For $\kappa = 0$, the result makes sense for $\gamma \in \mathbb{R}^+$ but becomes non-analytic if one tries to continue to the whole complex-$\gamma$ plane.\footnote{Although our treatment leads to a real valued $Z_{\kappa,\gamma}[S^3]$, it seems feasible that a more general regularisation can lead to an overall phase that depends on the choice of framing of $S^3$. It would be interesting to explore this both for the Abelian case and the non-Abelian extension that will be discussed in Section \ref{EuclideanGrav}.}  

\subsection{Euclidean edge-mode theory on the torus}
Since $A^\pm$ are now compact $\mathfrak{u}(1)$ gauge fields the boundary chiral bosons, which we denote as $\zeta^\pm$, are now compact. Following the procedure outlined in previous sections, and imposing the boundary condition 
\begin{equation}
A^\pm_\xi - \upsilon^\pm A^\pm_\varphi |_{\partial \mathcal{M}} = 0\,,
\end{equation}
with $ \upsilon^\pm \in \mathbb{R}$, we end up with an edge-mode theory governed by the action
\begin{equation}
S_{\text{edge}} = \frac{\kappa+i\gamma}{4\pi} \int d \xi d\varphi \, \partial_\varphi \zeta^+  \left( \partial_\xi  - \upsilon^+ \partial_\varphi  \right) \zeta^+ +  \frac{\kappa-i\gamma}{4\pi} \int d \xi d\varphi \, \partial_\varphi \zeta^-  \left( \partial_\xi  - \upsilon^- \partial_\varphi  \right) \zeta^- \, .
\end{equation}
The above theory, which we take to be in Euclidean signature, is non-unitary. Consequently, a Hilbert space interpretation of the theory is unclear. Nevertheless, we can try to compute the Euclidean torus partition function of $S_{\text{edge}}$. We take the periodicity of Euclidean time to be $\xi \sim \xi+\beta_T$, and we recall that $\varphi \sim \varphi+2\pi$. 
It is convenient to express the non-winding sector in terms of their respective Fourier modes, namely
\begin{equation}
\zeta^\pm(\xi,\varphi) = \frac{1}{\sqrt{2\pi}} \sum_{(m, n) \in \mathbb{Z}^2} \zeta^\pm_{m,n} e^{  2\pi i m \xi/ \beta_T +  i n \varphi}\,.
\end{equation}
The reality condition is $\zeta^\pm_{-m,-n} = \left( \zeta^\pm_{m,n}\right)^*$. The edge-mode action now reads
\begin{equation}\label{SEedge}
S_{\text{edge}} = \sum_{(m,n)\in \mathbb{Z}^2} \frac{\kappa+i\gamma}{4\pi}  \left( {2\pi m n} -\beta_T \upsilon^+  n^2\right) |\zeta^+_{m,n}|^2 + \frac{\kappa-i\gamma}{4\pi}  \left( {2\pi m n} - \beta_T \upsilon^-  n^2\right) |\zeta^-_{m,n}|^2\,,
\end{equation}
and the path integral becomes an integral over all the $\zeta^\pm_{m,n}$. We can compare the edge-mode action $i S_{\text{edge}}$ to the one stemming from Lorentzian edge-mode theory, i.e. $-S^{(E)}_{\text{edge}}$ in (\ref{SLEedgemodes}), discussed in the previous section. If we map $ (k,\lambda) = (2\kappa,2\gamma)$ and $(\upsilon,\bar{\upsilon}) = i (\upsilon^+,\upsilon^-)$, it follows that the deformed contour discussed below (\ref{deformed}) indeed leads to the Euclidean edge-mode theory (\ref{SEedge}). \\
\\
In order to render the path integral over $i S_{\text{edge}}$ convergent, we must again alter the integration contour $\mathcal{C}$ of $\zeta_{n,m}^\pm$. The path integration over the non-winding mode sector of $\zeta^\pm$ yields a functional determinant proportional to the Dedekind $\eta$-function (see Appendix \ref{Zthermal}). The sum over the winding sector must be similarly defined. The winding modes are given by $\zeta^\pm(t,\varphi) = m_\pm \varphi$ where $m_\pm$ are integers.\footnote{Given the periodicity of $\zeta^\pm$, one is also tempted to consider the vertex operators $\hat{\mathcal{O}}^\pm_n = :e^{i n \hat{\zeta}^\pm}:$ carrying $U(1)$ charge (\ref{U1charge}) $\mathcal{Q}^\pm_n = n/(\kappa\pm i\gamma)$. Although $\mathcal{O}^\pm_n$ bear some similarity to the anyonic operators (\ref{Psi}) of ordinary Abelian Chern-Simons theory, their physical meaning is more obscure at complex level. For instance, in the ordinary Abelian case $k \in \mathbb{Z}^+$ corresponds to the number of anyonic species, something hard to understand when the level is complex.} Consequently, we must perform the sum
\begin{equation}
\mathcal{S}_{\text{w}} = \sum_{m_\pm \in \mathbb{Z}} e^{-i \frac{ \kappa + i \gamma}{2} \upsilon^+ \beta_T m^2_+  - i \frac{ \kappa - i \gamma}{2}  \upsilon^- \beta_T m^2_-} =  \vartheta_3\left(0,q_+^{( \kappa+ i \gamma)/2}\right)  \vartheta_3 \left(0,q_-^{( \kappa- i \gamma)/2}\right) \,.
\end{equation}
The above sum exists provided $- ( { \kappa \pm  i\gamma} )\beta_T \upsilon^\pm/2 \pi$ is in the upper-half plane. In the last equality, we have defined ${q_{\pm} \equiv e^{-i \upsilon^{\pm}\beta_T}}$. 
Putting it all together, upon path integrating over $e^{iS_{\text{edge}}}$ we end up with the torus partition function
\begin{equation}
Z_{\text{edge}}\left[\beta_T\right] = \frac{\vartheta_3\left(0,q_+^{( \kappa+ i \gamma)/2}\right)  \vartheta_3 \left(0,q_-^{( \kappa- i \gamma)/2}\right)}{\eta(q_+) \eta(q_-)}\,.
\end{equation}
To render the Dedekind $\eta$-functions well-defined, we take $\upsilon^\pm$ to have a small negative imaginary part. Taking the $\beta_T\to 0^+$ limit we have 
\begin{equation}
\lim_{\beta_T \rightarrow 0^+}\log Z_{\text{edge}}\left[\beta_T\right] = \frac{\pi^2 }{6\beta_T i \upsilon^+} + \frac{\pi^2 }{6\beta_T i \upsilon^-} - \log |\kappa+i\gamma| \ldots
\end{equation}
The finite part agrees with the $S^3$ partition function (\ref{complexS3}). We further note that the above expression can be obtained from the Abelian edge-mode expression (\ref{edgefinal}) by a simple analytic continuation.

\section{Remarks on gravitation in three dimensions}\label{oneloop3dgrav}

The classical action of general relativity in three dimensions can be expressed as a Chern-Simons theory \cite{Achucarro:1987vz,Witten:1988hc}. The gauge group depends on the signature and sign of the cosmological constant $\Lambda$. We restrict to $\Lambda = +1/\ell^2 >0$. In Euclidean signature the gauge group is $SU(2)\times SU(2)$ and the level is complex. In Lorentzian signature the gauge group becomes $SL(2,\mathbb{C})$, which can be viewed as the complexification of $SU(2)$ (or $SL(2,\mathbb{R})$). These Chern-Simons theories are natural non-Abelian extensions of the complexified Abelian theories explored in sections \ref{complexCSL} and \ref{complexCS}. The purpose of this section is to comment briefly on how the properties of the complexified Abelian theories generalise, leaving a more detailed analysis to future work. 

\subsection{Euclidean signature}\label{EuclideanGrav}

We begin by considering the Euclidean theory whose action is given by $S_E = S^+_E + S^-_E$, where
\begin{equation}\label{3dgravE}
S^\pm_E[{A}^\pm_\mu] = \frac{\kappa \pm i \gamma}{4\pi}  \int_{\mathcal{M}} d^3 x  \, \varepsilon^{\mu\nu\rho} \, \text{Tr}\left({A}^\pm_\mu \partial_\nu {A}^\pm_\rho + \frac{2}{3} A^\pm_\mu A^\pm_\nu A^\pm_\rho \right)\,.
\end{equation}
The $A^\pm_\mu$ are $SU(2)$ gauge fields. In terms of the vielbein $e^i_\mu$ and spin-connection $\omega_{ij,\mu} = {\epsilon_{ijk}} {\omega^{k}}_\mu$,  where $\epsilon_{ijk}$ is the Levi-Civita symbol, one has
\begin{equation}
A^\pm_\mu =  \left( \omega_\mu^i \pm \frac{1}{\ell}e^i_\mu \right) T_i\,, \quad\quad \,T_i = \frac{1}{2} i \sigma_i\,,
\end{equation}
where the $\sigma_i$ are the Pauli matrices. Substituting $A^\pm$ into (\ref{3dgravE}) leads to Einstein gravity in three dimensions plus a parity-odd Chern-Simons type term for the spin connection. The imaginary part of the level, $\gamma \in \mathbb{R}^+$, is given by $\gamma = \ell/4 G$, while $\kappa \in \mathbb{Z}$ is the coupling of a parity-odd gravitational Chern-Simons term. \\
\\
The equations of motion stemming from (\ref{3dgravE}) are the flat connection equations, and the solutions are the space of flat connections modulo gauge transformations. For standard $SU(2)$ Chern-Simons theory on an $S^3$ we would discard the presence of multiple possible saddles due to the absence of non-trivial flat connections. However, when the Chern-Simons coupling is taken away from the integers one must exercise further caution. In particular, gauge transformations changing the winding number will no longer be trivial. So, the spaces of flat connections with differing winding number have different on-shell actions. Of these, most will not have a clear interpretation from the perspective of the gravitational theory since they will lead to non-invertible or otherwise non-standard vielbeins. This is a feature we did not have to confront in the Abelian toy models of the previous sections. We consider perturbative effects around the flat-connection $A^+ = g^{-1} dg$ and $A^- = 0$, where $g \in SU(2)$. Recalling that the group geometry of $SU(2)$ under the Haar metric is the three-sphere, this flat connection corresponds to the round $S^3$ in the gravitational theory as can be explicitly checked. Following reasoning analogous to the Euclidean $U(1)$ model with complexified levels, it has been argued that \cite{Gukov:2016njj,Anninos:2020hfj}
\begin{equation} \label{subleadingterms}
\mathcal{Z}_{\text{grav}} = e^{2\pi \gamma} \times e^{ i \varphi_{\text{grav}}} \left| \sqrt{\frac{2}{\kappa+i \gamma+2}} \sin {\frac{\pi}{\kappa+i \gamma + 2}} \right|^2\,,
\end{equation}
to all orders in a large-$\gamma$ perturbative expansion, and this expansion reproduces (\ref{logZgrav}).
According to Gibbons and Hawking \cite{Gibbons:1976ue}, $S_{\text{dS}} = \log \mathcal{Z}_{\text{grav}}$ calculates the quantum corrected entropy of the dS$_3$ horizon. To leading order in the large $\gamma$ expansion, $S_{\text{dS}} = 2\pi \gamma = \pi \ell/2 G$. The one-loop correction is given by the analytic continuation of the expression at integer level:
\begin{equation}\label{oneloopgrav}
e^{S^{(1)}_{\text{dS}}} = \left(\frac{1}{\text{vol} \, SU(2)} \right)^2 \times \left| \frac{4\pi^2}{\kappa+ i \gamma + 2} \right|^3\,.
\end{equation}
This is the natural non-Abelian extension of (\ref{complexS3}). It is challenging to reproduce the leading contribution $S_{\text{dS}}$ directly from the gravitational perspective (attempts include \cite{Maldacena:1998ih,Banados:1998tb,Govindarajan:2002ry}). In line with our discussion so far, we now consider the possibility of an edge-mode theory which may reproduce the subleading corrections to $S_{\text{dS}}$ from a Lorentzian perspective.

\subsection{Lorentzian signature}

We now turn to the Lorentzian theory
\begin{equation}\label{3dgravL}
S_L[\mathcal{A}_\mu] = \frac{k + i \lambda}{8\pi}  \int_{\mathcal{M}} d^3 x  \, \varepsilon^{\mu\nu\rho} \, \text{Tr} \left(\mathcal{A}_\mu \partial_\nu \mathcal{A}_\rho + \frac{2}{3}\mathcal{A}_\mu \mathcal{A}_\nu \mathcal{A}_\rho \right) + \text{c.c.}\,,
\end{equation} 
where $\mathcal{A}_\mu$ is an $SL(2,\mathbb{C})$ gauge field related to the vielbein and spin connection as
\begin{equation} 
\mathcal{A}_\mu = \left(\omega^i_\mu  + \frac{i}{\ell}  e^i_\mu\right)T_i\,, \quad \bar{\mathcal{A}}_\mu =\left(\omega^i_\mu  - \frac{i}{\ell}  e^i_\mu\right)T_i\,,
\end{equation}
where $(T_1,T_2,T_3) = \tfrac{1}{2}(i\sigma_2, \sigma_1, \sigma_3)$ are the real generators of $SL(2,\mathbb{R})$ obeying $\text{Tr}(T_i T_j) = \tfrac{1}{2} \eta_{ij}$ and $[T_i,T_j] = \varepsilon_{ijk}T^k$, with $T^i = \eta^{ij}T_j$. We further have that $k\in\mathbb{Z}^+$ and $\lambda\in\mathbb{R}$. \\
\\
As for the Abelian theory with $U_{\mathbb{C}}(1)$ gauge group, we would like to understand the edge-mode theory. We must pick a boundary condition for the $SL(2,\mathbb{C})$ gauge field. Any choice will break the full diffeomorphism group, for the same reason that boundary conditions break the gauge-symmetries of the Chern-Simons theory. Various proposals for boundary conditions have appeared in the literature \cite{Carlip:1994gy,Banados:1996ad,Arcioni:2002vv}. Our interest is to understand the diffeomorphism invariant one-loop correction $S^{(1)}_{\text{dS}}$ (\ref{oneloopgrav}), and more generally the subleading corrections to the tree-level de Sitter entropy $S_{\text{dS}}$ encoded in (\ref{subleadingterms}), from a Lorentzian perspective. In line with our previous discussions, we choose boundary conditions (\ref{gauge}), namely
\begin{equation}
\left( \mathcal{A}_t - \upsilon \mathcal{A}_\varphi \right) |_{\partial \mathcal{M}} = 0\,, 
\end{equation}
with $\upsilon \in \mathbb{C}$. Going through the same steps, the Lorentzian edge-mode theory is then described by a chiral $SL(2,\mathbb{C})$ WZW theory at level $k + i \lambda$ \cite{Maldacena:1998ih,Witten:1989ip,Arcioni:2002vv}
\begin{equation}\label{WZWsl2C}
S_{\text{edge}}[g] = \frac{k + i \lambda}{8\pi} \, \text{Tr} \int dt d\varphi  \left( g  \partial_\varphi  g^{-1} \right) \left( g \left(\partial_t - \upsilon \partial_\varphi \right) g^{-1} \right)  + \pi (k + i \lambda) S^{\text{WZ}}[g] + \text{c.c.} \,,
\end{equation}
where $g(t,\varphi)$ is an $SL(2,\mathbb{C})$ valued function, and 
\begin{equation}
S^{\text{WZ}}[g] = \frac{1}{24 \pi^2} \int_{\mathcal{B}} d^3 y \, \varepsilon^{\mu \nu \rho} \, \text{Tr}\left(g^{-1} \partial_\mu g g^{-1} \partial_\nu g  g^{-1} \partial_\rho g\right) \,,
\end{equation}
where $\mathcal{B}$ is a three-manifold whose boundary is the $t\varphi$-cylinder. As for the Abelian Lorentzian model, the Hamiltonian stemming from (\ref{WZWsl2C}) is real but unbounded from below. The theory satisfies an $SL(2,\mathbb{C})$ current algebra given by (see for instance \cite{Maldacena:1998ih})
\begin{equation}
\begin{split}
	 \left[ J^a_{n}, J^b_{m} \right] &= \tfrac{1}{2}(k + i \lambda ) n \delta_{ab} \delta_{n + m, 0} -  i f_{abc} \, J^c_{n+m}, \\
	 \left[ \tilde{J}^{a}_{n} , \tilde{J}^{b}_{m} \right] &= \tfrac{1}{2} (k - i \lambda ) n \delta_{ab} \delta_{n + m, 0} - i f_{abc}\, \tilde{J}^{c}_{n+m} \,,
\end{split}
\end{equation}
where the $SL(2,\mathbb{C})$ structure constants are $f_{abc} = \varepsilon_{abc}$, and $\varepsilon_{abc}$ is the Levi-Civita symbol. Using the above currents, the Sugawara construction yields a Virasoro algebra with central charge
\begin{equation}
c_{{SL}(2,\mathbb{C})} = \frac{3(k+i\lambda)}{(k+i\lambda+2)} +\frac{3(k-i\lambda)}{(k-i\lambda+2)} = \frac{6 \left(k (k+2) + \lambda ^2 \right)}{(k+2)^2+\lambda ^2}\,. 
\end{equation}
Some details of the derivation for $c_{{SL}(2,\mathbb{C})}$ are provided in Appendix \ref{centralcharge}. Notice that in the semiclassical limit $k\to\infty$, we have that $c_{{SL}(2,\mathbb{C})} \approx 6$ which is the number of field theoretic degrees of freedom in (\ref{WZWsl2C}). Interestingly, upon analytically continuing $k = -2$, the central charge is exactly $c_{{SL}(2,\mathbb{C})} = 6$, which might suggest that the theory is very simple at this point. Moreover, we see that $c_{\text{WZW}}$ is given by complexifying the level $k$ for the central charge $c_{{SU(2)}} = 3k/(k+2)$ of the $SU(2)$ WZW model. It also appears that the vacuum ($s=0$) $\widehat{\mathfrak{su}}(2)_k$ character (\ref{su2char}) can be analytically continued to complex level, giving the thermal partition function of this theory. \\
\\
As for the case of the Lorentzian complexified $U(1)$ model studied in Section \ref{complexCSL}, upon continuing the Lorentzian edge-mode theory (\ref{3dgravL}) to Euclidean signature, one can consider complexifying the $SL(2,\mathbb{C})$ contour leading to that of a WZW model with two copies of $SU(2)$ at complex level. In particular, to render the path integral better defined it is natural to pursue the same avenue as in Section \ref{complexL}. Upon complexifying the contour of $g$ to one where it is valued in $SU(2)\times SU(2)$ we can make contact with (\ref{oneloopgrav}) much like (\ref{nonabelianedge}) connects to the three-sphere partition function of $SU(2)$ Chern-Simons theory (\ref{zsu2s3}). A detailed treatment of the putative gravitational edge-mode theory (\ref{WZWsl2C}) and its thermodynamic properties is left for future work. \\
\\
As a final remark, we note that timelike Wilson lines piercing the origin of the spatial disc carry unitary irreducible representations of $SL(2,\mathbb{C})$ which, when non-trivial, contain an infinite number of states.\footnote{See \cite{Castro:2020smu} for a related discussion.} This is another indication of the non-compactness of $SL(2,\mathbb{C})$. It would be interesting to generalise the discussion in Section \ref{Comments on the non-Abelian case} to this case.

\section{\texorpdfstring{AdS$_4$/CFT$_3$ and $S^3$}{AdS4/CFT3 and S3}}\label{adscft}
\begin{figure}[ht]
	\centering
	\includegraphics[width=0.4\linewidth]{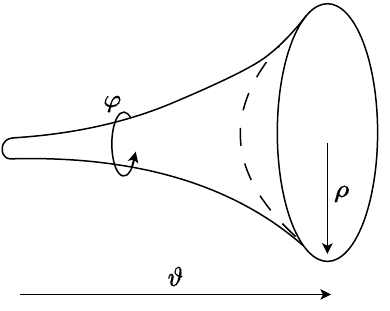}
	\caption{A constant $t$ slice of the geometry (\ref{LAdS}). The space is foliated by static patch hemispheres with exponentially increasing radius in the $\vartheta$ direction and the horizon at $\rho = \pi/2$ is the surface of this shape. }
	\label{fig:AdS4Spslicing}
\end{figure}

In this final section we would like to consider the AdS$_4$/CFT$_3$ correspondence for an AdS$_4$ spacetime whose asymptotic boundary is given by a Euclidean or Lorentzian three-dimensional de Sitter space. This allows us to explore the thermodynamic properties of strongly coupled conformal matter theories on a fixed de Sitter background and their geometrisation in the bulk of AdS$_4$.\\
\\
The metric of Euclidean AdS$_4$ is given by
\begin{equation}\label{EAdS}
\frac{ds^2}{\ell^2_{A}} = d\vartheta^2 + \sinh^2\vartheta \, d\Omega^2_3 \, ,
\end{equation}
with $\vartheta \in [0,\infty)$ and $d\Omega_3^2$ describing the round metric on a unit $S^3$. The four-dimensional cosmological constant is $\Lambda = -3/\ell_A^2$. The asymptotic boundary resides at $\vartheta_\infty = \infty$ and the induced metric at the boundary is the round metric on $S^3$. Recalling our previous discussion, we can consider a Lorentzian continuation of (\ref{EAdS}) to the following static spacetime
\begin{equation}\label{LAdS}
\frac{ds^2}{\ell^2_{A}} = d\vartheta^2 + \sinh^2\vartheta \left(-dt^2 \cos^2 \rho + d\rho^2 + \sin^2\rho \, d\varphi^2 \right) \, .
\end{equation}
This is a static Lorentzian anti-de Sitter universe whose constant-$\vartheta$ surfaces are given by the static patch of dS$_3$. A constant-$t$ slice of this geometry is shown in Figure \ref{fig:AdS4Spslicing}. The global geometry is given by replacing the three-dimensional static patch metric in (\ref{LAdS}) with the global one. The static geometry (\ref{LAdS}) has a horizon at $\rho = \pi/2$. The topology of the horizon is $S^1 \times \mathbb{R}^+$. The Bekenstein-Hawking entropy of the horizon is given by
\begin{equation}\label{SBH}
S_{\text{BH}} =  \frac{\pi \ell_A^2}{2\ell_{pl}^2}  \int_0^{-\log \varepsilon} d\vartheta \sinh \vartheta \approx  \frac{\pi \ell_A^2}{2\ell_{pl}^2}\left( \frac{1}{2 \varepsilon } - 1+ \ldots \right) \, ,
\end{equation}
where $\ell_{pl}$ is the four-dimensional Planck length (such that $G = \ell_{pl}^2$) and $\varepsilon$ is a small number cutting off $z = e^{-\vartheta}$. From the perspective of AdS/CFT the entropy $S_{\text{BH}}$ corresponds to the entropy across the dS$_3$ horizon of the dual CFT, and the $1/\varepsilon$ divergence corresponds to a local divergence in the CFT due to entanglement of modes localised on the $S^1$ horizon. 

\noindent It is useful to expand the geometry in small $z$
\begin{equation}
\frac{ds^2}{\ell^2_A} = \frac{dz^2}{z^2} + \frac{1}{4} \left( \frac{1}{z^2} - 2 + z^2  \right) d\Omega^2_3\,.
\end{equation}
From the above expansion it follows that the solution has vanishing boundary Brown-York stress tensor. Since the energy vanishes, we have that the logarithm of the thermal partition function is given entirely by the entropy. The thermal partition function is given by the Euclidean gravity path integral with an $S^3$ boundary, such that
\begin{equation}
\log Z[\text{EAdS}_4] \overset{?}{=} S_{\text{BH}}\,.
\end{equation}
In the semiclassical limit, we can calculate $\log Z[\text{EAdS}_4]$ by evaluating the on-shell Einstein action 
\begin{equation} \label{AdSonshell}
S_E[g_{ij}] = - \frac{1}{16\pi \ell_{pl}^2}\int_{\mathcal{M}} d^4x \sqrt{g} \left(R + \frac{6}{\ell_A^2} \right) - \frac{1}{8\pi \ell_{pl}^2}\int_{\partial\mathcal{M}} d^3x \sqrt{h} K\,,
\end{equation}
on the solution (\ref{EAdS}). The second term is the Gibbons-Hawking-York boundary term, where $K$ is the extrinsic curvature at the boundary $\partial \mathcal{M}$, and $h_{ij}$ is the induced metric at the boundary:
\begin{equation}
ds^2_{\text{bdy}} = \ell^2_{A} \sinh^2\vartheta_\infty \, d\Omega^2_3\,.
\end{equation}
\\
Using a similar regularisation prescription as in \eqref{SBH}, the on-shell action \eqref{AdSonshell} evaluates to 
\begin{equation} \label{UVdivs}
    S_E = \frac{\pi  \ell^2_{A}}{2 \ell_{pl}^2} - \frac{\pi \ell^2_{A}}{16\ell_{pl}^2 \varepsilon^3 } - \frac{3 \pi \ell^2_{A}}{16  \ell_{pl}^2 \varepsilon } \, . 
\end{equation}
In general, we can also add boundary counterterms which are built locally from the boundary metric $h_{ij}$. These will not affect the bulk equations of motion: 
\begin{equation}
S_{CT} = \frac{a_0}{\varepsilon^3} \int_{S^3} d^3 x \sqrt{\tilde{h}} + \frac{a_1}{\varepsilon} \int_{S^3} d^3 x \sqrt{\tilde{h}} R[\tilde{h}]\,,
\end{equation}
where $\tilde{h}$ is the boundary CFT metric in units of $\varepsilon$ with the conformal factor stripped off,
\begin{equation}
ds_{\text{CFT}}^2 = \frac{\ell^2_A}{4} d\Omega_3^2\,.
\end{equation}
The coefficients $a_0$ and $a_1$ are ambiguous and depend on our choice of terms localised at the boundary. In the dual quantum field theory these are related to ultraviolet divergences (\ref{divergence}) renormalising the boundary cosmological constant and Newton constant. One may choose the coefficients such as to cancel the UV divergences stemming from the on-shell action \cite{Balasubramanian:1999re}. Indeed, if our theory is supersymmetric, we expect that the cubic divergences associated to the cosmological constant term should vanish, and we therefore choose $a_0 = 1/4 \pi \ell_A \ell_{pl}^2$. Thus, in the semiclassical limit we have
\begin{equation}\label{logZAdS}
\log Z[\text{EAdS}_4] = - \frac{\pi \ell_A^2}{2\ell_{pl}^2} + \frac{3 \pi \ell^2_{A}}{16  \ell_{pl}^2 \varepsilon } - \frac{6 \pi^2 \ell_A a_1}{\varepsilon}\,.
\end{equation}
We see that the finite term indeed matches \eqref{SBH}. Rather than choosing $a_1$ to cancel the linear divergence in \eqref{logZAdS}, if we instead tune $a_1 = 7 \ell_A /96 \pi  \ell_{pl}^2$, we can also match the linearly divergent term in equation \eqref{SBH}.\footnote{Entanglement entropy in de Sitter was also studied holographically in \cite{Maldacena:2012xp}. There, the entropy of quantum fields cuts over many horizons at the late time boundary, whereas we consider the entanglement of gravitational fields within a single static patch.} \\
\\
For ABJM theory with $SU(N)_k\times SU(N)_{-k}$ gauge group, the partition function (\ref{logZAdS}) has been calculated exactly in \cite{Kapustin:2009kz} and matched to the bulk in \cite{Drukker:2010nc}. The finite part, in terms of the ABJM data, reads
\begin{equation}
\log Z_{\text{ABJM}}[S^3] = -  N^2 \frac{\pi \sqrt{2}}{3} \lambda^{-1/2}\,,
\end{equation}
where $\lambda = N/k$ is the 't Hooft parameter which is kept fixed and large in the large $N$ limit. The fact that $\log Z_{\text{ABJM}}[S^3]$ goes as $\sim N^2$ suggests that the theory is in a deconfined phase, reminiscent of the discussion in \cite{Witten:1998zw}. From our perspective, the calculations of \cite{Kapustin:2009kz, Drukker:2010nc} are a microscopic derivation of the Bekenstein-Hawking entropy of the horizon in (\ref{LAdS}). The Bekenstein-Hawking entropy (\ref{SBH}) predicts only linear ultraviolet divergences in the partition function. In fact, this is also true for the $S^3$ partition function calculated in \cite{Kapustin:2009kz}, since they regularise the theory in a way that preserves supersymmetry. The ultraviolet divergent piece of their calculation comes from the one-loop determinants of a transverse vector field and a gaugino on $S^3$. The absence of a cubic divergence is a result of the spectrum alone and is insensitive to any mass terms. \\
\\
Recalling (\ref{divergence}), the only other allowed divergence in a parity invariant three-dimensional quantum field theory is linear in $\ell^{-1}_{\text{uv}}$ and is indeed implicitly present in the localisation treatment of \cite{Kapustin:2009kz}. This is in agreement with (\ref{SBH}).  Interestingly, the calculations of \cite{Kapustin:2009kz,Drukker:2010nc} can be done for any value of the Chern-Simons level $k$ and rank $N$. They can be viewed as predicting string and loop corrections of the Bekenstein-Hawking entropy from the bulk perspective. In the perturbative limit, where $\lambda \ll 1$, it is found that 
\begin{equation}
\log Z_{\text{ABJM}}[S^3] \approx - N^2 \log \lambda^{-1} + \ldots\,,
\end{equation}
which is in agreement with the pure Chern-Simons partition function result (\ref{CSpert}) in the perturbative limit.\footnote{The relation between topological entanglement entropy and black holes in AdS/CFT was also studied in \cite{McGough:2013gka}.}\\
\\
One can also consider adding Wilson loops \cite{Kapustin:2009kz,Drukker:2010nc}. In the Euclidean picture, the simplest case is a Wilson loop that goes around the equator of the $S^3$.  To preserve supersymmetry, in addition to the gauge field, one also has additional matter fields on the loop. For instance, in ABJM, the partition function endowed with a $1/2$-BPS preserving Wilson loop in the fundamental representation is given as follows
\begin{equation}\label{WL}
\frac{Z_\text{ABJM}[S^3;W_{\mathcal{C}}]}{Z_{\text{ABJM}}[S^3]} = \frac{e^{i\pi/2} }{2} e^{\pi\sqrt{2\lambda}}\,,
\end{equation}
where the expression is given to leading order in the large $N$ limit at fixed and large 't Hooft coupling $\lambda = N/k$. From the bulk perspective, the above is computed by a string whose worldsheet intersects the $S^3$ boundary at the equatorial $S^1$. The phase is due to a type IIA B-field. We can  Wick rotate to a Lorentzian picture (\ref{LAdS}), where the equatorial $S^1$ at $\rho=0$ is now parameterised by the Lorentzian coordinate $t$. The worldsheet now intersects the boundary along $t$ at $\rho = 0$ and goes all the way to the horizon. In the thermodouble field picture, we can continue the worldsheet across the horizon such that the worldsheet intersects the second boundary static patch. As in our discussion of pure Chern-Simons theory, we might view the partition function with the insertion of the Wilson loop as computing a part of the entanglement entropy between the two static patches in the presence of an insertion at the origin of the two spatial discs. From the bulk perspective this is the contribution to the Bekenstein-Hawking entropy (\ref{SBH}) due to a worldsheet crossing the bulk horizon.

\chapter{An alternative edge-mode theory}
\label{GeneralBCs}
One might ask if the boundary conditions we have chosen for the disc (\ref{gauge}) are the most general we could consider. It might be possible that with an alternative choice of boundary conditions, we could engineer the Lorentzian edge-mode theory with complexified gauge group to have a spectrum which is bounded from below, hence avoiding the need to complexify the integration contour. In this chapter, we explore this possibility and show that the boundary condition (\ref{gauge}) does in fact lead to the most general edge-mode theory. This chapter is based on unpublished work that was done with Dionysios Anninos and K{\'e}vin Nguyen.

\section{General boundary conditions}
We will be interested again in the Lorentzian action \eqref{ABaction}, which is repeated here for convenience: 
\begin{equation} \label{realfieldaction}
    S_L = \frac{k}{4\pi} \int_{\mathcal{M}} d^3x \, \varepsilon^{\mu \nu \rho} (A_\mu \partial_\nu A_\rho - B_\mu \partial_\nu B_\rho ) - \frac{\lambda}{4\pi} \int_{\mathcal{M}} d^3x \, \varepsilon^{\mu \nu \rho} ( A_\mu \partial_\nu B_\rho + B_\mu \partial_\nu A_\rho) \, ,
\end{equation}
where $\mathcal{M}= D \times \mathbb{R}$. This action has a discrete symmetry under $A_\mu \leftrightarrow B_\mu$ and $k \rightarrow - k$. Instead of the condition \eqref{Complexgaugechoice}, we would like to impose the more general gauge condition
\begin{equation} \label{generalgauge}
   \begin{split}
A_t&=\gamma_1 A_\varphi + \gamma_2 B_\varphi\,,\\
B_t&=\gamma_3 A_\varphi + \gamma_4 B_\varphi \,,
\end{split} 
\end{equation}
where $\gamma_i \in \mathbb{R}$. If we take 
\begin{equation} \label{alphaoriginal}
    \begin{split}
        \gamma_1 &=\gamma_4=\upsilon_{re} \, , \\
        -\gamma_2 &= \gamma_3=\upsilon_{im} \, ,
    \end{split}
\end{equation}
then this is equivalent to the original gauge choice \eqref{Complexgaugechoice}. Upon imposing the boundary conditions, the discrete symmetry also requires us to map $\gamma_1 \leftrightarrow \gamma_4$ and $\gamma_2 \leftrightarrow \gamma_3$ . Varying the action, we find 
\begin{equation}
\begin{split}
    \delta S_L &= \frac{k}{4\pi} \int_{\mathcal{M}} d^3x \, \varepsilon^{\mu \nu \rho} (\delta A_\mu F^A_{\nu \rho} - \delta B_\mu F^B_{\nu \rho} + \partial_\nu (A_\mu \delta A_\rho - B_\mu \delta B_\rho)) \\
    &\quad- \frac{\lambda}{4\pi} \int_{\mathcal{M}} d^3x \, \varepsilon^{\mu \nu \rho} (\delta A_\mu F^B_{\nu \rho}  + \delta B_\mu F^A_{\nu \rho} + \partial_\nu (A_\mu \delta B_\rho + B_\mu \delta A_\rho) ) \, ,
\end{split}
\end{equation}
where $F_{\mu \nu}^A$ and $F_{\mu \nu}^B$ are the field strength tensors for $A_\mu$ and $B_\mu$ respectively. Requiring the total derivative boundary term to vanish leads us to
\begin{equation}
    k (A_t \delta A_\phi - A_\phi \delta A_t - B_t \delta B_\phi + B_\phi \delta B_t ) - \lambda (A_t \delta B_\phi - A_\phi \delta B_t + B_t \delta A_\phi - B_\phi \delta A_t) \stackrel{!}{=}0 \, .
\end{equation}
Upon imposing (\ref{generalgauge}), this is equivalent to the condition 
\begin{equation}
    k(\gamma_2 + \gamma_3) + \lambda (\gamma_1 - \gamma_4) = 0 \, . 
\end{equation}
We can therefore replace, for example, $\gamma_4$ in terms of the other $\gamma_i$ in order for the boundary term to vanish, and we will proceed to do this. It is interesting to note though that the condition is also satisfied at the fine tuned point 
\begin{equation}
    k = - \frac{\lambda(\gamma_1 - \gamma_4)}{\gamma_2 + \gamma_3} \, . 
\end{equation}
The associated gauge constraints are 
\begin{equation} \label{gaugeconstraint}
\begin{split}
    F^A_{ij} &= 0 \implies A_i = \partial_i \zeta_{re} \, , \\
    F^B_{ij} &= 0 \implies B_i = \partial_i \zeta_{im} \, , \qquad i \in (r,\phi) \, ,
\end{split}
\end{equation}
where $\zeta = \zeta_{re} + i \zeta_{im}$. Imposing the gauge and the gauge constraints on the action leads to the edge-mode theory
\begin{multline}
    S_{\text{edge}} = \frac{1}{4\pi} \int dt d\phi \, (\lambda \gamma_3 - k \gamma_1 ) (\partial_\phi \zeta_{re})^2 +(k \gamma_4 + \lambda \gamma_2 ) (\partial_\phi \zeta_{im} )^2 - 2\lambda \partial_t \zeta_{re} \partial_\phi \zeta_{im}\\
    +(k (\gamma_3 - \gamma_2) + \lambda (\gamma_1 + \gamma_4)) \partial_\phi \zeta_{re} \partial_\phi \zeta_{im}  + k( \partial_t \zeta_{re} \partial_{\phi} \zeta_{re} - \partial_t \zeta_{im} \partial_{\phi} \zeta_{im}) \, . 
\end{multline}
Plugging in the values \eqref{alphaoriginal}, this becomes the edge-mode theory that we considered Section \ref{complexCSL} of Chapter \ref{SoftDeSitter}. The discrete symmetry of (\ref{realfieldaction}) is mapped to $\zeta_{re} \leftrightarrow \zeta_{im}$, $k \rightarrow -k$, $\gamma_1 \leftrightarrow \gamma_4$ and $\gamma_2 \leftrightarrow \gamma_3$ in the edge-mode theory. In terms of $\zeta$ and $\bar{\zeta}$, the action is 
\begin{multline} \label{Generaledgemodetheory}
    S_{\text{edge}} = \frac{1}{16\pi} \int dt d\phi \, (\lambda \gamma_3 - k \gamma_1 ) (\partial_\phi \zeta + \partial_\phi\bar{\zeta})^2 - ( k\gamma_4 + \lambda \gamma_2) (\partial_\phi \zeta - \partial_\phi \bar{\zeta} )^2 \\
    - 2i (k\gamma_3  + \lambda \gamma_1 ) ((\partial_\phi \zeta)^2 - ( \partial_\phi \bar{\zeta})^2) + 2( k + i\lambda ) \partial_t \zeta \partial_\phi \zeta + 2( k - i \lambda ) \partial_t \bar{\zeta} \partial_\phi \bar{\zeta}\, . 
\end{multline}
We can then express the fields in terms of the mode expansions
\begin{equation} \label{zetamodeexp}
    \begin{split}
        \zeta(t,\phi) &= \frac{1}{\sqrt{2\pi}} \sum_{n \in \mathbb{Z}/\{0\}} \alpha_n (t) e^{i n \phi} \, , \\
        \bar{\zeta}(t,\phi) &= \frac{1}{\sqrt{2\pi}} \sum_{n \in \mathbb{Z}/\{0\}} \bar{\alpha}_n (t) e^{- i n \phi} \,,
    \end{split}
\end{equation}
where $\alpha_n = a_n + i b_n$, with $a_n$ and $b_n$ real-valued. The Hamiltonian is then
\begin{multline}
        \hat{H} = \frac{1}{16 \pi}\sum_{n \in \mathbb{Z}/\{0\}} n^2 \bigg[ (k \gamma_1 - \lambda \gamma_3 ) (\alpha_n \alpha_{-n} + 2\alpha_n \bar{\alpha}_n + \bar{\alpha}_n \bar{\alpha}_{-n}) \\
        + (k \gamma_4 + \lambda \gamma_2 )(\alpha_n \alpha_{-n} - 2 \alpha_n \bar{\alpha}_n + \bar{\alpha}_n \bar{\alpha}_{-n}) + 2 i (k \gamma_3 + \lambda \gamma_1 ) ( \alpha_n \alpha_{-n} - \bar{\alpha}_n \bar{\alpha}_{-n}) \bigg] \,.
\end{multline}
Since the kinetic term in \eqref{Generaledgemodetheory} is unchanged by the new boundary conditions, we can use the same coordinate transformation as in \eqref{ABrels} in the previous chapter to bring the Hamiltonian to the form 
\begin{multline} \label{HamiltonianAB}
    \hat{H} = - \frac{\sqrt{k^2 + \lambda^2}}{32 \pi \lambda} \sum_{n \in \mathbb{Z}/\{0\}} n^2 \bigg[ (\gamma_2 + \gamma_3 ) \sqrt{k^2 + \lambda^2} ( \hat{A}_n^+ \hat{B}_n^+ + \hat{A}_n^- \hat{B}_n^- + \hat{A}_n^+ \hat{A}_n^- + \hat{B}_n^+\hat{B}_n^- ) \\
     + ( k(\gamma_2 + \gamma_3 ) + 2 \lambda \gamma_1 ) ( \hat{B}_n^+ \hat{B}_n^- - \hat{A}_n^+ \hat{A}_n^- ) + i (\gamma_2 - \gamma_3) \lambda ( \hat{A}_n^+ \hat{B}_n^+ - \hat{A}_n^- \hat{B}_n^- ) )\bigg] \, ,
\end{multline}
where the final term comes from a normal ordering ambiguity from using the commutators. Note here that the Hamiltonian does not contain structures like $\hat{A}_n^+ \hat{B}_n^{-}$ or $\hat{A}_n^- \hat{B}_n^{+}$. The discrete symmetry in these variables is 
\begin{equation}
    \begin{split}
        \hat{A}_n^+ &\leftrightarrow i \hat{A}_n^- \, , \\
        \hat{B}_n^+ &\leftrightarrow -i \hat{B}_n^- \, , \\
        k \rightarrow -k \, , \quad \gamma_1 &\leftrightarrow \gamma_4,  \quad \gamma_2 \leftrightarrow \gamma_3 \, .
    \end{split}
\end{equation}
If we take \eqref{alphaoriginal} for the $\gamma_i$ in equation \eqref{HamiltonianAB}, the first term drops out and this reverts back to the Hamiltonian that we considered in the previous chapter, which comprised of the remaining two terms. Therefore, we should investigate whether diagonalising these new terms can produce an energy spectrum that is bounded from below, such that if we switch off the remaining terms, we might have a sensible edge-mode theory. 

\section{Bogoliubov transformation}
We will employ the Bogoliubov transformation to diagonalise the Hamiltonian. We will begin by applying this technique, as reviewed in \cite{Bogoliubovnotes}, for our original Hamiltonian from Section \ref{complexCSL}.
\subsection{Original edge-mode theory}
First, focus on the Hamiltonian (\ref{HamiltonianAB}) with \eqref{alphaoriginal}:
\begin{equation} 
\begin{split}
    \hat{H} &=  \frac{\sqrt{k^2 + \lambda^2}}{16 \pi} \sum_{n \in \mathbb{Z}/\{0\}} n^2 \bigg[\upsilon_{re} ( \hat{A}_n^+ \hat{A}_n^- -\hat{B}_n^+ \hat{B}_n^-) + i \upsilon_{im} ( \hat{A}_n^+ \hat{B}_n^+ - \hat{A}_n^- \hat{B}_n^- ) )\bigg] \,.
\end{split}
\end{equation}
In matrix form, this reads
\begin{equation}
    \hat{H} =  \frac{\sqrt{k^2 + \lambda^2}}{16 \pi} \sum_{n \in \mathbb{Z}/\{0\}} \frac{n^2}{2}   (\hat{A}_n^+ \hat{B}_n^- \hat{A}_n^- \hat{B}_n^+) 
    \begin{pmatrix}
  \upsilon_{re}  & i \upsilon_{im}   & 0 & 0 \\
-i \upsilon_{im}  & -\upsilon_{re} & 0 & 0 \\
 0 & 0 &  \upsilon_{re} & - i \upsilon_{im}    \\
 0 & 0 &  i \upsilon_{im}  & - \upsilon_{re}  \\
\end{pmatrix}
    \begin{pmatrix}
    \hat{A}_n^- \\
    \hat{B}_n^+ \\
    \hat{A}_n^+\\
    \hat{B}_n^- 
    \end{pmatrix}   \, . 
\end{equation}
We will take the transformation 
\begin{equation} \label{ABtoCBtrans}
\begin{pmatrix}
    \hat{A}_n^- \\
    \hat{B}_n^+ \\
    \end{pmatrix} = 
\begin{pmatrix}
 u_1 & v_1 \\
 v_2 & u_2 \\
\end{pmatrix}
\begin{pmatrix}
    \hat{C}_n^- \\
    \hat{D}_n^+ \\
    \end{pmatrix} \, ,
\end{equation}
where $u_1, u_2, v_1 , v_2 \in \mathbb{C}$. Given the commutation relations
\begin{equation}
    [\hat{A}_n^+, \hat{A}_n^- ] = [\hat{B}_n^+, \hat{B}_n^- ] = \frac{8\pi}{n\sqrt{k^2 + \lambda^2}} \,,
\end{equation}
and imposing that 
\begin{equation}
\begin{gathered}
     \relax [ \hat{C}_n^+, \hat{C}^-_n ] = [\hat{D}_n^+, \hat{D}^-_n ] = \frac{8\pi}{n\sqrt{k^2 + \lambda^2}} \, ,  \\
     [\hat{C}_n^\pm,\hat{D}_n^\pm]=0 
\end{gathered}
\end{equation}
we have that
\begin{equation}
    \begin{split}
        |u_i |^2 - |v_i |^2 = 1 , \qquad i \in \{1,2\} \, . 
    \end{split}
\end{equation}
In order to have $[\hat{A}_n^\pm,\hat{B}_n^\pm]=0$, we also require 
\begin{equation} \label{uvcondition}
v_1u_2^*-u_1 v_2^*=0\,.
\end{equation}
We can therefore parametrise the matrix coefficients as 
\begin{equation} \label{uvparams1}
    \begin{split}
        u_i &= e^{i \theta_i} \cosh r_i \, ,\\
        v_i &= e^{i \phi_i} \sinh r_i \, ,
    \end{split}
\end{equation}
with $\{ \theta_i, \, \phi_i, \, r_i \} \in \mathbb{R}$. Then the condition \eqref{uvcondition} reads
\begin{equation}
\tanh r_1\, e^{i(\phi_1-\theta_2)}=\tanh r_2\, e^{i(\theta_1- \phi_2)}\,.
\end{equation}
Therefore, we must have
\begin{equation}
\label{constraint1}
r_1=r_2\,, \qquad \theta_1+\theta_2=\phi_1+\phi_2+2n \pi\,, \quad n\in \mathbb{Z}\,.
\end{equation}
Focusing on the top left block of the Hamiltonian,
\begin{equation}
\left( \hat{C}_n^+ , \hat{D}_n^- \right)
\begin{pmatrix}
 u_1^* & v_2^* \\
 v_1^* & u_2^* \\
\end{pmatrix}
 \begin{pmatrix}
\upsilon_{re}  & i \upsilon_{im}  \\
-i \upsilon_{im}  & -\upsilon_{re} \\
\end{pmatrix}
\begin{pmatrix}
 u_1 & v_1 \\
 v_2 & u_2 \\
\end{pmatrix}
\begin{pmatrix}
    \hat{C}_n^- \\
    \hat{D}_n^+ \\
    \end{pmatrix} \, ,
\end{equation}
and expanding out the matrices, we find that the off-diagonal terms are complex conjugates of each other. After imposing \eqref{constraint1}, in order for the transformation to produce a diagonal Hamiltonian we must have
\begin{equation}
    e^{2 i \phi_1} \sinh^2 r =  e^{2 i \theta_2} \cosh^2 r \, . 
\end{equation}
Thus,
\begin{equation}
    \phi_1 = \theta_2 , \qquad \tanh^2 r = 1 \, ,
\end{equation}
which requires $r \rightarrow \infty$. This solution is singular, and therefore our Hamiltonian represents a system at an unstable equilibrium, which is reflected in the fact that the spectrum \eqref{Loreenergies} found in the previous chapter is unbounded from below.

\subsection{Diagonalising the new term}
We may simplify our general Hamiltonian \eqref{HamiltonianAB} to just comprise of the term not seen with the previous choice of boundary conditions by choosing the coefficients 
\begin{equation}
\gamma_2=\gamma_3\equiv \gamma\,, \qquad \gamma_1=-\frac{k}{\lambda}\gamma \, . 
\end{equation}
Then in matrix form, the Hamiltonian is 
\begin{equation} \label{NewHpiece}
\begin{split}
      \hat{H} &= - \frac{(k^2 + \lambda^2) \gamma}{32 \pi \lambda} \sum_{n \in \mathbb{Z}/\{0\}} n^2  (\hat{A}_n^+ \hat{B}_n^- \hat{A}_n^- \hat{B}_n^+) 
    \begin{pmatrix}
 1  & 1  & 0 & 0 \\
 1  & 1 & 0 & 0 \\
 0 & 0 & 1 & 1  \\
 0 & 0 & 1 & 1  \\
\end{pmatrix}
    \begin{pmatrix}
    \hat{A}_n^- \\
    \hat{B}_n^+ \\
    \hat{A}_n^+\\
    \hat{B}_n^- 
    \end{pmatrix}  \, .
\end{split}
\end{equation}
We can attempt to diagonalise this term by taking the same transformation as in equation \eqref{ABtoCBtrans}, which requires us to again impose the constraints \eqref{uvparams1} and thus \eqref{constraint1}. Then by following the same steps as before, we find that the condition required for the Hamiltonian to be diagonal is 
\begin{equation}
     0 = e^{i \theta_2} \sinh r+e^{i \phi _1} \cosh r  \, .
\end{equation}
The solution of this is given by
\begin{equation}
\tanh r=-e^{i(\phi_1-\theta_2)}\,.
\end{equation}
Since $r \in \mathbb{R}$, this can only be satisfied if $\tanh r = \pm 1$, where $r \rightarrow \pm \infty$. We see that this solution is again singular, and hence does not represent a sensible Bogoliubov transformation. The Hamiltonian \eqref{NewHpiece} will also represent a system in an unstable equilibrium, and so including this term in our theory does will not result in a bounded energy spectrum. We have therefore shown that the edge-mode theory considered in Section \ref{complexCSL} is, in fact, of the most general form for our purposes, and that considering more general boundary conditions for the disc does not provide a richer edge-mode theory. 

\chapter*{Conclusion}
\addcontentsline{toc}{chapter}{\Large Conclusion}
In this thesis we have delved into various approaches for characterising quantum physics in de Sitter space, often by making use of the topological nature of low-dimensional gravity. In Part \ref{Part1} we found that in two-dimensional gravity, it is possible to have thermodynamically stable geometries that interpolate between de Sitter and anti-de Sitter regions. Specifically, for geometries that interpolate to an asymptotically AdS boundary, we postulated the existence of boundary theories whose microscopic degrees of freedom should describe the expanding region in the interior holographically. We suggested two such possible dual theories; a matrix model with a density of eigenstates that dips slightly downwards in comparison to the pure AdS energy density, and an SYK-like theory that includes relevant deformations and, at least na\"ively, may have to be non-unitary \cite{Anninos:2020cwo,Anninos:2022qgy}. We then changed tact and investigated the behaviour of correlation functions in a fixed de Sitter background, working for the most part in two dimensions, but ultimately extending our results to general spacetime dimensions. We showed that in order to reproduce the correct asymptotic form of the two-point function of a massive scalar field, it is crucial to include saddles of complex length in the geodesic approximation. This became evident after analytically continuing the relevant geodesics from the sphere and resulted in an oscillatory behaviour in the two-point function. \\
\\
In Part \ref{Part2}, we explored contributions to the entropy of the de Sitter horizon stemming from topological entanglement entropy. We studied de Sitter in three dimensions, where the Euclidean Chern-Simons path integral produces the exact, all-loop quantum corrected de Sitter entropy. There ought to be an intrinsically Lorentzian Hilbert space interpretation for this entropy as a counting of microscopic states. We proposed that this should take the form of an edge-mode theory living close to the de Sitter horizon and found that, after complexifying the contour of integration in the edge-mode theory, it was possible to reproduce the form of the entropy in an Abelian toy model. Difficulty arises for the full three-dimensional gravity theory due to the non-compactness of the gauge group. We then attempted to tame this unboundedness by exploring more general boundary conditions for the near-horizon region, but found that the theory considered was in fact of the most general type. 

\section*{Future directions}
\addcontentsline{toc}{section}{Future directions}
Having proposed dual matrix model theories to the interpolating geometries in two dimensions, in the future it would be valuable to test these proposals. Recently, much interesting work has been done on generalising the matrix model description to incorporate dilaton gravity theories away from the pure AdS$_2$ limit \cite{Witten:2020wvy, Maxfield:2020ale, Eberhardt:2023rzz}. It would therefore be fruitful to try and use such methods to pin down the precise form that the matrix model with density of states \eqref{rhocompact} takes. Another way to test this duality would be to incorporate matter in the theory and compare correlation functions of such matter in the bulk and boundary theories \cite{Jafferis:2022wez}. For correlators in the bulk theory, one can use the geodesic approximation discussed in Chapter \ref{ComplexGeodesics} to reduce the problem to one of computing geodesic lengths, and in the same way that we saw for pure de Sitter, it will likely become crucial to include saddles of complex length in order to do so. Ultimately, one would hope to isolate the de Sitter piece of the ultraviolet theory, and it may be possible to do so by taking a $T\overline{T}$-like deformation \cite{Gross:2019ach, Gross:2019uxi}.\\
\\
Furthermore, geodesics are not the only interesting extended extremal objects that are used in holography. For instance, co-dimension 2 surfaces are related to entanglement entropy \cite{Rangamani:2016dms}, and co-dimension 1 (or zero), to holographic complexity
\cite{Susskind:2018pmk,Chapman:2021jbh}. It is reasonable to expect the existence of complex surfaces in the context of de Sitter \cite{Fischetti:2014uxa}; see also \cite{Jorstad:2022mls}. However, it would be hard to interpret complex areas or volumes as measures of these naturally real quantities. As we have seen, in the case of the two-point function, the complex geodesics combine such that the final answer is real. It would be interesting to understand the more general role that complex surfaces might play in holography. It may also be interesting to study the role of complex geodesics in interacting quantum field theories in de Sitter, perhaps along the lines of the K\"{a}llen-Lehmann representation \cite{Bros:1990cu}, or in other backgrounds with positive cosmological constant, such as de Sitter black holes.\\
\\
As previously mentioned, it has been advocated that the dual theory to de Sitter might live on a stretched horizon \cite{Susskind:2021dfc, Shaghoulian:2021cef, Shaghoulian:2022fop}. The microscopic candidate theory is an SYK model in a particular double-scaled limit \cite{Susskind:2021esx, Lin:2022nss, Susskind:2022bia, Rahman:2022jsf}. We computed correlation functions of bulk scalar fields anchored at opposite stretched horizons in Section \ref{lorentzian_props}. The conclusion is that real geodesics exist only up to some critical time $t_c$, that depends on the position of the stretched horizon. After this time, the correlator starts exhibiting oscillations, and the geodesic length becomes complex. Using holography, we would expect the boundary theory to know about this time scale. It looks hard to envision how a standard SYK model would incorporate this scale. Other proposals relating non-Hermitian SYK models to de Sitter include \cite{Anninos:2020cwo, Garcia-Garcia:2022adg, Anninos:2022qgy}. \\
\\
From the three-dimensional perspective, in future work it would be interesting to attempt to converge on the full, non-Abelian three-dimensional gravity theory from the perspective of a large-$N$ limit. Chern-Simons theory with gauge group $SU(N)$ describes higher spin gravity and, upon taking a 't Hooft limit with $N \rightarrow \infty$ and large level, admits a description in terms of a topological string theory \cite{Gopakumar:1998ki}. Presumably, this must be related to a corresponding Lorentzian picture by way of a large-$N$ Wess-Zumino-Witten model with gauge group $SU(N)_k$. It has been suggested that this theory reduces to one consisting of free fermions \cite{Kiritsis:2010xc}. Therefore, one hopes this will provide another simplified environment in which to study contributions to the de Sitter horizon entropy stemming from topological entanglement entropy. Furthermore, given that JT gravity can be rewritten in the first order formalism as a BF theory, which in turn can be found as a dimensional reduction from Chern-Simons theory in three dimensions, it may be worthwhile to explore combining the $2$D and $3$D perspectives described in this work. One can ask if there is a notion of topological entanglement entropy in $2$D gravity \cite{Joung:2023doq, Mertens:2022ujr}, and therefore whether it is possible to replicate the setup explored in Part \ref{Part2} in the lower dimensional theory. It would be interesting to use such a setup to see if there might be an entanglement nature to the quantum entropy of the de Sitter horizon in two dimensions as well. \\
\\
One of the major motivations for pursing a quantum theory of de Sitter in low-dimensional gravity was the hope that our findings would elucidate some features of the more physical, four-dimensional theory. This could occur quite directly in the two-dimensional case, due to the fact that the near-horizon limit of the Nariai black hole is dS$_2 \times S^2$. Thus, if one could successfully isolate the ultraviolet theory that describes dS$_2$, then it may be possible to use this to describe the microscopic degrees of freedom of the four-dimensional black hole in de Sitter, at least in some limit. The influence of our three-dimensional results on the $4$D case may be somewhat more philosophical, in the sense that if it is possible to converge on a Lorentzian edge-mode theory that correctly reproduces the exact Gibbons-Hawking entropy, then this suggests that we should take seriously the proposal that the de Sitter horizon entropy is counting microstates. We saw in three dimensions that entanglement entropy was a key contribution to this entropy, stemming from the fact that we can rewrite gravity as a gauge theory precisely in this case. However, diffeomorphisms act as gauge symmetries in gravity in any dimensions, and so this implies that there should be some entanglement nature of the entropy in higher dimensional theories as well.

\begin{appendices}

\chapter{Thermodynamics of finite boundary geometries} \label{appAdS2}
In this appendix we consider pure JT gravity and the dilaton-gravity model arising from the dimensional reduction of four-dimensional Einstein gravity in the presence of a Dirichlet boundary. 

\section{\texorpdfstring{AdS$_2$}{AdS2} JT gravity}
The dilaton potential that gives rise to an AdS$_2$ geometry is $V(\phi) = 2 \phi$. We can see that this potential results in the Euclidean AdS$_2$ black hole metric (which is the Poincar\'e disc) by using equation (\ref{V-Areleation}):
\begin{equation}
    \frac{ds^2}{\ell^2} = (r^2 - r_h^2 ) d\tau^2 + \frac{dr^2}{(r^2 - r_h^2)}\,, \qquad r \in [r_h,\infty) \,,
\end{equation}
where $r_h>0$ and $\tau \sim \tau + \beta$. The on-shell action is 
\begin{equation}
\begin{split}
    \log Z_{\text{AdS}_2} 
    &= 2 \pi  \phi_0  +  \beta \, \phi_b^2 \,,
\end{split}
\end{equation}
with the Euclidean time periodicity $\beta = 2\pi/r_h$. The Tolman temperature is
\begin{equation}
    \beta_T  = \frac{2\pi}{\phi_h} \sqrt{r_b^2 - \phi_h^2}\,,
\end{equation}
where, as previously, $\phi_h = r_h$. One thus finds \cite{Lemos:1996bq}
\begin{equation}
	 - \beta_T F_{\text{AdS}_2} = 2 \pi \phi_0 + \phi_b \sqrt{4 \pi^2 + \beta_T^2}\, . 
\end{equation}
Using the expressions (\ref{energy}), (\ref{entropy}), and (\ref{heatcapacity}), we find the following thermodynamic quantities
\begin{equation}
    E_{\text{AdS}_2}  = - \sqrt{\phi_b^2 - \phi_h^2}\,, \quad\quad S_{\text{AdS}_2}  = 2 \pi \phi_0 + 2 \pi \phi_h\,, \quad\quad C_{\text{AdS}_2}  = \frac{4 \pi^2   \phi_b \, \beta_T^2}{\left(\beta_T^2+4 \pi^2\right)^{3/2}}\,.
\end{equation}
We note that $C_{\text{AdS}_2}$ is manifestly positive since $\phi_b > 0$. If we take $\phi_b \gg 1$, we find the leading order expressions
\begin{equation}
    E_{\infty}  = -\phi_b + \frac{2\pi^2}{\phi_b\beta^2}\,, \quad\quad S_{\infty}  = 2 \pi \phi_0 + \frac{4 \pi^2}{\beta}\,, \quad\quad C_{\infty}  = \frac{4\pi^2}{\beta}\,,
\end{equation}
in agreement with the known expressions (see, for example,  \cite{Witten:2020ert}) for asymptotically near-AdS$_2$, up to a physically inconsequential shift in $E_{\infty}$. We note that in this limit thermodynamical variations are with respect to $\beta$ which is defined entirely at the horizon. 

\section{Schwarzschild dilaton gravity}
The dilaton potential that appears in the dimensional reduction of the four-dimensional Schwarzschild solution is \cite{Cavaglia:1998xj}
\begin{equation}
    V(\phi) = \frac{1}{2 \sqrt{\phi}} \, , \qquad \phi \geq 0 \,. 
\end{equation}
Using (\ref{V-Areleation}) we find the two-dimensional Euclidean metric
\begin{equation}
    \frac{ds^2}{\ell^2} = (\sqrt{r} - \sqrt{r_h}) \, d\tau^2 + \frac{dr^2}{(\sqrt{r} - \sqrt{r_h})}\,. 
\end{equation}
The on-shell Euclidean action yields
\begin{equation}
\begin{split}
    \log Z_{\text{flat}} &= 2 \pi \phi_0 + \beta \left( \sqrt{\phi_b} -\frac{3}{4} \sqrt{\phi_h}  \right) \,, 
\end{split}
\end{equation}
where $\beta = 8 \pi \sqrt{r_h}$ is the periodicity in Euclidean time and $\phi_h = r_h$. The Tolman temperature is 
\begin{equation}
    \beta_T  =  8 \pi \sqrt{\phi_h} \sqrt{  \sqrt{r_b} - \sqrt{\phi_h}} \,.
\end{equation}
The thermodynamic properties of Schwarzschild in two dimensions with a finite boundary are found to be
\begin{equation}\label{SchwHC}
    E_{\text{flat}} = - \sqrt{  \sqrt{\phi_b} - \sqrt{\phi_h}}\,, \quad S_{\text{flat}} = 2 \pi \left( \phi_0 +  \phi_h \right)\,, \quad C_{\text{flat}} = \frac{8 \pi \phi_h ( \sqrt{\phi_b} - \sqrt{\phi_h})}{3 \sqrt{\phi_h} - 2 \sqrt{\phi_b}}\,.
\end{equation}
Here we see further evidence of a finite boundary resulting in a transition from positive to negative heat capacity. Since $\phi_b > \phi_h >0$, the numerator of $C_{\text{flat}}$ in (\ref{SchwHC}) is always positive, but the sign of the denominator depends on the location of the wall with respect to $\phi_h$. This is the phenomenon observed by York for Schwarzschild black holes in a Dirichlet box \cite{Hawking:1982dh, York:1986it}. 

\chapter{Double interpolating geometry with finite boundary} \label{Double Interpolating Geometry finite}
In this appendix, we consider the Dirichlet problem,  with  $\beta_T$ and $\phi_b$ fixed, for the dilaton-gravity theory with potential (\ref{SandPot}).
\newline\newline 
For $\phi_b>0$, the temperatures of each saddle are 
\begin{eqnarray}
	\beta_T^1 &=& \frac{4\pi}{|2 \phi_1 + \tilde{\phi} - 4x|} \sqrt{r_b^2 + \tilde{\phi} \, r_b - 2x^2 + \phi_1 (4x - \phi_1 - \tilde{\phi})} \,, \\
	\beta_T^2 &=& \frac{4\pi}{|\tilde{\phi} - 2 \phi_2|} \sqrt{r_b^2 + \phi_2^2 + \tilde{\phi}(r_b - \phi_2)} \,, \\
	\beta_T^3 &=& \frac{4\pi}{|\tilde{\phi} + 2 \phi_3|} \sqrt{r_b^2 - \phi_3^2 + \tilde{\phi}(r_b - \phi_3)} \,.
\end{eqnarray}
\subsection*{Case 1: $\tilde{\phi} \ge 0$}
We require $\phi_b > 0$ in order to have a geometry containing the full, double interpolating solution (\ref{Doubleintepsaddle1}). The free energy of the first saddle at $\phi_1$ is 
\begin{equation} \label{F1}
	-\beta_T F_{1} =  2 \pi \left( \phi_0 + 2 x  - \frac{\tilde{\phi}}{2} \right) + \frac{1}{2} \sqrt{(\beta_T^2 + 4 \pi^2 )\left((2 \phi_b + \tilde{\phi} )^2 + 8 x^2 - 8 x \tilde{\phi}\right) }  \,.
\end{equation}
The properties of the second and third saddles have been described in Section \ref{Interpolating Thermodynamic properties}, with their free energies given by (\ref{Finterp1}) and (\ref{FA}) respectively. From equation (\ref{heatcapacity}), the heat capacity of the double interpolating saddle is
\begin{equation} \label{C1}
	C_1 = \frac{2 \pi^2 \beta _T^2 \sqrt{(2 \phi_b + \tilde{\phi})^2+8 x^2 -8 x \tilde{\phi} }}{(\beta_T^2 + 4 \pi^2)^{3/2}}> 0\,,
\end{equation}
and hence the saddle is stable, similarly to the case with an asymptotic AdS$_2$ boundary. The heat capacity of the second saddle $C_2$ will be either (\ref{CInterpunstable}) or (\ref{CInterpstable}) depending on the ranges of $\beta_T$ and $\phi_b$ as described in Section \ref{Interpolating Thermodynamic properties}. The third saddle $C_3$ will have specific heat (\ref{CA}). \\
\\
First take the case $ \tfrac{4\pi}{\tilde{\phi}} \sqrt{\phi_b( \phi_b  + \tilde{\phi})} \ge \beta_T \ge 2\pi$ and $\phi_b \ge \chi_-$ such that the second saddle has heat capacity (\ref{CInterpunstable}) and is unstable. Then comparing the free energies of the first and third saddle we find
\begin{multline} \label{Fdiff31}
	\Delta F = F_{3} - F_{1}\\
	= \frac{1}{2\beta_T} \left(8 \pi x - \sqrt{\beta_T^2 + 4 \pi^2} \left(2 \phi_b + \tilde{\phi}\right)+  \sqrt{\left(\beta_T^2+ 4 \pi^2\right) \left((2 \phi_b + \tilde{\phi} )^2+8 x^2 - 8 x \tilde{\phi} \right)}\right) \,.
\end{multline}
This can either be positive or negative depending on the range of $x$. The difference is positive if 
\begin{equation} \label{xcond}
	x < \frac{1}{\beta_T^2 - 4 \pi^2} \left( (\beta_T^2 +4 \pi^2)\tilde{\phi} - 2 \pi \sqrt{(\beta_T^2 + 4 \pi^2)(4 \phi_b^2 + 4 \phi_b \tilde{\phi} + \tilde{\phi}^2)} \right)\,, 
\end{equation}
and negative otherwise. For example, take $\tilde{\phi}= 1, \, \phi_b = 1$ and $\beta_T = 4\pi$ to satisfy the conditions on $\phi_b$ and $\beta_T$. Then if $x< \tfrac{5}{3} - \sqrt{5}$ then (\ref{Fdiff31}) will be positive and otherwise it will be negative. Thus, for certain temperatures, the double interpolating saddle will be dominant.\\
\\
When instead we take $ \tfrac{4\pi}{\tilde{\phi}} \sqrt{\phi_b( \phi_b  + \tilde{\phi})} < \beta_T < 2\pi$ and $\phi_b < \chi_-$, we have a stable saddle at $\phi_2$ and no longer have a saddle at $\phi_3$. In this case the difference in free energies is 
\begin{multline}  \label{Fdiff21}
	\Delta F = F_{2} - F_{1} = \frac{1}{2 \beta_T} \left(4\pi (2x - \tilde{\phi}) + \right. \\ \left.
	  \sqrt{(\beta_T^2 + 4 \pi^2 )(8 x^2 - 8 x \tilde{\phi} + (2 \phi_b + \tilde{\phi})^2)} + \sqrt{(4\pi^2 - \beta_T^2)(\tilde{\phi}^2 - 4 \phi_b \tilde{\phi} - 4 \phi_b^2)}\right) \,,
\end{multline}
which is negative for this range of $\beta_T$. Therefore, the interpolating saddle will dominate over the double interpolating one.

\subsection*{Case 2: $\tilde{\phi} < 0$}
In this case, the free energy and heat capacity will still be given by (\ref{F1}) and (\ref{C1}) respectively. To ensure that these expressions are real, we must have $ (2 \phi_b + \tilde{\phi})^2 > 8x ( \tilde{\phi}- x)$. As in the case $\tilde{\phi} < 0$ in Section \ref{Interpolating Thermodynamic properties}, the change in sign of $\tilde{\phi}$ restricts the possible ranges of $\beta_T$. Again, the first case is when $\sqrt{2}\pi \left|\tfrac{2 \phi_b + \tilde{\phi}}{\tilde{\phi}}\right| \ge \beta_T \ge 2 \pi$ and $\phi_b \ge - \chi_+$ where the $\phi_2$ saddle is again unstable.  The difference between the free energies of the third and first saddle is (\ref{Fdiff31}). For $\tilde{\phi} <0$ this can again be either positive if the condition (\ref{xcond}) is satisfied, or negative otherwise.\\
\\
If instead, $\sqrt{2}\pi \left|\tfrac{2 \phi_b + \tilde{\phi}}{\tilde{\phi}}\right| < \beta_T < 2 \pi$ and $|\tilde{\phi}|/2< \phi_b < - \chi_+$, then all three saddles at $\phi_1$, $\phi_2$ and $\phi_3$ are stable. As seen in equation (\ref{FdiffInterp}), in this case the AdS$_2$ saddle at $\phi_3$ is thermodynamically favoured over the interpolating one at $\phi_2$. The difference in the free energies between $\phi_3$ and $\phi_1$ is again given by (\ref{Fdiff31}), which in this case will be negative, and so the AdS$_2$ saddle at $\phi_3$ is favoured. However, if $0 < \phi_b < |\tilde{\phi}|/2$ then there is no $\phi_3$ saddle and the  difference between the $\phi_1$ and $\phi_2$ saddles is given by (\ref{Fdiff21}), which again is negative under the assumptions we have here. Therefore, the interpolating saddle will again dominate over the double interpolating one.


\chapter{WKB approximation}
\label{app_WKB}
The Klein-Gordon equation as a function of $P\equiv P_{X,Y}$ can be solved systematically, order by order in the large mass expansion, through a WKB approach. A similar procedure is followed in Appendix B of \cite{Balasubramanian:2019stt} for the case of timelike geodesics in AdS.\\
\\
To start, we recall here equation (\ref{kg_P}),
\begin{equation}
\label{eq:exa}
(1-P^2) \partial_{P}^2 G(P) - d P \partial_{P} G(P) - m^2 \ell^2 G(P) = 0 \,.
\end{equation}
We take the ansatz,
\begin{equation}
G(P) \propto \exp \left( -m \ell X(P)  \right) \,,
\end{equation}
where $X$ is assumed to be independent of $m$. At leading order in the large mass limit, it satisfies the equation,
\begin{equation}
\left(P^2-1\right) X'(P)^2+1=0 \,.
\end{equation}
This equation is solved by
\begin{eqnarray}
\label{eq:wkbsol1}
X_\pm^{(1)}(P) &=& c_\pm^{(1)} \pm \arccos P \,,  \qquad \textrm{for} \, P<-1 \, , \\
\label{eq:wkbsol2}
X_\pm^{(2)}(P) &=& c_\pm^{(2)} \pm \arccos P \,,  \qquad \textrm{for} \, -1<P<1  \,, \\
\label{eq:wkbsol3}
X_\pm^{(3)}(P) &=& c_\pm^{(3)} \pm \arccos P \,,  \qquad \textrm{for} \,  P>1  \,.
\end{eqnarray}
The six constants $c_\pm^{(i)}$  will be fixed later. To next order, we add functions $A_\pm^{(i)}(P)$, such that now,
\begin{equation}
G^{(j)}(P) = m^\alpha  \sum_{i=\pm}A_i^{(j)}(P) \exp \left( -m \ell \, X_i^{(j)}(P)  \right) \,,
\end{equation}
where $\alpha$ is an arbitrary constant, and we again assume that the functions $A_\pm^{(i)}(P)$ are independent of $m$. Plugging this back in the Klein-Gordon equation, we obtain
\begin{equation}
(d-1) P A_i^{(j)} +2 \left(P^2-1\right) \partial_{P} A_i^{(j)}  = 0 \,,
\end{equation}
which is solved by
\begin{equation}
A_i^{(j)}(P) = a_i^{(j)}(1-P^2)^{\frac{1-d}{4}}\, ,
\end{equation}
where $a_i^{(j)}$ are constant factors.  Although $a_i^{(j)}$ and $c_\pm^{(i)}$ are a priori not independent, they represent multiplicative factors with different powers of $m$, and thus they have to be treated separately. The general solution we obtain for the propagator at this order is given by
\begin{equation}
G^{(j)}(P) = \frac{m^\alpha}{(1-P^2)^{\frac{d-1}{4}}} \left[ a^{(j)}_+ e^{-m  \ell (c^{(j)}_+ + \arccos P)} +   a^{(j)}_- e^{-m \ell (c^{(j)}_- - \arccos P)} \right] \,.
\end{equation}
The solutions (\ref{eq:wkbsol1})-(\ref{eq:wkbsol3}) break down near $P=\pm 1$ when $1-P^2 \sim (m\ell)^{-2}$. 
In order to connect the solutions in the three disjoint domains, we need to solve (\ref{eq:exa}) exactly near $P=\pm 1$ and then match the solution to the adjacent WKB solutions. \\
\\
Let us start with $P=-1$. In the vicinity of this point, (\ref{eq:exa}) simplifies to
\begin{equation}
2(P+1) \partial_{P}^2 G(P) + d \partial_{P} G(P) - m^2 \ell^2 G(P) = 0 \,,
\end{equation}
which is solved by
\begin{equation}
  \label{eq:pm1ex}
  G(P) = (P+1)^{{\frac{2-d}{4}}} \left[ a I_{\frac{d}{2}-1}\left(m \ell \sqrt{2P+2}\right)+
  b K_{\frac{d}{2}-1}\left(m \ell \sqrt{2P+2}\right) \right] \, ,
\end{equation}
where $a$ and $b$ are two constants and $I$ and $K$ are modified Bessel functions. In order to recover (\ref{t_zero_d}), we need to set
\begin{equation}
  a=2^{\frac{1}{2} - \frac{3d}{4}} \left( \frac{\pi \ell}{m}\right)^{1-\frac{d}{2}} e^{-m \ell \pi} \, , \qquad b=0 \, .
\end{equation}
This completely fixes the form of the correlator. We now proceed to match this locally exact solution to the WKB solutions in (\ref{eq:wkbsol1})-(\ref{eq:wkbsol2}). For $P<-1$ taking the large-$m$ limit of (\ref{eq:pm1ex}) gives
\begin{equation}
  \label{eq:lm1}
  G(P) = -2^{\frac{3}{4}(1-d)} m^{\frac{d-3}{2}} (\pi \ell)^{\frac{1-d}{2}}  e^{-m \ell \pi} (-1-P)^{\frac{1 - d}{4}} \sin\left( m \ell \sqrt{-2-2P}- \frac{\pi (d+1)}{4}\right) \,.
\end{equation}
Using $\arccos P \approx \pi - i \sqrt{2}\sqrt{-1-P}$, 
(\ref{eq:wkbsol1}) can be matched to (\ref{eq:lm1}). This fixes the constants to be
\begin{equation}
  \nonumber
    \alpha = \frac{d-3}{2} \, , \quad a_-^{(1)} = i^{d-1} 2^{-\frac{d+1}{2}} (\pi \ell)^{\frac{1-d}{2}} \, , \quad  a_+^{(1)} = 2^{-\frac{d+1}{2}} (\pi \ell)^{\frac{1-d}{2}}  \, , \quad c_-^{(1)}= 2\pi\, , \quad c_+^{(1)} = 0  \, .
\end{equation}
Plugging these back into (\ref{eq:wkbsol1}), we obtain the simple expression
\begin{equation}
  G^{(1)}(P) = m^{\frac{d-3}{2}}  (2\pi \ell )^{\frac{1-d}{2}} \textrm{Re}\left[ \frac{e^{-m \ell\arccos P }}{(1-P^2)^{\frac{d-1}{4}}} \right]  \, , \quad \textrm{valid for} \ P<-1 \,.
\end{equation}
This expression recovers (\ref{P_less}), which has been obtained from a large-$m$ limit of the exact (hypergeometric) solution to (\ref{eq:exa}). Similarly, for $P>-1$ taking the large-$m$ limit of (\ref{eq:pm1ex}) gives
\begin{equation}
  \label{eq:lm2}
  G(P) =  2^{-\frac{3d+1}{4}}  m^{\frac{d-3}{2}}(\pi\ell)^{\frac{1-d}{2}} (1+P)^{\frac{1-d}{4}} e^{m \ell \sqrt{2+2P}-m\ell \pi} \, .
\end{equation}
Using $\arccos P \approx \pi - \sqrt{2}\sqrt{P+1}$, this can be matched to the solution in (\ref{eq:wkbsol2}) with the plus sign,
\begin{equation}
  \nonumber
  a_-^{(2)}=0 \, , \qquad a_+^{(2)} =    2^{-\frac{d+1}{2}} (\pi\ell)^{\frac{1-d}{2}}  \, , \qquad c_+^{(2)} = 0 \, .
\end{equation}
Note that the only difference compared to the $P<-1$ solution is that the $X_-$ component is exponentially suppressed and therefore no longer present. The WKB solution to the correlator   becomes
\begin{equation}
 \label{eq:g2sol}
  G^{(2)}(P) = \frac{1}{2} \times m^{\frac{d-3}{2}}  (2\pi\ell)^{\frac{1-d}{2}}  \frac{e^{-m \ell \arccos P }}{(1-P^2)^{\frac{d-1}{4}} }\, , \quad \textrm{valid for} \ -1<P<1  \,. 
\end{equation}
The expression recovers (\ref{P_greater}). In order to connect the solution in the middle domain to that in $P>1$, we need to investigate the $P\approx 1$ region.
Here we have the locally exact solution
\begin{equation}
  \label{eq:pp1ex}
  G(P) = (1-P)^{ \frac{2-d}{4}} \left[ \tilde a I_{\frac{d}{2}-1}\left(m\ell \sqrt{2-2P}\right) +
  \tilde b K_{\frac{d}{2}-1}\left(m \ell \sqrt{2-2P}\right) \right] \, .
\end{equation}
Matching it to (\ref{eq:g2sol}) fixes the constants
\begin{equation}
  \tilde a = 0 \, , \qquad \tilde b = 2^{\frac{1}{2}- \frac{3d}{4}} \left(\frac{\ell}{m}\right)^{1-\frac{d}{2}} \pi^{-\frac{d}{2}} \, .
\end{equation}
These coefficients can also be independently verified by matching \eqref{eq:pp1ex} to the flat space short distance singularity \eqref{P1}, and using the asymptotic expansion of the modified Bessel K function,
\begin{equation}
      K_{\frac{d}{2} - 1}\left( z\right) \sim
\begin{cases}
 -\log \left( \dfrac{z}{2}\right) + \Gamma'(1) & \qquad d = 2 \, , \\[10pt]
 \dfrac{\Gamma \left(\frac{d}{2} - 1\right)}{2} \left(\dfrac{2}{z}\right)^{\frac{d}{2} - 1} & \qquad d>2 \, ,
\end{cases}
\end{equation}
which is valid for for $0< |z| \ll \sqrt{\tfrac{d}{2}}$. The $d=2$ case reproduces \eqref{d=2flat} up to an overall additive constant. We note that matching (\ref{eq:pp1ex}) to (\ref{eq:wkbsol3}) implies that (\ref{eq:g2sol}) remains valid in the $P>1$ region,
\begin{equation}
  G^{(3)}(P) = G^{(2)}(P)  \, , \quad \textrm{valid for} \  P>1 \,. 
\end{equation}
Finally, note that in this WKB approximation the equation is solved order-by-order in $m$, and so this method validates the subtle limit of the hypergeometric function that we took in the main text.

\chapter{Quantum mechanical path integral} 
\label{path_integral_app}
In this appendix, we would like to evaluate the following (Euclidean) path integral, 
\begin{equation}
Z (\Phi_0, \Phi_N) = \int^{\delta \Theta (\Phi _N) = 0}_{\delta \Theta (\Phi_0) = 0} D \delta \Theta \exp \left( - m \int_{\Phi_0}^{\Phi_N} d\Phi L(\Phi, \delta \Theta , \delta \dot{\Theta}) \right) \,,
\end{equation}
for a generic quadratic Lagrangian of the form
\begin{equation}
L(\Phi, \delta \Theta , \delta \dot{\Theta}) = \frac{\ell}{2} \left(\delta \dot{\Theta}^2 + \alpha(\Phi) \delta \Theta^2 \right)\,,
\end{equation}
with $\alpha(\Phi)$ an arbitrary function of $\Phi$. 
This path integral is the quadratic correction to the saddle-point solution, and so we want the fluctuations to vanish at the endpoints. This is a textbook path integral that can be solved by discretising the $\Phi$ interval, see for instance \cite{Schulten2}. By definition, we would like to compute
\begin{multline}
Z (\Phi_0, \Phi_N) = \lim_{N\to \infty} \left( \frac{m \ell}{2\pi \Delta \Phi} \right)^{N/2} \\
\int d\delta \Theta_1 \cdots \int d\delta \Theta_{N-1} \exp \left( - m \ell \Delta \Phi \sum_{j=0}^{N-1} \left( \frac{ (\delta \Theta_{j+1} - \delta \Theta_j)^2}{2 \Delta \Phi^2}  + \frac{\alpha_j}{2} \delta \Theta_j^2\right) \right) \,,
\end{multline}
where $\Phi_j = \Phi_0 + j \Delta \Phi $ and $ \alpha_j = \alpha (\Phi_j)$. The exponent in this equation can be written in a quadratic form,
\begin{equation}
 \left( - m \ell \Delta \Phi \sum_{j=0}^{N-1} \left( \frac{ (\delta \Theta_{j+1} - \delta \Theta_j)^2}{2 \Delta \Phi^2}  + \frac{\alpha_j}{2} \delta \Theta_j^2\right) \right) = - \sum_{j,k=1}^{N-1} \delta \Theta_j \, a_{jk} \, \delta \Theta_k \,,
\end{equation}
where $a_{jk}$ are the matrix elements of the following $(N-1) \times (N-1)$ matrix,
\begin{equation}
\left( a_{jk} \right) =  \frac{m \ell}{2 \Delta\Phi} \begin{bmatrix} 
    2 & -1 & 0 & \dots  & 0 & 0 \\
    -1 & 2 & -1 & \dots & 0 & 0 \\
    0 & -1 & 2 & \dots & 0 & 0 \\
    \vdots & \vdots & \vdots & \ddots & \vdots & \vdots \\
      0 & 0& 0 &\dots  & 2 & -1  \\
    0 & 0& 0 &\dots  & -1 & 2 
    \end{bmatrix}
    + \frac{m \ell \Delta\Phi}{2} 
  \begin{bmatrix} 
    \alpha_1 & 0 & 0 & \dots  & 0 & 0 \\
    0 & \alpha_2 & 0 & \dots & 0 & 0 \\
    0 & 0 & \alpha_3 & \dots & 0 & 0 \\
    \vdots & \vdots & \vdots & \ddots & \vdots & \vdots \\
       0 & 0& 0 &\dots  & \alpha_{N-2} & 0  \\
    0 & 0& 0 &\dots  & 0 & \alpha_{N-1} 
    \end{bmatrix} \,.
\end{equation}
If $\det (a_{jk}) \neq 0$, then we can do the multiple Gaussian integrals to obtain, 
\begin{equation}
\begin{split}
Z (\Phi_0,  \Phi_N) &= \lim_{N\to \infty} \left( \frac{m \ell }{2\pi \Delta \Phi} \right)^{N/2} \left( \frac{\pi^{N-1}}{\det a_{jk}} \right)^{1/2} \\
&= \lim_{N\to \infty} \left( \frac{m \ell }{2 \pi} \frac{1}{\Delta \Phi \left( \frac{2 \Delta \Phi}{m \ell } \right)^{N-1} \det a_{jk} } \right)^{1/2} \,. \label{Ztheta}
\end{split}
\end{equation}
It is convenient to define the following function, 
\begin{equation}
f(\Phi_0,\Phi_N) \equiv \lim_{N \to \infty}  \left(  \Delta \Phi \left( \frac{2 \Delta \Phi}{m \ell} \right)^{N-1} \det a_{jk} \right) \,. \label{def_f}
\end{equation}
In order to take the $N \to \infty$ limit, we first define the discrete determinant,
\begin{equation}
\begin{split}
D_{N-1} &\equiv \left( \frac{2 \Delta \Phi}{m \ell} \right)^{N-1} \det a_{jk} \\
&= \begin{vmatrix} 
    2 + \Delta\Phi^2 \alpha_1 & -1 & 0 & \dots  & 0 & 0 \\
    -1 & 2+ \Delta\Phi^2 \alpha_2 & -1 & \dots & 0 & 0 \\
    0 & -1 & 2 + \Delta\Phi^2 \alpha_3 & \dots & 0 & 0 \\
    \vdots & \vdots & \vdots & \ddots & \vdots & \vdots \\
      0 & 0& 0 &\dots  & 2 + \Delta\Phi^2 \alpha_{N-2}& -1  \\
    0 & 0& 0 &\dots  & -1 & 2 + \Delta\Phi^2 \alpha_{N-1} 
    \end{vmatrix}
\,.
\end{split}
\end{equation}
Now we assume that the dimension of the matrix is variable, so $n = N-1$ can vary. We take the determinant using the last column to obtain the following recursion relation,
\begin{equation} \label{recursionrel}
D_n = \left( 2 + \Delta\Phi^2 \alpha_n \right) D_{n-1} - D_{n-2} \,.
\end{equation}
The first two determinants are given by $D_1 = 2 + \Delta\Phi^2 \alpha_1$ and $D_2 = 3 + 2 \Delta \Phi^2 (\alpha_1 + \alpha_2) + \Delta \Phi^4 \alpha_1 \alpha_2$. Using these along with (\ref{recursionrel}), we can analytically continue the discrete determinant to $n \in \mathbb{Z}_{\leq 0}$, finding, for example, $D_0 = 1$ and $D_{-1} = 0$. Rewriting the recursion relation in a more suggestive way, we obtain
\begin{equation}
\frac{D_{n+1} - 2 D_n + D_{n-1}}{\Delta \Phi^2} = \alpha_{n+1} D_n \,.
\end{equation}
But now we are interested in the continuum limit of this expression, where $\Delta\Phi \to 0$ (or $N \to \infty$), so we may interpret the above recursion relation as a second order differential equation in the variable $\Phi = N \Delta\Phi + \Phi_0$. Note that the extra factor of $\Delta \Phi$ in (\ref{def_f}) cancels since the above equation is linear in $D$. We then get,
\begin{equation}
\frac{d^2 f(\Phi_0, \Phi)}{d\Phi^2} = \alpha (\Phi) f(\Phi_0, \Phi) \,.
\end{equation}
The boundary conditions $D_0 = 1$ and $D_{-1} = 0$ give,
\begin{eqnarray}
f(\Phi_0, \Phi_0) & = & \lim_{N \to \infty} \Delta\Phi D_{-1} = 0 \,, \\
\left. \frac{df (\Phi_0, \Phi)}{d\Phi} \right|_{\Phi = \Phi_0} & = &  \lim_{N \to \infty} \Delta \Phi \frac{(D_0 - D_{-1})}{\Delta\Phi}  = 1 \,. 
\end{eqnarray}
Now that we have $f$, we can go back to (\ref{Ztheta}) to find the final result for the path integral, which reads
\begin{equation}
Z (\Phi_0, \Phi_N) = \sqrt{\frac{m\ell}{2\pi f(\Phi_0, \Phi_N)}} \,.
\end{equation}
In Chapter \ref{ComplexGeodesics} we had $\alpha(\Phi) = -1$, so we get $f(\Phi_0, \Phi_N) = \sin (\Phi_N - \Phi_0)$. This is the result we used in the main text to compute the first correction to the geodesic length, see (\ref{one_loop_path_integral}).

\chapter{Geodesic equation in higher dimensions} 
\label{app_ddim}
Here we show that the geodesic equations for the $d$-dimensional sphere are solved by $\Theta_i =0$. The metric of dS$_d$ can be written compactly as  
\begin{equation}
	\frac{ds^2}{\ell^2} = \sum_{a=1}^d \left( \prod_{m = 1}^{a-1} \cos^2 \Theta_m \right) d \Theta_a^2 \, ,
\end{equation}
where $ -\tfrac{\pi}{2} \leq \Theta_i \leq \tfrac{\pi}{2}$ for $1 \leq i < d - 1$ and $0 \leq \Theta_{d}  < 2 \pi$ with $\Theta_{d}  \equiv \Phi$. As in the two-dimensional case, we can use $\Phi$ to parameterise the geodesic so that it will follow a path $(\Theta_1(\Phi), \cdots, \Theta_{d-1} (\Phi))$, that extremises the length functional,
\begin{equation}  \label{ddimlength}
		\tilde{L}^d = \ell \int d\Phi \sqrt{\ \sum_{a=1}^{d} \left( \prod_{m = 1}^{a-1} \cos^2 \Theta_m \right) \dot{\Theta}_a^2 } \, ,
\end{equation}
where again the dot represents a derivative with respect to $\Phi \equiv \Theta_d$, so that $\dot{\Theta}_d = 1$. The Euler-Lagrange equations are
\begin{equation}
	\frac{\partial \mathcal{L}}{\partial \Theta_i} - \frac{d}{d\Phi} \left( \frac{\partial \mathcal{L}}{\partial \dot{\Theta}_i} \right) = 0 \, ,
\end{equation}
where $\mathcal{L}$ is the integrand of (\ref{ddimlength}). We find that  
\begin{equation}\label{c4c4}
	\begin{split}
		\frac{\partial \mathcal{L}}{\partial \dot{\Theta}_i} &= \frac{\dot{\Theta}_i}{\mathcal{L}} \prod_{m=1}^{i-1} \cos^2 \Theta_{m} \,,  \\
		\frac{\partial \mathcal{L}}{\partial \Theta}_i &= - \frac{\sin \Theta_i \cos \Theta_i}{\mathcal{L}} \left(\prod_{n=1}^{i -1} \cos^2 \Theta_n  \right) \left( \sum_{a=i+1}^{d} \left( \prod_{m = i +1}^{a-1} \cos^2 \Theta_m \right) \dot{\Theta}_a^2 \right) \, .
	\end{split}
\end{equation}
with $ i = 1, \ldots, d -1$.  It can be shown that 
\begin{equation}
	\frac{d}{d \Phi} \left( \frac{\partial \mathcal{L}}{\partial \dot{\Theta}_i} \right) =   \left( \frac{\ddot{\Theta}_i}{\mathcal{L}} - \frac{\dot{\Theta}_i}{\mathcal{L}^2} \frac{d \mathcal{L}}{d\Phi} -  \frac{2 \dot{\Theta}_i}{\mathcal{L}} \sum_{j=1}^{i-1} \tan \Theta_j  \dot{\Theta}_j \right) \prod_{n=1}^{i-1} \cos^2 \Theta_n 
	\, . 
\end{equation}
The equation of motion will therefore have an overall factor of $\prod_{n=1}^{i -1} \cos^2 \Theta_n$. We will choose the endpoints of the geodesics to lie at $\Theta_i (\Phi_1)= 0 $ and $\Theta_i (\Phi_2)= 0 $ for all $ i = 1, \ldots , d-1$. Therefore, although it appears that $\Theta_i = \pm \tfrac{\pi}{2}$ would solve the equation of motion for $i = 2, \ldots d-1$, this solution does not obey the boundary conditions and so we neglect it. For $i=1$ this product evaluates to $1$, and so we can focus on the equation without this overall factor. Therefore, the equation of motion is 
\begin{multline} \label{ddimeom}
 \bigg(\sin \Theta_i \cos \Theta_i  \left( \sum_{a=i+1}^{d} \left( \prod_{m = i +1}^{a-1} \cos^2 \Theta_m \right) \dot{\Theta}_a^2 \right) 
	+  \ddot{\Theta}_i - \frac{\dot{\Theta}_i}{\mathcal{L}} \frac{d \mathcal{L}}{d\Phi} -  2 \dot{\Theta}_i \sum_{j=1}^{i-1}  \tan \Theta_j  \dot\Theta_j\bigg)  = 0 \, ,
\end{multline}
where 
\begin{equation}
	\frac{d \mathcal{L}}{d\Phi} = 
	\frac{\partial \mathcal{L}}{\partial \dot{\Theta}_i} \ddot{\Theta}_i+
	\frac{\partial \mathcal{L}}{\partial \Theta}_i
	\dot{\Theta}_i  \,,
\end{equation}
with the various terms above defined in equation \eqref{c4c4}. Therefore, we can see that $\Theta_i = 0$, for $i = 1 ,\ldots d-1$ is a solution to equation (\ref{ddimeom}). 

\chapter{Details on the stretched horizon} 
\label{app:details}
\begin{figure}[H]
        \centering
        \includegraphics[height=5cm]{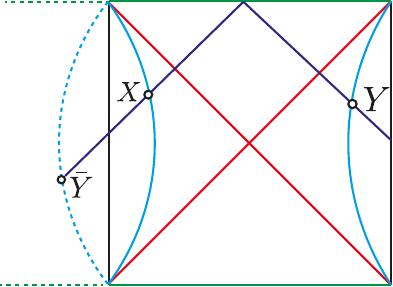}
\caption{Illustration of the null rays involved in fixing the critical static time $t_c$, after which symmetric geodesics between the two stretched horizons do not exist anymore.}\label{fig:transitiontc}
\end{figure}

To find the geodesics between complimentary stretched horizons, it is useful to relate the global \eqref{globalembedding} and static coordinates \eqref{SPembedding} in embedding space. These are related to the three-dimensional Minkowski coordinates by
\begin{eqnarray}
		X_0 =& \ell \sinh T\qquad &= \ell \sqrt{1 - r^2} \sinh t \, , \nonumber\\   
		X_1 =& \ell \cos \varphi \cosh T &= \ell r\, ,   \\
		X_2 =& \ell \sin \varphi \cosh T &= \ell \sqrt{1-r^2} \cosh t \,  .\nonumber \label{Global_Static_rels}
\end{eqnarray}
Rearranging, we can write the global coordinates in terms of the static ones as
\begin{align}
T(r,t) &= \text{arcsinh} \left(\sqrt{1-r^2} \sinh t  \right) \,, \\
\varphi(r,t)  &= \arccos \left( \frac{r}{\sqrt{\cosh^2 t - r^2\sinh^2 t }} \right) . \label{T_phiSP}
\end{align}
It is clear from this that the points opposite each other with a fixed $r = r_{st}$ will have the same $T_0$ and their angles will be at $\varphi_0$ and $- \varphi_0$. This gives the de Sitter invariant distance written in the main text,
\begin{equation} 
    P_{X,Y}^{st} = \frac{1}{\ell^2} \left[-(X_0)^2+(X_1)^2-(X_2)^2\right] = \left(r_{st}^2-1\right) \cosh 2 t+r_{st}^2 \,.
\end{equation} 
We can also recover the critical time from geometric arguments. Starting from $t=0$ and increasing in time, note that there are two final geodesics at a time $|t|= t_c$ that are almost null. After this, there are no more real geodesics. Let us compute $P_{X,Y}$ for the endpoints of the final geodesics. Recall that in $d=2$ the full Penrose diagram is two copies of the usual square. If we call $Y$ the point on the right of the Penrose diagram, it is easy to see that its antipodal point $\bar{Y}$ will be outside the Penrose diagram and will be null separated to $X$, the left endpoint of the geodesic, see Figure \ref{fig:transitiontc}. If $X$ is null separated from the antipodal point of $Y$, then the last geodesics have $P_{X,Y} = -1$. To compute $t_c$ we just need to use the metric in static coordinates (\ref{ddimStatic}), and find a null ray that passes through the point $(r=0, t=0)$. That ray intersects $r=r_{st}$ at exactly the $t_c$ given by (\ref{critical_t}). So, we have recovered this time scale from a geometric point of view.

\chapter{Chern-Simons determinant on \texorpdfstring{$S^3$}{S3}} \label{S3det}
In this appendix, we consider the determinant that stems from the quadratic part of Chern-Simons theory on $S^3$. We first recall that the canonical mass dimension of $A_\mu(x)$ is zero, as is the mass dimension for the gauge parameter $\alpha(x)$. We define the normalisation of our path integration measures to be
\begin{eqnarray}\label{norm}
1 &\equiv& \int \mathcal{D} \alpha \, e^{-\ell_{\text{uv}}^{-3}\int d^3 x \sqrt{g} \alpha(x)^2/2}\,, \\ 
1 &\equiv& \int \mathcal{D} A^T_\mu \, e^{-\ell_{\text{uv}}^{-1}\int d^3 x \sqrt{g} g^{\mu\nu} A^T_\nu(x) A^T_\mu(x)/2}\,,  \\
1 &\equiv& \int \mathcal{D} \bar{c}  \mathcal{D} c \, e^{\ell_{\text{uv}}^{-2}\int d^3 x \sqrt{g}   \bar{c}(x) {c}(x)}\,, 
\end{eqnarray}
where $\nabla^\mu A^T_\mu(x) = 0$ and $\alpha(x)$ is a real scalar, while $\bar{c}(x)$ and $c(x)$ are Grassmann valued fields. To render the exponents dimensionless, we multiply by the appropriate powers of the UV cutoff length scale $\ell_{\text{uv}} = 1/\Lambda_{\text{uv}}$. In fixing the above normalisation, we fix any ultraviolet ambiguities stemming from field rescalings. The fields can be expanded in a complete basis of eigenfunctions of the Laplacian on the round three-sphere. For instance, 
\begin{equation}
\alpha(x) = \sum_l \alpha_l \phi_l(x)\,, \quad\quad -\nabla^2 \phi_l = \xi_l \phi_l\,, \quad\quad \int d^3 x \sqrt{g} \phi_l(x) \phi_{l'}(x) = \delta_{l l'}\,,
\end{equation}
such that mass dimension of $\alpha_l$ is $-3/2$. Following similar steps for the other fields, we obtain the following path integration measures
\begin{equation}
\mathcal{D} \alpha \equiv  \prod_{l} {\Lambda_{\text{uv}}^{3/2}} \,  \frac{d\alpha_l}{{\sqrt{2\pi}}}\,, \quad \mathcal{D} \bar{c} \mathcal{D} c \equiv  {\prod}'_{l} \,{\Lambda_{\text{uv}}^{-2}} d\bar{c}_l d c_l\,,  \quad \mathcal{D} A^T_\mu \equiv \prod_l \Lambda_{\text{uv}}^{1/2} \frac{dA_l}{\sqrt{2\pi}}\,.
\end{equation}
With the above definitions of the path integration measures, and the Chern-Simons action normalised as
\begin{equation}
S_{\text{CS}} = \frac{1}{2} \int d^3x  \, \varepsilon^{\mu\nu\rho} A_\mu \partial_\nu A_\rho\,,
\end{equation} 
we can proceed via the Fadeev-Popov gauge fixing procedure. We take the metric on $S^3$ to be
\begin{equation}
ds^2 = \ell^2 \left(d\theta^2 + \sin^2\theta d\Omega_2^2 \right)\,,
\end{equation}
with volume $\text{vol}\, S^3 = 2\pi^2 \ell^3$. Working in the Lorenz gauge $\nabla_\mu A^\mu = 0$, one finds 
\begin{equation}\label{fulldet}
|Z_{U(1)_k}[S^3] | = \sqrt{\frac{2\pi}{k}}  \times \frac{\sqrt{2\pi}(\text{vol} \, S^3 \Lambda_{\text{uv}}^3)^{-1/2}}{\text{vol} \, U(1)} \times \sqrt{\frac{{\det}'[ -\nabla^2 / \Lambda_{\text{uv}}^2]}{{\det[ L L^\dag /\Lambda_{\text{uv}}^2]}^{1/2}}}\,,
\end{equation}
where $\text{vol} \, U(1) = 2\pi$. The term $\sqrt{2\pi}(\text{vol} \, S^3 \Lambda_{\text{uv}}^3)^{-1/2}/\text{vol} \, U(1)$ follows from the zero-mode contribution $d\alpha_0$. It originates from the residual part of the gauge group volume that is not cancelled by the Lorenz gauge.  \\
\\
After the dust settles, we are left to evaluate the following ratio of functional determinants
\begin{equation}
r^2 =  \frac{{\det}'\left(-\nabla^2 / \Lambda_{\text{uv}}^2\right)}{\sqrt{\det L L^\dag /\Lambda_{\text{uv}}^2}}\,,
\end{equation}
where the Laplacian $-\nabla^2$ acts on scalar functions on $S^3$ and $L L^\dag$ acts on transverse vector fields on $S^3$. The prime indicates we are dropping the zero mode of $-\nabla^2$. The respective spectra are well known \cite{Rubin:1984tc}. For the scalar Laplacian we have
\begin{equation}
\lambda_n = \frac{1}{\ell^2}\times n(n+2)\,, \quad\quad d_n = (n+1)^2\,, \quad\quad n = 0,1,\ldots
\end{equation}
For the $L L^\dag$ operator, the eigenvalues and degeneracies are given by
\begin{equation}
\lambda_n = \frac{1}{\ell^2}\times (n+1)^2\,, \quad\quad d_n = 2n(n+2)\,, \quad\quad n = 1,2,\ldots
\end{equation}
Using a heat kernel regularisation,\footnote{For a $\zeta$-function regularisation scheme, see \cite{Fliss:2017wop}.} we have that
\begin{equation}
-\log {\det}'\left(-\frac{\nabla^2}{\Lambda_{\text{uv}}^2}\right) = \sum_{n=1}^\infty (n+1)^2 \, \int_{0}^\infty \frac{d\tau}{\tau} e^{-\frac{\varepsilon^2}{4\tau}}  e^{-\tau n(n+2)}\,,
\end{equation}
where the dimensionless cutoff parameter is taken to be $\varepsilon \equiv 2 e^{-\gamma}/\ell\Lambda_{\text{uv}}$ with $\gamma$ being the Euler constant. It is convenient to rewrite the above in the following form
\begin{equation}
-\log {\det}'\left(-\frac{\nabla^2}{\Lambda_{\text{uv}}^2}\right) = \int_{\mathcal{C}} \frac{du}{\sqrt{u^2 + \varepsilon^2}} e^{-i \sqrt{u^2+\varepsilon^2}} \left( \frac{1+e^{i u}}{1-e^{i u}} \frac{e^{i u}}{(1-e^{i u})^2} - e^{i u} \right)\,.
\end{equation}
The contour $\mathcal{C} = \mathbb{R}+i\delta$ is parallel to the real axis but has a small positive imaginary part $0<\delta < \varepsilon$. We can deform the contour to go down the branch cut from the left and up from the right. We find
\begin{equation}
-\log {\det}'\left(-\frac{\nabla^2}{\Lambda_{\text{uv}}^2}\right) = \int_{\varepsilon}^\infty \frac{dt}{\sqrt{t^2 - \varepsilon^2}} \left( \frac{1+e^{-t}}{1-e^{-t}} \frac{e^{-t}}{(1-e^{-t})^2} - e^{-t} \right) \left( e^{\sqrt{t^2-\varepsilon^2}} + e^{-\sqrt{t^2-\varepsilon^2}} \right)\,.
\end{equation}
Similar considerations lead to 
\begin{equation}
-\frac{1}{2} \log {\det} \, L L^\dag = 2 \int_{\varepsilon}^\infty \frac{dt}{\sqrt{t^2 -\varepsilon^2}} \frac{3 e^{-2 t} -e^{-3 t} }{\left(1-e^{-t}\right)^3}\,.
\end{equation}
The ultraviolet behaviour is encoded in the small-$t$ regime of the integrands. Setting $\varepsilon = 0$ in the integrand and performing a small-$t$ expansion we find that there is no $\sim 1/\varepsilon^3$ divergence, and moreover that the leading divergence goes as $\sim 1/\varepsilon$. There is also a logarithmic $\sim \log \varepsilon$ divergence. Since we are in odd dimensions there should be no logarithmic divergences contributing to $\log Z[S^3]$. Recalling the expression (\ref{fulldet}) we see that this is indeed the case once we take into account the overall factor $\ell_{\text{uv}}^{-3/2}$. Evaluating the $t$-integral in the heat kernel regularisation in the small-$\varepsilon$ limit leads to 
\begin{equation}
\log r^2 = - \frac{3 \pi }{2 \varepsilon } -3 \log \frac{e^{\gamma} \varepsilon}{2} +\log 2 \pi ^2 = - \frac{3 \pi }{2 \varepsilon } + 3 \log \ell\Lambda_{\text{uv}} +\log 2 \pi ^2 \,.
\end{equation} 
Combining the above with (\ref{fulldet}) and recalling that $\text{vol}\, S^3 = 2\pi^2 \ell^3$, we find
\begin{equation}\label{finaldet}
|Z_{U(1)_k}(S^3)| = \sqrt{\frac{1}{k}} e^{-\frac{3\pi}{4\varepsilon}}\,.
\end{equation}
A similar picture holds for Chern-Simons theories with more general gauge group in the perturbative, large $k$ regime. 
The coefficient of the $1/\varepsilon$ term in (\ref{finaldet}) can be tuned by adding a local background dependent term
\begin{equation}
S_{\text{b}} = \Lambda_b \int d^3 x \sqrt{g} R = 12\pi^2 \ell \Lambda_b\,,
\end{equation}
where we recall that $R = 6/\ell^2$. For the particular choice $\Lambda_b =  -{e^{\gamma} \Lambda_{\text{uv}} }/{32 \pi }$ we can set the divergent term to zero. \\
\\
As a final remark, we note that the choices of normalisation (\ref{norm}) serve another physical purpose, namely to set the partition function of the theory on $S^1 \times S^2$ equal to unity (up to $1/\varepsilon$ divergences, which can be absorbed into local counterterms). This agrees well with the fact that Chern Simons theory quantised on a spatial $S^2$ has a unique state with vanishing energy. To compute the partition function in this case, one must take into account the non-trivial moduli space of flat connections due to the presence of a non-contractible cycle \cite{Rozansky:1993rt}. The volume of this moduli space is crucial to cancel the $k$ dependence of the partition function. 

\chapter{Euclidean path-integral of Abelian edge-mode theory} \label{Zthermal}
Here we provide a derivation of the thermal partition function (\ref{zedge}) from the Euclidean path integral. We must Wick rotate the edge-mode theory (\ref{edge}) to Euclidean signature by taking $t = -i \xi$ and imposing the thermal periodicity condition $\xi \sim \xi + \beta_T$. Notice that although the original Chern-Simons theory, being topological, is insensitive to the signature of spacetime, the edge-mode theory can sense it. The reason for this is that from the gauge theory perspective we should also continue $A_t = i A_\xi$ when continuing $t = -i \xi$. This would violate the reality conditions of the gauge-fixing condition (\ref{gauge}) unless we also continue $\upsilon = - i \upsilon_E$. If we also analytically continue $\upsilon$, the edge-mode theory would remain unchanged, and hence unaware of the underlying signature. For the purposes of computing the thermal partition function, we keep $\upsilon$ unchanged upon continuation to Euclidean signature.\\
\\
It is convenient to expand the non-winding sector in a Fourier basis
\begin{equation}
\zeta(\xi,\varphi) = \frac{1}{\sqrt{2\pi}} \sum_{(m,n) \in \mathbb{Z}^2} e^{2\pi m i \xi/\beta_T +  i n \varphi} \zeta_{m,n}\,,
\end{equation}
with $\bar{\zeta}_{m,n} = \zeta_{-m,-n}$. Performing the Gaussian path integral over the non-winding sector, we are led to evaluate
\begin{equation}
Z_{\text{non-winding}}[\beta_T] = \mathcal{N} \prod_{m, n=1}^\infty \frac{1}{1 + (\beta_T\upsilon n/2\pi m)^2}\,.
\end{equation}
We will be rather cavalier about the overall normalisation $\mathcal{N}$ as we can determine it by imposing that there is a unique vacuum state and hence $\lim_{\beta_T\to\infty} \left(1- \beta_T\partial_{\beta_T} \right) \log Z_{\text{non-winding}}[\beta_T] = 0$. Using the infinite product representation of $\sinh z$, we can express the above as
\begin{equation}
Z_{\text{non-winding}}[\beta_T] = e^{-\beta_T\upsilon \epsilon_0} \times e^{ \beta_T \upsilon/24} \prod_{n=1}^\infty \frac{1}{1- e^{-\beta_T\upsilon n}} = \frac{e^{-\beta_T \upsilon \epsilon_0}}{\eta(q)}\,,
\end{equation}
where we have absorbed any divergences of the overall energy scale into $\epsilon_0$ and $q = e^{-\upsilon\beta_T}$. If we further incorporate the winding mode sector wrapping the $\varphi$ cycle, we get an additional contribution given by 
\begin{equation}
Z_{\text{winding}}[\beta_T] = \sum_{n\in\mathbb{Z}} e^{-k \upsilon \beta_T n^2/2} = \vartheta_3(0,q^{k/2})\,.
\end{equation}
Combining the two, we find agreement with (\ref{zedge}).

\chapter{Lorentzian edge-mode Hamiltonian}
\label{LedgeApp}
\begin{figure}[H]
	\centering
	\includegraphics[width=0.5\linewidth]{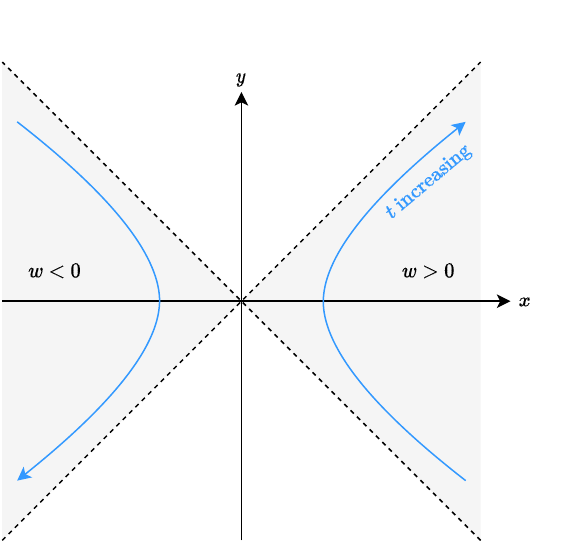}
	\caption{Plot showing the shaded regions $|x| > |y|$, where the Hamiltonian (\ref{xgryH}) is valid. The blue arrows indicate increasing $t = \text{arctanh}(y/x)$. By gluing this solution along $|x| = |y|$ with the one in the $|y|>|x|$ region, we arrive at the solution (\ref{Loreenergies}).}
	\label{fig:xygraphSdS}
\end{figure}

In this appendix, we show how the Lorentzian edge-mode Hamiltonian reduces to that of a conformal quantum mechanics. We start with the Hamiltonian in the form (\ref{LedgeH}). Since the modes carrying fixed momentum $n$ on the circle decouple, it is sufficient to study the Hamiltonian of a single mode number $n$: 
\begin{equation}
\hat{\mathcal{Q}}^{(n)}_t = \frac{\sqrt{k^2+\lambda^2}}{8\pi} n^2 \left( \upsilon_{re} \left( \hat{A}_n^+ \hat{A}_n^- - \hat{B}_n^+ \hat{B}_n^- \right) + i  \upsilon_{im} \left( \hat{A}_n^+ \hat{B}_n^+ - \hat{A}_n^- \hat{B}_n^-  \right) \right)\,.
\end{equation}
where $\hat{\mathcal{Q}}_t  = \tfrac{1}{2} \sum_n \hat{\mathcal{Q}}^{(n)}_t$. We represent the raising and lowering operators as 
\begin{equation}
\hat{A}^\pm_n = \left( \frac{4\pi}{n\sqrt{k^2 + \lambda^2}}  \right)^{1/2} \left( x_n \pm \frac{d}{dx_n} \right)\,, \quad \hat{B}^\pm_n =  \left( \frac{4\pi}{n\sqrt{k^2 + \lambda^2}}  \right)^{1/2} \left( y_n \pm \frac{d}{dy_n} \right)\,.
\end{equation}
In these new coordinates, the Hamiltonian is 
\begin{equation}
\hat{\mathcal{Q}}^{(n)}_t = \frac{n}{2} \left[ \upsilon_{re} \left( - \left(\frac{d}{dx_n}\right)^2 + \left(\frac{d}{dy_n}\right)^2 + x_n^2 - y_n^2\right) + 2i  \upsilon_{im} \left( x_n \frac{d}{dy_n}  + y_n \frac{d}{dx_n} \right) \right]\,.
\end{equation}
We then make the following change of coordinates: 
\begin{equation}
\begin{split} \label{HyperboilicCoordinates}
x_n = w_n \cosh t_n , \quad\quad  y_n = w_n \sinh t_n\,,  \quad\quad (t_n,w_n) \in \mathbb{R}^2\,,\\
\end{split}
\end{equation}
covering the region $|x|>|y|$, shown in Figure \ref{fig:xygraphSdS}. In terms of $(w_n, t_n)$, our Hamiltonian is 
\begin{equation} \label{xgryH}
\hat{\mathcal{Q}}^{(n)}_t = \frac{n}{2} \left[ \upsilon_{re} \left( - \frac{d^2}{dw_n^2 } - \frac{1}{w_n} \frac{d}{dw_n} + \frac{1}{w_n^2} \frac{d^2}{dt_n^2}+ w_n^2\right) + 2i  \upsilon_{im} \frac{d}{dt_n} \right]\,.
\end{equation}
For $|x|<|y|$ the Hamiltonian is given by
\begin{equation}
\hat{\mathcal{Q}}^{(n)}_t = - \frac{n}{2} \left[ \upsilon_{re} \left( - \frac{d^2}{dw_n^2 } - \frac{1}{w_n} \frac{d}{dw_n} + \frac{1}{w_n^2} \frac{d^2}{dt_n^2}+ w_n^2\right) + 2i  \upsilon_{im} \frac{d}{dt_n} \right]\,.
\end{equation}
We can then find the Schr\"{o}dinger equation for each mode. By expanding the wavefunction in a Fourier basis as 
\begin{equation}
\Psi (w_n,t_n) = \int_{\mathbb{R}} \frac{dp_n}{2\pi} \,  \psi_n (w_n) e^{i p_n t_n}, \quad\quad p_n \in \mathbb{R}\,,
\end{equation}
and taking $|x|>|y|$, we find
\begin{equation}
\frac{1}{2}\left( - \frac{d^2}{dw_n^2 } - \frac{1}{w_n} \frac{d }{dw_n} - \frac{p_n^2}{w_n^2} + w_n^2\right)\psi_n   = \frac{1}{\upsilon_{re} }\left(\frac{E_n}{n}+   \upsilon_{im} p_n \right)\psi_n .
\end{equation} 
To get this in the Shr\"{o}dinger form, we then make the substitution $\psi_n \rightarrow \frac{1}{\sqrt{w_n}} \psi_n$, and this becomes 
\begin{equation} \label{ComformalQuantumMechaics}
\frac{1}{2}\left(- \frac{d^2}{dw_n^2 } +w_n^2 - \frac{1/4 + p_n^2}{w_n^2}\right) \psi_n = \frac{1}{\upsilon_{re} } \left( \frac{E_n}{n} +  \upsilon_{im} p_n\right) \psi_n ,
\end{equation}
which is the conformal quantum mechanics problem studied in \cite{deAlfaro:1976vlx, Anous:2020nxu}. We need to patch this solution to the one for $|y| > |x|$ which can be found by exchanging $x_n \leftrightarrow y_n$ in (\ref{HyperboilicCoordinates}) and following the same steps. The result is a change of sign in the left-hand side of (\ref{ComformalQuantumMechaics}). 

\chapter{\texorpdfstring{$SL(2,\mathbb{C})$}{SL(2,C)} WZW central charge}
\label{centralcharge}
To calculate the current algebra and central charge of the chiral $SL(2,\mathbb{C})$ WZW model (\ref{WZWsl2C}), we first study the non-chiral WZW model and then take the anti-holomorphic sector to describe the current algebra of the chiral theory. This is because the chiral theory is invariant under 
\begin{equation}
	g(z,\bar{z}) \rightarrow g(z,\bar{z}) \bar{\Omega}^{-1}(\bar{z}) \,, 
\end{equation}
where $(z,\bar{z})$ are coordinates on the complex plane. The $SL(2,\mathbb{C})$ WZW model is given by 
\begin{equation}
	S = \frac{k + i \lambda}{8\pi} \int d^2x \, \text{Tr} \left( \partial^\mu g^{-1} \partial_\mu g \right) - \frac{i (k + i \lambda)}{12 \pi} \int d^3 y \, \varepsilon^{\mu \nu \rho} \, \text{Tr} \left( g^{-1} \partial_\mu g g^{-1} \partial_\nu g g^{-1} \partial_\rho g\right) + \text{c.c.} \,,
\end{equation}
where $g \in SL(2,\mathbb{C})$. By computing the equations of motion for this action, we find the following conserved currents 
\begin{equation}
	\begin{split}
		J_z = - \tfrac{1}{2} (k + i \lambda ) \partial_z g g^{-1} \,, &\qquad  J_{\bar{z}} = \tfrac{1}{2} (k + i \lambda ) g^{-1} \partial_{\bar{z}} g \,,\\
		\tilde{J}_{\bar{z}} = - \tfrac{1}{2} (k - i \lambda ) \partial_{\bar{z}} \bar{g} \bar{g}^{-1}\, , &\qquad  \tilde{J}_{z} = \tfrac{1}{2} (k - i \lambda ) \bar{g}^{-1} \partial_{z} \bar{g} \,,
	\end{split}
\end{equation}
Here the subscripts label which currents are holomorphic/ anti-holomorphic. We can calculate the current algebra (using the procedure described for instance in \cite{DiFrancesco:1997nk}) and find
\begin{equation}
	\begin{split}
		\left[ J^a_{z,n}, J^b_{z,m} \right] &= \tfrac{1}{2}(k + i \lambda ) n \delta_{ab} \delta_{n + m, 0} + \sum_c i f_{abc} J^c_{z,n+m} \,,\\
		\left[ J^a_{\bar{z},n}, J^b_{\bar{z},m} \right] &= \tfrac{1}{2} (k + i \lambda ) n \delta_{ab} \delta_{n + m, 0} - \sum_c i f_{abc} J^c_{\bar{z},n+m}\,, \\
		\left[ \tilde{J}^{a}_{\bar{z},n}, \tilde{J}^{b}_{\bar{z},m} \right] &= \tfrac{1}{2} (k - i \lambda ) n \delta_{ab} \delta_{n + m, 0} - \sum_c i f_{abc} \tilde{J}^{c}_{\bar{z},n+m} \,,\\
		\left[ \tilde{J}^{a}_{z,n}, \tilde{J}^{b}_{z,m} \right] &= \tfrac{1}{2} (k - i \lambda ) n \delta_{ab} \delta_{n + m, 0} + \sum_c i f_{abc} \tilde{J}^{c}_{z,n+m} \,.
	\end{split}
\end{equation}
The anti-holomorphic sector is then given by the second and third of these. Following the Sugawara construction, we find the anti-holomorphic energy-momentum tensor
\begin{equation}
	\bar{T}(\bar{z}) = \frac{1}{k + i \lambda +2} \sum_a (J^a_{\bar{z}} J^a_{\bar{z}} ) (\bar{z}) + \frac{1}{k - i \lambda +2} \sum_a (\tilde{J}^a_{\bar{z}} \tilde{J}^a_{\bar{z}} ) ({\bar{z}}) \,,
\end{equation}
where we have used $f_{abc} f_{cbd} = \epsilon_{abc} \epsilon_{cbd} = 2 \delta_{ad}$. From the OPE
\begin{equation}
	\bar{T}(z) \bar{T}(w) =  \left(\frac{3(k+i \lambda)}{2(k + i \lambda + 2)} + \frac{3(k-i \lambda) }{2(k - i \lambda + 2)} \right)  \frac{1}{(\bar{z}-\bar{w})^4} + \frac{2\bar{T}(\bar{w})}{(\bar{z}-\bar{w})^2} + \frac{\partial \bar{T}(\bar{w})}{\bar{z}-\bar{w}}\,,
\end{equation}
we can find the central charge 
\begin{equation}
	c_{{SL}(2,\mathbb{C})} = \frac{3(k+i\lambda)}{(k+i\lambda+2)} +\frac{3(k-i\lambda)}{(k-i\lambda+2)} = \frac{6 \left(k (k+2) + \lambda ^2 \right)}{(k+2)^2+\lambda ^2}\,. 
\end{equation}

\end{appendices}

\bibliography{bibliography}{}
\bibliographystyle{unsrturl}

\end{document}